\documentclass[aps,rmp,twocolumn,superscriptaddress,floatfix]{revtex4-1}
\pdfoutput=1
\usepackage{epsfig}
\usepackage{graphicx}
\usepackage{amsmath}
\usepackage{amssymb}
\usepackage{color}
\usepackage{amscd}
\usepackage{bm}  
\usepackage{enumerate}
\usepackage{mathtools}
\usepackage{amscd}
\usepackage{epstopdf}
\usepackage{verbatim}
\usepackage{xspace}
\usepackage{multirow}
\usepackage{hyperref}

\def\nabvec{\mbox{\boldmath $\nabla$}}

\newcommand{\beq}{\begin{equation}}
\newcommand{\eeq}{\end{equation}}
\newcommand{\beqa}{\begin{eqnarray}}
\newcommand{\eeqa}{\end{eqnarray}}
\newcommand{\bea}{\begin{eqnarray}}
\newcommand{\eea}{\end{eqnarray}}
\newcommand{\bem}{\begin{math}}
\newcommand{\eem}{\end{math}}

\newcommand{\bsf}[1]{\textsf{\textbf{#1}}}

\newcommand{\qo}{{{\bf q},\omega}}
\newcommand{\xhat}{{\bf \hat{x}}}
\newcommand{\yhat}{{\bf \hat{y}}}

\newcommand{\Tr}{\mbox{Tr}}

\newcommand{\bfr}{{\bf r}}

\newcommand{\bfp}{{\bf p}}
\newcommand{\bfq}{{\bf q}}

\newcommand{\bfv}{{\bf v}}

\newcommand{\bnabla}{{\bm \nabla}}

\newcommand{\bdelta}{{\bm \delta}}

\newcommand{\bTheta}{{\bm \Theta}}
\newcommand{\bomega}{{\bm \omega}}
\newcommand{\bnu}{\hat{\bm \nu}}

\newcommand{\aver}[1]{\left\langle {#1}\right\rangle}

\begin{document}
\title{Soft active matter}

\author{M.C. Marchetti}
\affiliation{Physics Department and Syracuse Biomaterials Institute
Syracuse University Syracuse NY 13244 USA}
\author{J.F. Joanny}
\affiliation{Physicochimie Curie (CNRS-UMR168 and Universit\'e Pierre et Marie Curie) Institut Curie Section de Recherche 26 rue d'Ulm 75248 Paris Cedex 05 France}
\author{S. Ramaswamy}
\affiliation{Department of Physics Indian Institute of Science Bangalore 560
12 India}
\affiliation{TIFR Centre for Interdisciplinary Sciences 
21 Brundavan Colony Narsingi
Hyderabad 500 075 India}
\author{T.B. Liverpool}
\affiliation{Department of Mathematics University of Bristol Bristol BS8
1TW UK}
\author{J. Prost}
\affiliation{Physicochimie Curie (CNRS-UMR168) Institut Curie Section
de Recherche
26 rue d'Ulm 75248 Paris Cedex 05 France}
\affiliation{E.S.P.C.I 10 rue Vauquelin 75231 Paris Cedex 05 France}
\author{Madan Rao}
\affiliation{Raman Research Institute Bangalore 560 080 India}
\affiliation{National Centre for Biological Sciences (TIFR) Bangalore
560065 India}
\author{R. Aditi Simha}
\affiliation{Department of Physics Indian Institute of Technology Madras
Chennai 600 036 India}

\date{\today}

\begin{abstract}
In this review we summarize theoretical progress in the field of active matter, placing it in the context of recent experiments. Our approach offers a unified framework for the mechanical and statistical properties of living matter: biofilaments and molecular motors in vitro or in vivo, collections of motile microorganisms, animal flocks, and chemical or mechanical imitations. A major goal of the review is to integrate the several approaches proposed in the literature, from semi-microscopic to phenomenological. In particular, we first consider ÒdryÓ systems, defined as those where momentum is not conserved due to friction with a substrate or an embedding porous medium, and clarify the differences and similarities between two types of orientationally ordered states, the nematic and the polar. We then consider the active hydrodynamics of a suspension, and relate as well as contrast it with the dry case. We further highlight various large-scale instabilities of these nonequilibrium states of matter. We discuss and connect various semi-microscopic derivations of the continuum theory, highlighting the unifying  and generic nature of the continuum model. Throughout the review, we discuss the experimental relevance of these theories for describing bacterial swarms and suspensions, the cytoskeleton of living cells, and vibrated granular materials. We suggest promising extensions towards greater realism in specific contexts from cell biology to ethology, and remark on some exotic active-matter analogues. Lastly, we summarize the outlook for a quantitative understanding of active matter, through the interplay of detailed theory with controlled experiments on simplified systems, with living or artificial constituents.  
\end{abstract}

\date{\today}
\maketitle
\tableofcontents

\section{Introduction}
\label{sec:introduction}

\begin{figure}[h]
\centering
\includegraphics[width=0.6\columnwidth]{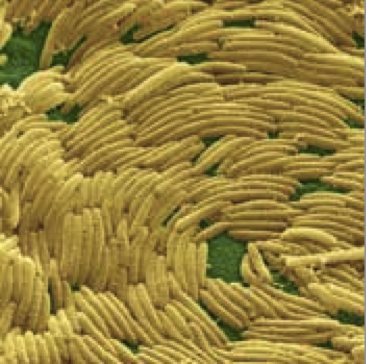}
\caption{(color online) Liquid-crystalline order in a myxobacterial flock.
Image courtesy of Gregory Velicer (Indiana University Bloomington) and Juergen
Bergen (Max-Planck Institute for Developmental Biology).}
\label{fig:myxo}
\end{figure}

\begin{figure}
\centering
\includegraphics[width=0.99\columnwidth]{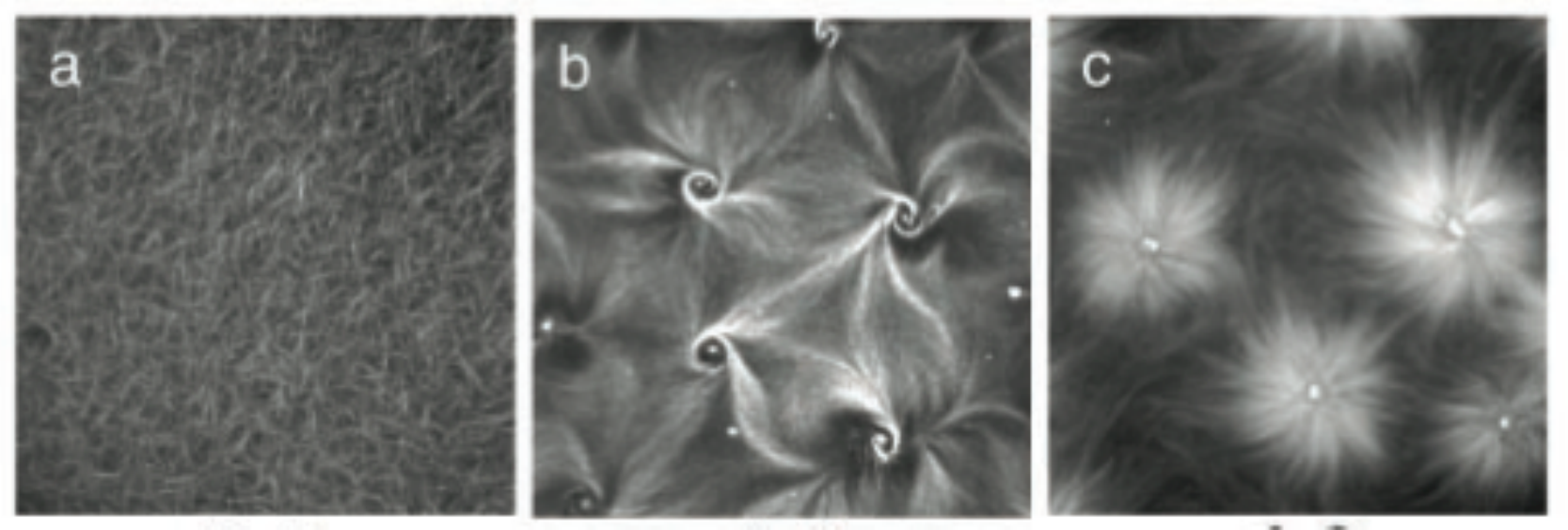}
\caption{ Patterns organized in vitro by the action of multimeric
kinesin
complexes on microtubules, imaged by dark field microscopy.
The concentration of motor proteins increases from left to right. Image (a)
shows a disordered array of microtubules. The other two images display
motor-induced organization in
spiral (b) and aster (c) patterns. The bright spots in the images correspond to
the minus end of microtubules. These remarkable experiments from
\textcite{Surrey2001} led the way to the
study of pattern formation in active systems. Adapted with permission from
\textcite{Surrey2001}.}
\label{fig:Surrey}
\end{figure}

\begin{figure}
\centering
\includegraphics[width=0.7\columnwidth]{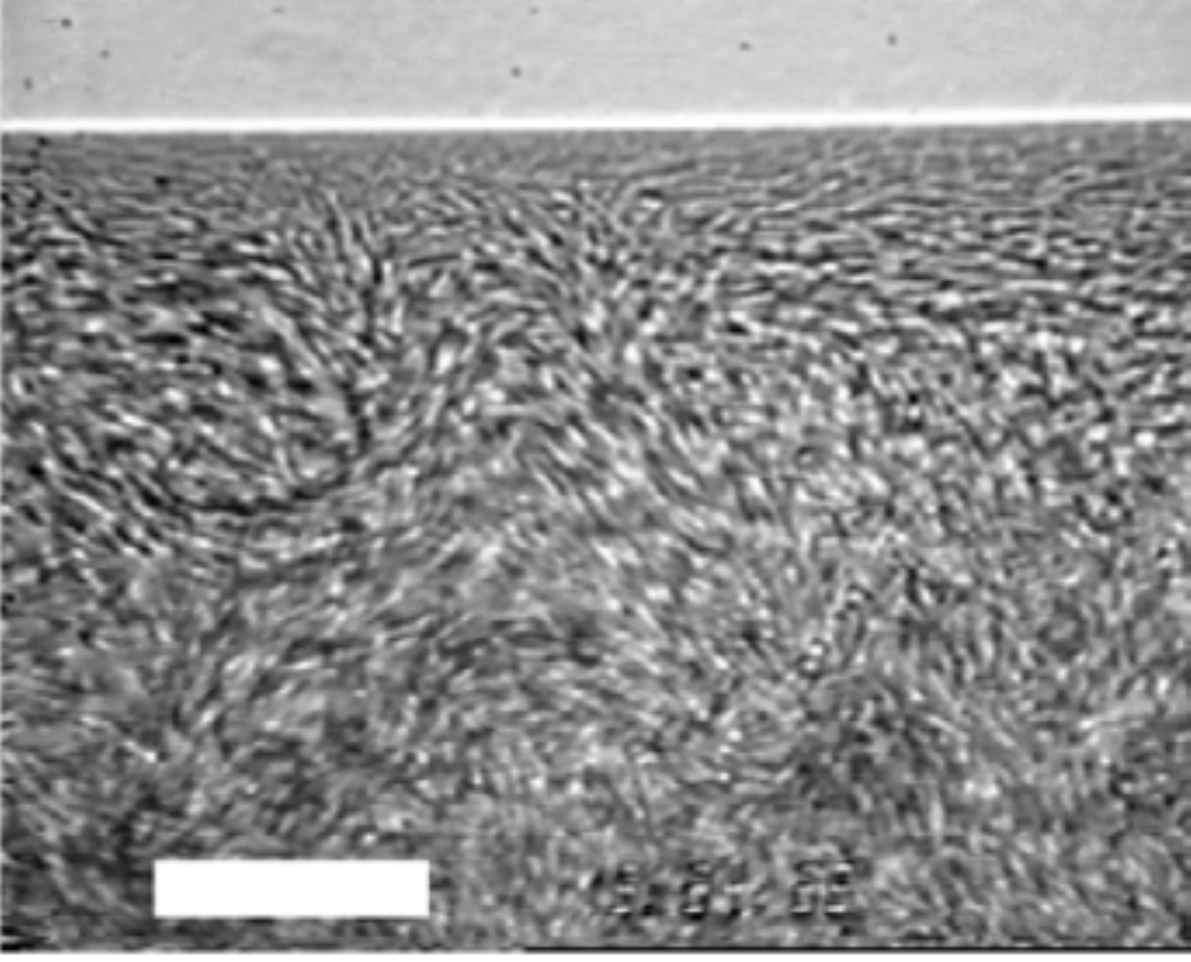}
\caption{Bacterial ``turbulence" in a sessile drop of Bacillus subtilis viewed
from
below through the bottom of a petri dish.
Gravity is perpendicular to the plane of the picture, and the horizontal white
line near the top is the air-water-plastic contact line. 
The central fuzziness is due to collective motion, not quite captured at the
frame rate of $1/30 s$. The scale bar is $35\mu m$. 
Adapted with permission from \textcite{Dombrowski2004}}
\label{fig:Dombrowski}
\end{figure}

\begin{figure}
\centering
\includegraphics[width=0.99\columnwidth]{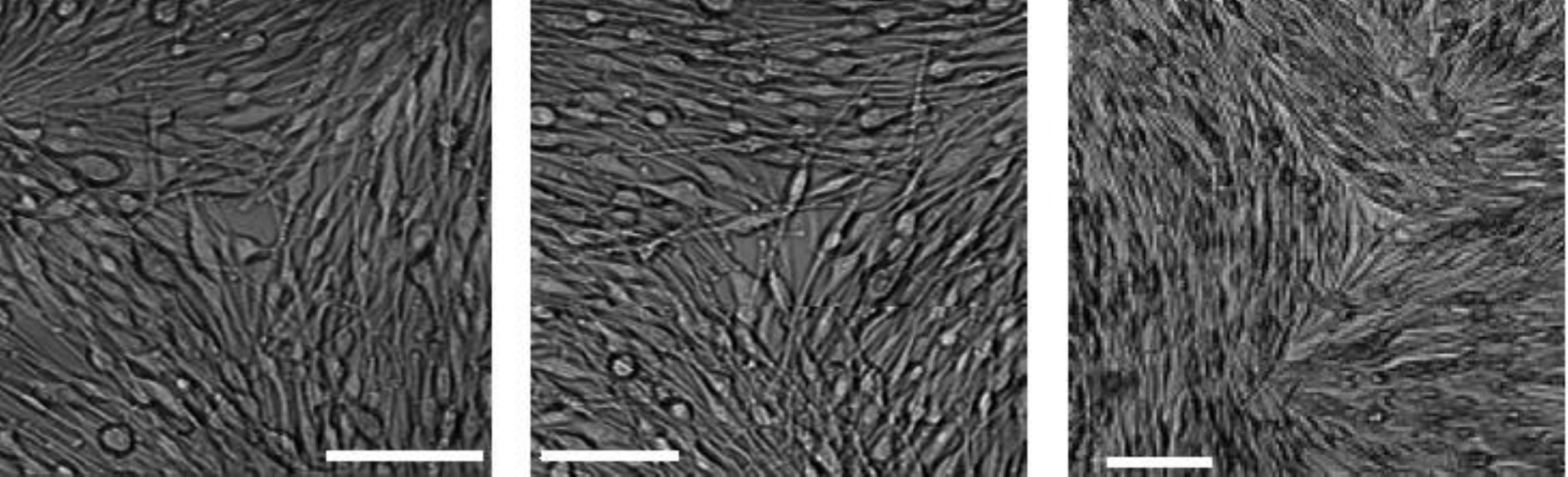}
\caption{A disclination defect of strength $m=-1/2$  formed by human melanocytes on a plastic
surface. The bars are $100\mu m$. The three images show three different
situations: (i) the core of the defect is an area free of cells (left); (ii) the
core of the disclination is an area with isotropically distributed cells
(center); and (iii) the core of the defect is occupied by a star-shaped cell
(right). The cells that form the nematoid order are in an elongated bipolar
state. Reproduced with permission from \textcite{Kemkemer2000}.}
\label{fig:melano}
\end{figure}

\begin{figure}
\centering
\includegraphics[width=0.8\columnwidth]{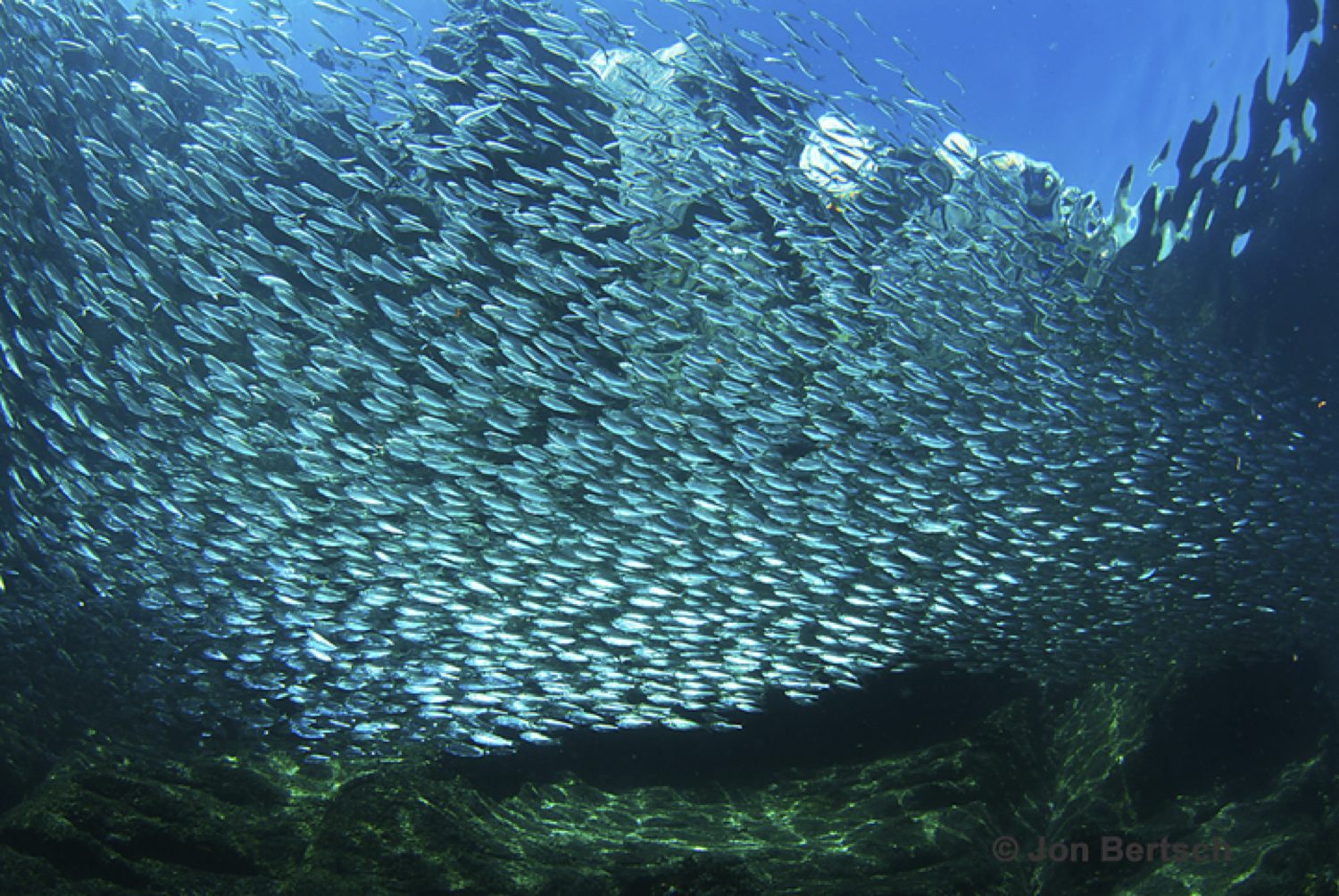}
\caption{(color online)  A remarkable demonstration of polar order in a sardine
school (Courtesy of Jon Bertsch,
from underwater images form the Sea of Cortez:
http://www.thalassagraphics.com/blog/?p=167).}
\label{fig:sardines}
\end{figure}

\begin{figure}
\centering
\includegraphics[width=0.7\columnwidth]{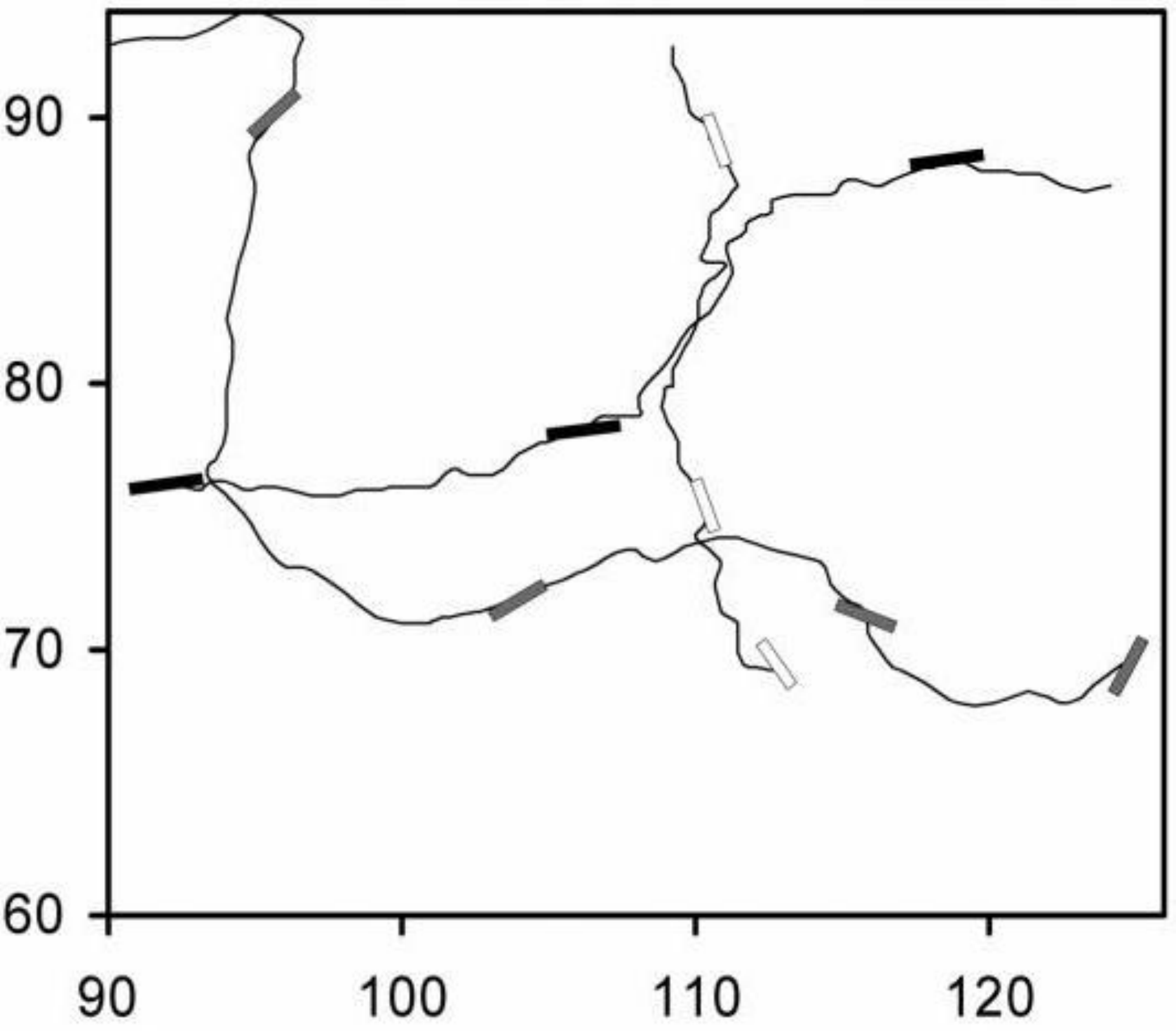}
\caption{(color online) Colloids that swim: Hydrogen peroxide is oxidized to
generate protons in solution and electrons in the wire on the platinum end of
micron-sized gold-platinum wires. The protons and electrons are then consumed
with the reduction of $H_2O_2$ on the gold end. The resulting ion flux induces
an electric field and motion of the particle relative to the fluid, propelling
the particle towards the platinum end at speeds of
tens of microns a second. On longer timescales the rods tumble due to thermal or
other noise; the resulting motion is thus a persistent random walk.  Reprinted
with permission from \textcite{Paxton2004}.}
\label{fig:sen}
\end{figure}

The goal of this article is to introduce the reader to a general framework and 
viewpoint for the study of the mechanical and statistical properties of living
matter and of some remarkable non-living imitations, on length scales from
sub-cellular to oceanic. The ubiquitous nonequilibrium condensed systems that
this review is concerned with
~\citep{Toner2005,Julicher2007,Joanny2009,Vicsek2012} have
come to be known
as \textit{active matter}
~\citep{Ramaswamy2010}.
Their unifying characteristic is that they are composed of self-driven units
-- active particles -- each capable of converting stored or ambient free energy
into systematic movement \citep{Schweitzer2003}. The
interaction of active particles with each other
and with the medium they live in gives rise to highly correlated collective
motion and mechanical stress. Active particles are generally elongated and their
direction of self-propulsion is set by their own anisotropy, rather than fixed
by an external field. Orientational order is thus a theme that runs through much
of the active-matter narrative, as can be seen for instance in the image
of a swarm of myxobacteria, shown in Fig. \ref{fig:myxo}.
The biological systems of our interest include
\textit{in vitro} mixtures of cell extracts with bio-filaments and associated
motor proteins (Fig.~\ref{fig:Surrey}), the whole cytoskeleton of living cells,
bacterial suspensions (Fig.~\ref{fig:Dombrowski}),
cell layers (Fig.~\ref{fig:melano}), and terrestrial, aquatic (Fig.~\ref{fig:sardines}) and aerial flocks.
Non-living active matter arises in layers of vibrated granular rods, colloidal
or
nanoscale particles propelled through a fluid by catalytic activity at their
surface (Fig. \ref{fig:sen}), and collections of robots.
A distinctive -- indeed, defining -- feature of active systems compared to
more familiar nonequilibrium systems is the fact that the energy input that
drives the system out of equilibrium is local, at the level of each particle,
rather than at the system's boundaries as in a shear flow, for example. 
Each
active particle consumes and dissipates energy going through a cycle that fuels
internal changes, generally leading to motion. Active systems exhibit a wealth
of intriguing nonequilibrium properties, including emergent structures with
collective behavior qualitatively different from that of the individual
constituents, bizarre fluctuation statistics, nonequilibrium order-disorder
transitions,  pattern formation on mesoscopic scales, unusual
mechanical and rheological properties, and wave propagation and sustained
oscillations even in the absence of inertia in the strict sense. 

Living systems of course provide the preeminent example of active matter, 
exhibiting extraordinary properties such as reproduction, adaptation,
spontaneous motion, and dynamical organization including the ability to
generate and to respond in a calibrated manner to forces. 

A theoretical
description of the general properties of living matter is not currently
achievable because of its overall complexity, with the detailed state of a cell
determined by a hopelessly large number of variables. However, in a given living
organism there are at most 300 different cell types, which an optimist could
view as a very small number given the immensity of the accessible parameter
space. Perhaps, therefore, global principles such as conservation laws and
symmetries constrain the possible dynamical behaviors of cells or, indeed, of
organisms and populations, such as collections of bacteria
(Fig. \ref{fig:Dombrowski}), fish schools (Fig. \ref{fig:sardines})
and bird flocks. Quantifying the spontaneous dynamical
organization and motion of living systems is a first step toward understanding
in a generic way some of these principles, by focusing on specific questions
that are accessible to theory. This has proved to be the case for the long
wavelength behavior of active membranes  \citep{Prost1996,
Ramaswamy2000,Rao2001,Manneville2001,Ramaswamy2001};
the general theory of flocking \citep{Toner1995,Toner1998},
\citep{Toner2005}; 
or the
macroscopic mechanical properties of the cytoskeleton as an active gel
\citep{Kruse2005,Julicher2007}.
Agent-based models offer a minimal approach to the study of active systems,
with an emphasis on order and fluctuations rather than forces and mechanics. In
such 
a setting, seminal studies of flocking as a phase transition were first carried
out by~\textcite{Vicsek1995} [see also the remarkable computer-animation work
of~\textcite{Reynolds1987}], with important modifications and
extensions
by~\textcite{Gregoire2004} and~\textcite{Chate2006,Chate2008}.
These models
describe point particles with fixed speed moving on an inert
background. The direction of motion changes according to a
noisy local rule that
requires particles to align with their neighbors at each time step. This family
of models displays a well-defined transition from a disordered to an ordered
phase with decreasing noise strength or increasing density. In the
context of the cytoskeleton activated by motor proteins,
detailed simulations of
ensembles of semi-flexible filaments on which motor bundles can exert force
dipoles have also been carried out by several
authors \citep{Mogilner1996,Nedelec1997,Pinot2009,Head2011}.

\begin{figure}
\centering
\includegraphics[width=0.99\columnwidth]{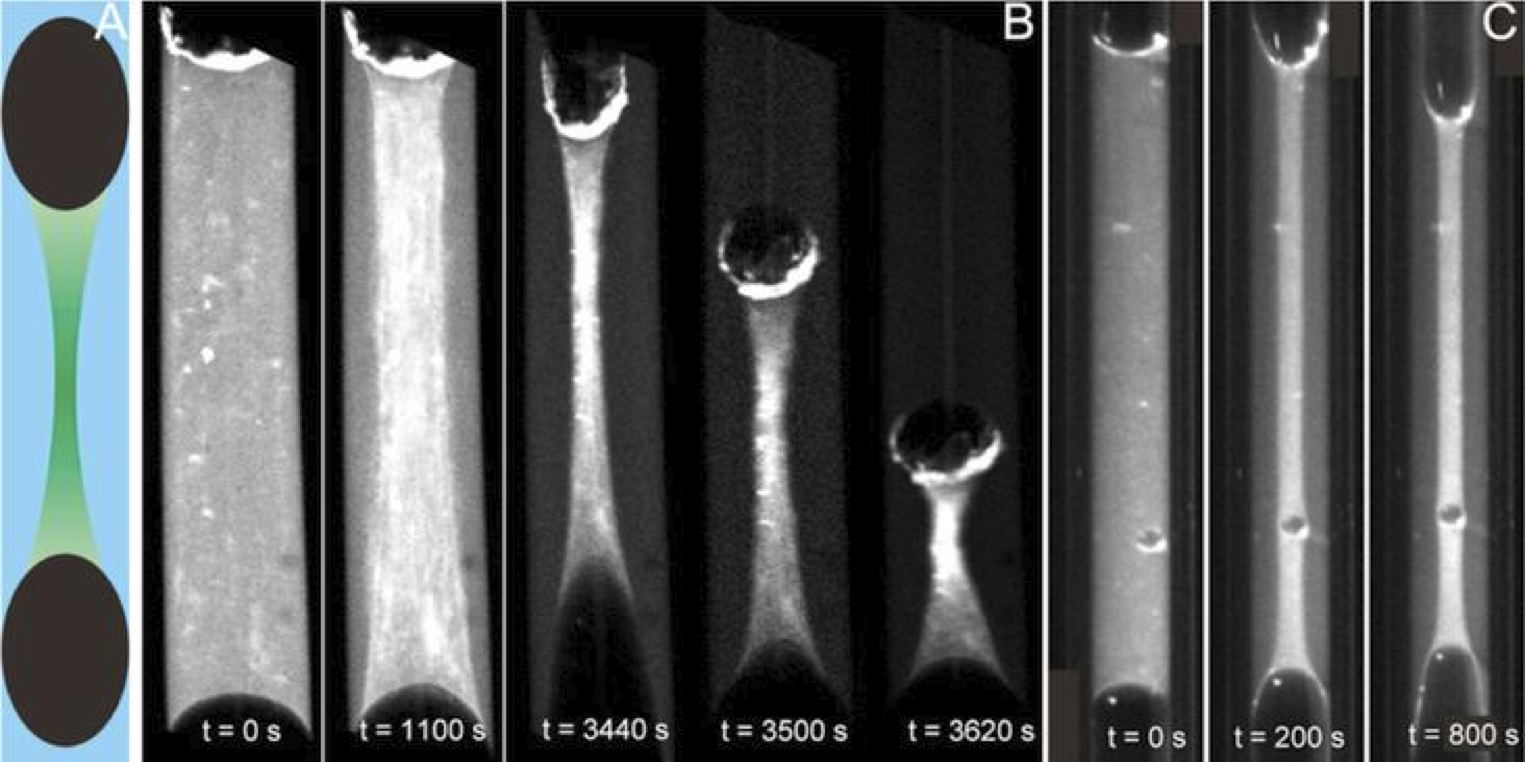}
\caption{(color online) Measurement of contractile stresses in cell (Xenopus
oocyte) extracts and a reconstituted acto-myosin gel made using skeletal muscle
myosin II thick filaments: (a) Schematic of active gel in a narrow glass
capillary (inner diameter, $400 \mu m$) sandwiched between two oil droplets. (b)
Confocal fluorescence images of the network at five different time points shows
that the active gel first pulls away from the capillary walls and deforms the
oil droplets as it contracts before reaching a breaking point when the gel
collapses. ( c) Dark-field images of contracting cell extract shows the same
behaviour. Analysis of the deformation of the oil droplet gives an estimate of
the contractile force which is around $1\mu N$ or $100 pN$ per actin bundle.
Adapted with permission from \textcite{Bendix2008}.}
\label{fig:contraction}
\end{figure}

In this article we will not review the agent-based models and the wealth of
results obtained by numerical simulations, but rather focus on identifying
generic aspects 
of the large scale behavior of active systems and characterizing their material
properties.
Many of the macroscopic properties of active systems are universal in the sense 
that systems operating at widely differing length scales, with significant
differences in their detailed dynamics at the microscopic level, display broadly similar properties. Visually similar
flocking phenomena are seen in fish shoals \citep{Parrish1997} and collections
of keratocytes \citep{Szabo2006}. Contractile stresses are evident
on a sub-cellular scale in the cytoskeleton (Fig.~\ref{fig:contraction})
\citep{Bendix2008,Joanny2009}, as well as on a
scale of many cells in swimming algae \citep{Rafai2010}.
The hope is to be able
to classify active matter in a small number of universality classes,
based on considerations of symmetry and conservation laws, 
each with a well-defined macroscopic behavior. We consider here four
classes of active matter according to the nature of the broken symmetry
of the ordered phase and the type of momentum damping. First the broken
symmetry: elongated self-propelled objects are in general polar entities with
distinct heads and tails, which can cooperatively order either in a polar
(ferromagnetic) phase
or in a nematic phase. In a polar phase, all the microscopic objects are on
average aligned in the same direction; this is the case for bacteria or fish.
The polar order is described by a vector order parameter ${\bf p}$,
known as the polarization. Nematic ordering can be obtained in two ways, either
in systems where polar self-propelled objects are parallel but with
random
head-tail orientations or in systems where the self-propelled particles are
themselves head-tail symmetric such as the melanocytes that distribute pigment
in the skin. The polar/apolar distinction in the context of living matter is
brought out clearly by~\textcite{Gruler1999} and~\textcite{Kemkemer2000}, as
shown in 
Fig. \ref{fig:melano}.
Nematic order
is described by a tensor order parameter, ${\bsf{Q}}$, known as the alignment tensor. Next, the nature of damping: in a bulk fluid, active or
otherwise, viscosity damps the relative motion of neighboring regions, which
means that the \textit{total} momentum of the system is conserved. In systems in
contact with a substrate, e.g., restricted to lie on a surface or between
two closely spaced walls, or moving through a porous medium, the
drag due to the substrate dominates, either simply through no-slip or through
binding and unbinding of constituents of the systems to the confining
medium,
and 
the momentum of the fluid is not conserved.  Table~\ref{Table:classification}
summarizes
 various examples of active systems classified according
to symmetry and presence/absence of momentum conservation (wet/dry). Of course even in
``dry'' systems, where
momentum is not conserved, hydrodynamic or
medium-mediated interactions may in some cases be important,
depending on the length scale of interest~\citep{Schaller2011a}.
For instance, for active particles in a viscous fluid of viscosity $\eta$ and
experiencing a frictional drag $\gamma$, hydrodynamic flows 
can be neglected only on
length scales larger than $\sqrt{\eta/\gamma}$. 
\begin{table}
\label{Table:classification}
\centering
\begin{ruledtabular}
\begin{tabular}{|c|p{3.5cm}|p{3.5cm}|}
& nematic & polar\\ \hline
 & melanocytes  & migrating animal herds\\
dry& vibrated granular rods &  migrating cell layers\\
& & vibrated asymmetric granular particles\\ \hline
 &  & cell cytoskeleton\\
& bipolar catalytic rods in suspension& cytoskeletal extract in bulk
suspension\\
wet& & swimming bacteria in bulk\\
& & Au-Pt catalytic colloids
\end{tabular}
\end{ruledtabular}
\caption{Examples of active systems classified according
to symmetry and presence/absence of momentum conservation (wet/dry).}
\end{table}

A useful theoretical framework to describe the macroscopic properties of active
matter is provided 
by the methods of nonequilibrium statistical mechanics. In a generalized
hydrodynamic approach, a coarse-grained description of the large-scale,
long-time 
behavior of the system is given in terms of a small number of continuum fields. 
The evolution of these fields is written in terms of a set of continuum or
hydrodynamic equations that  modify the well-known liquid crystal hydrodynamics
\citep{Gennes1993,Martin1972} to include new nonequilibrium terms that arise
from the activity. Generalized hydrodynamic theories have been very
successful in the description of many condensed matter systems, such as
superfluids
\citep{Dzyaloshinskii1980}, liquid crystals \citep{Martin1972}, polymers
\citep{Milner1993}, as well as of course simple fluids.

One approach to obtain a hydrodynamic theory of active systems
is to start from a microscopic model and use the tools of statistical physics
to coarse-grain the model and obtain the long-wavelength, long-time 
scale equations
\citep{Liverpool2003,Ahmadi2005,Ahmadi2006,Aranson2005,Kruse2003}.
This task is difficult if the microscopic description is realistic and it can
only 
be carried out at the cost of approximations, such as low density or weak
interactions. It allows one to relate the parameters in the macroscopic
equations to specific physical mechanisms (albeit in a model-dependent manner)
and to estimate them in terms of experimentally accessible quantities. The
low-density limit has been worked out for several models,
initially without and later including the effect of the embedding solvent. Some
specific examples are discussed in section \ref{microscopic}.

An alternative, more pragmatic, approach is to directly write hydrodynamic
equations for the macroscopic fields including all terms allowed by symmetry, as
was pioneered for dry flocks by
\textcite{Toner1995,Toner1998}, and extended to the case of self-propelled
particles suspended in a fluid by \textcite{Simha2002} and more generally
to active-filament solutions \citep{Hatwalne2004}. As one might expect in
hindsight, novel terms, of a
form ruled out for thermal equilibrium systems, appear in these
``pure-thought'' versions of the equations of active hydrodynamics. 

A more systematic implementation of the phenomenological hydrodynamic approach
is to treat the nonequilibrium steady state of an active system as arising
through the imposition of a non-vanishing but small driving force on a
well-defined parent \textit{thermal equilibrium} state whose existence is not in
question~\citep{Groot1984}. For example,  in the biological case of systems composed of
cytoskeletal filaments and motor proteins, 
such as the cell cytoskeleton, the driving force is the difference $\Delta
\mu$ between ATP and its hydrolysis products. If $\Delta \mu$
is assumed to be
small, the macroscopic hydrodynamic equations can be derived in a systematic way
following the Onsager procedure. One identifies thermodynamic fluxes and forces
and writes the most general linear relation between them that respects the
symmetries of the problem. The novel terms mentioned above that arose by 
directly writing down the hydrodynamic equations of active matter are then
seen to be a consequence of off-diagonal Onsager coefficients and an imposed
constant nonzero $\Delta \mu$. Generalized hydrodynamic theories have been very
successful in the description of complex and simple  fluids  \citep{Martin1972}.
The advantage is that the
equations are expanded around a well defined state. The drawback is 
that many active systems,  biological ones being particularly pertinent
examples,  are far from equilibrium and 
one might for example miss some important physics by the restriction to
the linear nonequilibrium regime. This phenomenological approach can, however,
be very useful, especially when coupled to microscopic derivations for specific
model systems.

Regardless of the choice of hydrodynamic framework, the reasons for the choice
of \textit{variables} remain the same. When an extended system is disturbed
from equilibrium by an external perturbation, its relaxation is controlled by
the microscopic interactions among constituents. It is useful to divide the
relaxation processes into fast and slow, and to build a theory of the slow
dynamics in which the fast processes enter as noise and damping.  Such a
division is unambiguous when there are collective excitations with relaxation
rates $\omega(q)$ that vanish as the wavevector $q$ goes to zero; these are the
hydrodynamic modes of the systems \citep{Martin1972,Forster1975}. Familiar
examples are diffusion and sound waves in fluids. A formulation in terms of
hydrodynamic fields provides a generic description of the nonequilibrium large
scale physics that relies only on general properties and local thermodynamics,
and does not depend strongly on microscopic details of the interactions. In
order to build a hydrodynamic theory, whether phenomenologically or by deriving
it from a microscopic model, the  first task is to identify the slow variables,
which are the local densities of conserved quantities, the ``broken-symmetry''
variables which have no restoring force at zero wavenumber, and,  in the
vicinity of a continuous phase transition, the amplitude of the order parameter
\citep{Martin1972,Forster1975}. From there on, the procedure is well defined
and systematic. Conserved quantities are fairly easy to identify. Momentum
conservation for ``wet'' systems means that the momentum density is a slow,
conserved
variable, while for ``dry'' systems it is a fast variable. Identifying the
type of order -- polar or nematic, in the cases that we will examine-- yields
both
the amplitude of the order parameter and the broken-symmetry modes.

Several useful reviews on various aspects of the behavior of
flocks and
active systems have appeared in the recent literature
\citep{Toner2005,Julicher2007,Joanny2009,Ramaswamy2010,Vicsek2012}, as well as on the
properties of swimmers~\citep{Lauga2009,Ishikawa2009,Koch2011} and bacterial
suspensions and
colonies~\citep{Ben-Jacob2000,Murray2003,Cates2012}. The present review aims
at highlighting the unity of various approaches and
at providing a classification of  active systems. It demonstrates the link
between microscopic derivations and continuum models, showing that the
hydrodynamic equations derived 
from different microscopic models with the same general symmetry have the same
structure and only differ in the details of parameter values, and that the
continuum models proposed and used in the literature can all be formulated in a
unified manner.
 It is hoped that this review will provide 
a useful and self-contained starting point for new researchers entering the
field. 

This review is organized as follows. We first consider dry active systems
focusing on the ordering transition, the properties of the ordered phase
and the differences between polar and nematic active matter. We also discuss
in section \ref{dry} a system first analyzed 
theoretically by Baskaran and Marchetti
\citep{Baskaran2008,Baskaran2008a} where the self-propelled
particles are polar in their movement but nematic in their interaction and hence
in the nature of their macroscopic order. In section \ref{wet}, we discuss
orientable active particles suspended in a fluid, or active gels, for short. We
give a systematic derivation of the constitutive equations of a nematic active
gel, for a system weakly out of equilibrium, and dwell briefly on the effects of
polar order, viscoelasticity and the presence of multiple components. In section
\ref{applications}, we present some applications of the hydrodynamic theory of
active matter to specific geometries that could arise in experiments. We focus
on instabilities of thin films and on the  rheological properties of active
matter. Section \ref{microscopic} gives a brief account of microscopic theories
of active matter. We offer examples to show how microscopic
theories allow one to determine in principle the phenomenological transport
parameters introduced in the hydrodynamic theories and to estimate
their order of magnitude. The concluding section \ref{conclusions}
presents open questions related to hydrodynamic theories of active matter and
gives some perspectives.

\section{``Dry" Active Matter}
\label{dry}
In this section we consider active systems with
no momentum conservation. This class includes bacteria gliding on a rigid
surface~\citep{Wolgemuth2002}, animal herds on land \citep{Toner1998} or, in the
artificial realm, vibrated granular particles on a plate
\citep{Ramaswamy2003,Yamada2003,Aranson2006,Narayan2007,
Kudrolli2008,Deseigne2010}
in all of which momentum is damped by friction with the substrate. It is also
plausible  that models without an explicit
ambient fluid contain the main physics of concentrated collections of
swimming bacteria~\cite{Drescher2011} and motor-filament suspensions~\cite{Liverpool2003}, 
where steric and stochastic effects could for most purposes
overwhelm hydrodynamic interactions. Minimal flocking models have also been
tested against observations on aerial displays by large groups of birds
where, despite the much lower concentration in comparison to bacterial
suspensions,  (presumably inertial) hydrodynamic effects  seem negligible  \citep{Ballerini2008,Ginelli2010}.

\begin{figure}
\centering
\includegraphics[width=0.9\columnwidth]{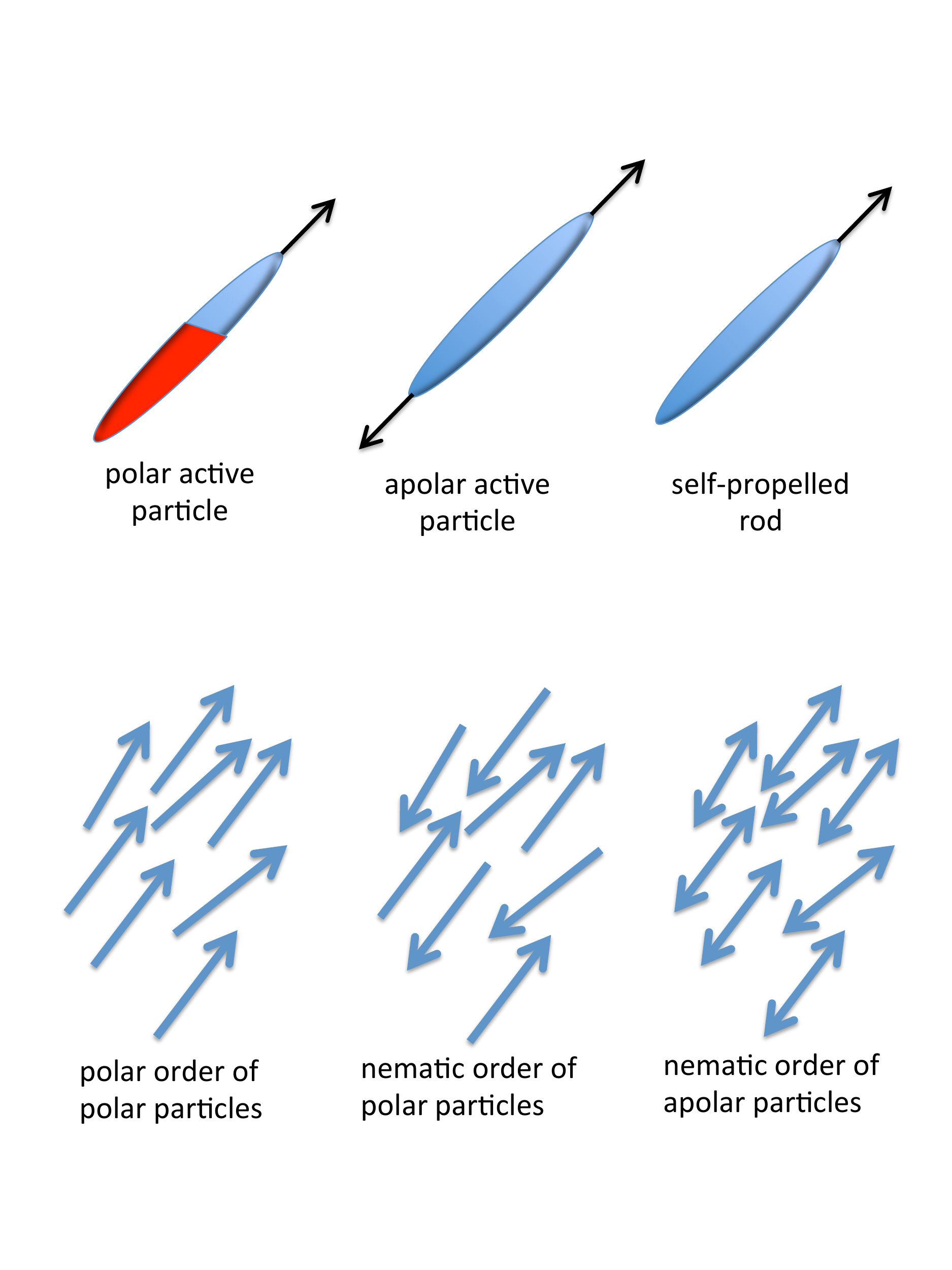}
\caption{(color online) Schematic of the various types of active particles and
orientationally 
ordered states. Polar active particles (top left image), such as bacteria or
birds, have a head
and a tail and are generally self-propelled along their long axis. They can
order in polar
states (bottom left) or nematic states (bottom center). The polar state is also
a moving 
state with a nonzero mean velocity. Apolar active particles (top center image)
are head-tail 
symmetric and can order in nematic states (bottom right). Self-propelled rods
(top right
image) are head-tail symmetric, but each rod is self-propelled in a specific
direction along its long axis. The self-propulsion renders the particles polar,
but for
exclusively apolar interactions (such as steric effects), self-propelled rods
can only order 
in nematic states (bottom center image). }
\label{fig:symmetry-cartoon}
\end{figure}

We refer to these systems as ``dry" active systems. In this case the only
conserved quantity is the number of particles (neglecting of course cell
division and death) and the associated hydrodynamic field is the local density
of active units. Over-damped active particles that can order in states with
polar
and nematic symmetries are described below. The first class consists of polar
or self-propelled units with interactions that tend to promote polar order,
i.e., explicitly align the particles head to head and tail to tail, as in the
classic Vicsek model~\citep{Vicsek1995}. The second class consists of active
particles that may themselves be apolar, such as melanocytes, the cells that
distribute pigments in the skin, where activity induces non-directed motion on
each cell~\citep{Gruler1999} or polar, such as self-propelled hard
rods~\citep{Baskaran2008,Peruani2006,Yang2010}, but with interactions that
tend to
align particles regardless of their polarity, so that the ordered state, if
present, has nematic symmetry.  A cartoon of the different cases is shown in
Fig.~\ref{fig:symmetry-cartoon}.

\subsection{Polar Active Systems: Toner and Tu continuum model of flocking}
\label{sec:polardry}
A continuum effective theory for the flocking
model of~\textcite{Vicsek1995} was proposed in 1995
by~\textcite{Toner1995,Toner1998} [see also~\citep{Toner2005}]. The Vicsek model
describes a collection of self-propelled particles with fixed
speed and noisy polar aligning interactions, and displays a
nonequilibrium phase
transition from a disordered state at low density or
high noise
to an ordered, coherently moving state at high density or low
noise strength. Toner and Tu formulated  the continuum model
phenomenologically solely on the
basis of symmetry considerations. Recently, the Toner-Tu
model was derived by~\textcite{Bertin2006,Bertin2009} and~\textcite{Ihle2011} by
coarse-graining
the microscopic Vicsek model. 
These derivations provide a microscopic basis for
the hydrodynamic theory, yielding explicit values for
essentially all the parameters in the Toner-Tu model, except for the noise
strength in the latter. The Boltzmann-equation approach of \textcite{Bertin2006}
and \textcite{Ihle2011} leads to a \textit{deterministic} coarse-grained
description, with the stochasticity of the Vicsek model reflected in an average
sense through the diffusion and relaxation terms. In this section we introduce
the continuum equations in
their simplest form and analyze their consequences. 

Since the particles are moving on a frictional substrate, the only conserved
field is the number density $\rho({\bf r},t)$ of active particles. In addition,
to describe the possibility of states with polar orientational order, one must
consider the dynamics of a polarization vector field, ${\bf p}({\bf r},t)$.
These continuum fields can be defined in terms of the position ${\bf r}_n(t)$
of each active particle and a unit vector $\hat{\bm{\nu}}_n(t)$
denoting the
instantaneous orientation of the velocity of each particle as
\begin{subequations}
\begin{gather}
\label{rho}\rho({\bf r},t)=
\sum_n\delta({\bf r}-{\bf r}_n(t))
\;,\\
\label{p}{\bf p}({\bf r},t)=\frac{1}{\rho({\bf r},t)}
\sum_n\hat{\bm\nu}_n(t)\delta({\bf r}-{\bf r}_n(t))
\;.
\end{gather}
\end{subequations}
Although we are dealing with a nonequilibrium system, it is convenient and
instructive to write the dynamical equations in a form in which terms
that can be viewed as arising from a free energy functional $F_p$ are separated
out: 
\begin{subequations}
\label{polar-eqs}
\begin{gather}
\label{rho-eq}\partial_t\rho=-\bm\nabla\cdot(v_0\rho{\bf p})\;,\\
\label{P-eq}\partial_t{\bf p}+\lambda_1({\bf p}\cdot\bm\nabla){\bf p}=-
\frac{1}{\gamma}
\frac{\delta F_p}{\delta{\bf p}}+{\bf f}\;,
\end{gather}
\end{subequations}
where $v_0$ is the self-propulsion speed of the active particles,
$\gamma$ a rotational viscosity,
and $\lambda_1$ a parameter controlling the
strength of the advective term on the left hand side of Eq.~\eqref{P-eq}. The
last term on the right hand side of Eq.~\eqref{P-eq} captures the fluctuations
and is taken to be white, Gaussian
noise, with zero mean and correlations
\begin{equation}
\langle f_\alpha({\bf r},t) f_\beta({\bf
r'},t')\rangle=2\Delta\delta_{\alpha\beta}\delta({\bf r}-{\bf r'})
\delta(t-t')\;.
\label{noise}
\end{equation}
For simplicity, and consistency with~\textcite{Toner1998} and with
the derivation of \textcite{Bertin2006} of the Toner-Tu equations from the model of 
\textcite{Vicsek1995}, we have neglected
the diffusive current and the associated noise in the density
equation in \eqref{polar-eqs}. In
\eqref{polar-eqs}  and everywhere in the following we have taken the noise terms
to be
purely additive, ignoring dependence on the local values of $\rho$
(see, e.g., \citep{Dean1996})
or of the order parameter (here ${\bf p}$). This approximation is adequate for
the present
purposes of calculating two-point correlators in a linearized theory, but is
actively under discussion \citep{Mishra2009,Mishra2012,Gowrishankar2012a} for situations of strong
inhomogeneity such as the coarsening of active nematics.

Note that in the continuum flocking model ${\bf p}$ plays a dual role: on
the one hand ${\bf p}$ is  the orientational order parameter of the system, on
the
other $v_0{\bf p}$ represents the particle velocity field. This duality is
crucial in determining the large scale behavior of this nonequilibrium system.
The free energy functional used in \eqref{polar-eqs} is given by
\begin{align}
\label{Fp}
{F_p}
=&  \int_{\bf r}\,\Big\{
\frac{\alpha(\rho)}{2}|{\bf p}|^{2}+\frac{\beta}{4}|{\bf p}|^{4}+\frac{ K}
{2}\left(\partial_\alpha p_\beta\right)\left(\partial_\alpha
p_\beta\right)\nonumber\\
&- v_1\bm\nabla\cdot{\bf
p}\frac{\delta\rho}{\rho_0}+\frac{\lambda}{2}|{\bf
p}|^{2} \bm\nabla\cdot{\bf p}
\Big\}\,.
\end{align}
where
$\rho_0$ is the average density and
$\delta \rho=\rho-\rho_0$.
The first two terms on the right hand side of Eq.~\eqref{Fp} control the
mean-field continuous order-disorder transition  that takes
place as the parameter $\alpha$ goes through zero. In the derivation of 
\textcite{Bertin2006}, $\alpha$ depends on local density $\rho$ and the noise
strength in the underlying microscopic model, and turns negative at sufficiently
large $\rho$. A reasonable phenomenological approach to
describe the physics near the transition is to take 
$\alpha(\rho)=D_r\left(1-\rho/\rho_c\right)$, changing sign at a characteristic
density $\rho_c$. For stability reasons the
coefficient 
$\beta$ is positive. The ratio $D_r/\gamma$ has dimensions of frequency. 
In the following we
will use it to set our unit of time by letting $D_r/\gamma=1$. We also divide
implicitly 
all the coefficients in the free energy \eqref{Fp} by $D_r$. 
The third term in the free energy describes
the energy cost for a  spatially inhomogeneous deformation of the order
parameter. The Frank constant $K$ is positive. For simplicity we employ a
one-elastic
constant approximation and use the same value for the stiffness associated with
splay and bend deformations of the order parameter field, as also for gradients
in its magnitude. Splay/bend anisotropy can however, play a very important role
in active systems, as demonstrated for instance by~\textcite{Voituriez2006}. The
last two terms are allowed only in systems with polar symmetry [see, e.g.,
\citep{Kung2006}]; they give the  density and $|\mathbf{p}|^2$
contributions to
the spontaneous splay.
Upon integration by parts, they can be seen to enable gradients in $\rho$ or in
the magnitude $|\mathbf{p}|$ of the order parameter to provide a local aligning
field for $\mathbf{p}$. Alternatively, the $\lambda$ term can be viewed as a
splay-dependent correction to $\alpha(\rho)$: splay of one sign enhances, and
of the other sign reduces the local tendency to order. 

By using Eq.~ \eqref{Fp} for the free energy, the equation for ${\bf p}$ takes
the form

\begin{align}
\label{P-eq-2}\partial_t{\bf p}+&\lambda_1({\bf p}\cdot\bm\nabla)
{\bf p}=-\left[\alpha(\rho)+\beta|{\bf p}|^{2}\right]{\bf p}+K\nabla^2{\bf p}
\nonumber\\
&-v_1\bm\nabla\frac{\rho}{\rho_0}+\frac{\lambda}{2}\bm\nabla |{\bf p}|^2-\lambda
{\bf p}(\bm\nabla\cdot{\bf p})+{\bf f}\;,
\end{align}
It is instructive to compare Eq.~\eqref{P-eq-2} with the Navier-Stokes equation
for a simple fluid. The term proportional to $\lambda_1$ is the
familiar advective nonlinearity. Unlike the fluid, the flocking model does not
possess Galilean invariance as the particles are moving relative to a substrate.
As a result,
$\lambda_1 \neq v_0$ [which would have read $\lambda_1 \neq 1$ had we not
extracted a velocity scale in (\ref{rho-eq})].
If we think of ${\bf p}$ as proportional to a flow
velocity, the second term on the right hand side represents the effects of
viscous forces.
The third and fourth terms of Eq.~\eqref{P-eq-2} can be rewritten in an
approximate
pressure-gradient form as $-({1}/{\rho_0}) \bm \nabla P$ with a pressure
$P(\rho, |{\bf p}|) \simeq {v_1}\rho - ({\lambda\rho_0}/{2}) |{\bf p}|^{2}$.
This highlights the parallel and the contrast with the Navier-Stokes equation:
in the latter, an equation of state determines the thermodynamic pressure in
terms of density and temperature and  not the velocity field.
The first term on the right hand side of Eq.~\eqref{P-eq-2} arises from the role
of ${\bf p}$ as order parameter for a polarized state and does not have an
analogue in equilibrium fluid flow. There is, however, an equilibrium analogue in the context of  
electrostriction, a property of all dielectric materials that causes them to change their shape under the application of an electric field, although in this case polarization does not correspond to velocity, but rather to reactive polarity.
General symmetry arguments given by~\textcite{Toner2005,Mishra2010}
allow for more
general polar terms in Eq.\eqref{P-eq-2}:
$\frac{\lambda_3}{2}\bm\nabla|{\bf p}|^2+\lambda_2{\bf p}(\bm\nabla\cdot{\bf
p})$
The derivation based on the free energy \eqref{Fp}, using  \eqref{P-eq}
yields $\lambda_3=-\lambda_2=\lambda$.  For the
out-of-equilibrium system considered here the coefficients of these two terms
are not generally related, but the microscopic derivation
of~\textcite{Bertin2009} also yields $\lambda_3=-\lambda_2$ as well as
$\lambda_i\sim v_0^2$ and $v_1= v_0/2$.

\subsubsection{Homogeneous steady states}
The dynamical model described by Eqs.~\eqref{rho-eq} and \eqref{P-eq} exhibits
by construction, in a mean-field treatment of homogeneous
configurations, a continuous transition from a disordered to an ordered state.
For $\alpha>0$, corresponding to an equilibrium density $\rho_0<\rho_c$, the
homogeneous steady state of the system is disordered or isotropic, with ${\bf
p}=0$ and a corresponding zero mean velocity. For $\alpha<0$,  corresponding to
$\rho_0>\rho_c$, the system orders in a state with uniform orientational order,
with $|{\bf p}_0|=\sqrt{-\alpha_0/\beta}$, where $\alpha_0=\alpha(\rho_0)$. In the
ordered state, which is also a moving state, with ${\bf v}=v_0{\bf
p}_0$, continuous rotational symmetry is spontaneously broken.
This mean-field analysis survives fluctuation corrections even in two
dimensions \citep{Toner1995,Toner1998}, evading the Mermin-Wagner theorem
forbidding -- at equilibrium -- the spontaneous breaking of a continuous
symmetry in two spatial dimensions
\citep{Mermin1966,Hohenberg1967,Chaikin2000}. In the
present
nonequilibrium
system, however, the theorem does not apply.

The escape from Mermin-Wagner is primarily due to the advective
nonlinearities in \eqref{polar-eqs} which effectively generate long-range
interactions in the system, even if density variations are ignored as shown
recently by \textcite{Toner2012}. The ordered phase itself, especially near
the transition, is exceedingly complex as a result of coupling to the  density.
Although the \textit{mean} density is an important control parameter for the
transition to a flock, the \textit{local} propensity to order depends, via
$\alpha(\rho)$ in \eqref{Fp}, on the \textit{local} density
$\rho(\mathbf{r},t)$. This is analogous to the density-dependent exchange
coupling in a compressible magnet in~\textcite{Milovsevic1978}. The analogy ends
there: unlike in the magnet, the order parameter feeds back strongly into the
density dynamics via the current on the right hand side of Eq.
(\ref{polar-eqs}b). See section \ref{ordered_dry_props} of
\textcite{Toner1998}, and the review by
\textcite{Toner2005} for a more complete discussion of this point.

\subsubsection{Properties of the isotropic state}
To study the linear stability of the homogeneous steady states we examine the
dynamics 
of spatially inhomogeneous fluctuations $\delta\rho=\rho-\rho_0$ and $\delta{\bf
p}=
{\bf p}-{\bf p}_0$ from the values $\rho_0$ and ${\bf p}_0$ in each of the 
homogeneous states. The isotropic state is characterized by $\rho_0<\rho_c$ and 
${\bf p}_0=0$. The equations for the dynamics of linear fluctuations about these
values are 
\begin{subequations}
\label{lineq-rho-P}
\begin{gather}
\label{rho-lineq-iso}\partial_t\delta\rho=-v_0\rho_0\bm\nabla\cdot{\bf p}\;,\\
\label{P-lineq-iso}\partial_t{\bf p}=-\alpha_0{\bf
p}-\frac{v_1}{\rho_0}\bm\nabla
{\delta\rho}  +K\nabla^2{\bf p}+{\bf f}\;,
\end{gather}
\end{subequations}
where $\alpha_0>0$.
Fourier-transforming (\ref{lineq-rho-P}) in space and time to look at modes
of the form $\exp(i \mathbf{q} \cdot \mathbf{r} - i \omega t)$ leads to
dispersion relations between frequency $\omega$ and wavevector $\mathbf{q}$ for
small fluctuations about the uniform isotropic phase.
Polarization fluctuations transverse to $\mathbf{q}$ decouple
and decay at a rate $\alpha_0+Kq^2$. The relaxation of coupled fluctuations of
density and longitudinal polarization is controlled by coupled hydrodynamic
modes with frequencies 
\begin{equation}
\omega^I_\pm(q)=-\frac{i}{2}(\alpha_0+Kq^2)\pm\frac{i}{2}\sqrt{
(\alpha_0+Kq^2)^2-4v_0v_1q^2}.
\end{equation}
Linear stability requires that fluctuations
decay at long times so that $Im[\omega(q)]<0$. The isotropic state is linearly
stable 
for all parameter
values, provided $v_0v_1>0$. The parameter $v_1$ enters (\ref{lineq-rho-P})
like an effective compressional modulus. Microscopic derivations of
the continuum model have yielded $v_1=v_0/2$ at low
density \citep{Baskaran2008,Bertin2009,Ihle2011}. In this case $v_1>0$ and the
isotropic state is always stable. At high density, however, caging effects can
result in a density dependence of $v_1$, which can in turn lead to a density
instability of the isotropic state as shown by \textcite{Tailleur2008,Fily2012}. Here we restrict our discussion to the situation where the
isotropic state is linearly stable. Even in this case, it has
unusual properties: approaching the mean-field ordering transition, i.e.,
decreasing $\alpha_0$, is like decreasing friction. As argued by
\textcite{Ramaswamy1982} for equilibrium fluids on a substrate, the dispersion
relations can acquire a real part for $\alpha_0\leq v_0v_1/K$ and a range of
wavevectors, $q_{c-}\leq q\leq q_{c+}$, with \citep{Baskaran2008}
\begin{equation}
q_{c\pm}^2=\frac{2v_0v_1}{K^2}\left[1-\frac{K\alpha_0}{2v_0v_1}\pm\sqrt{1-\frac{
K\alpha_0}{v_0v_1}}\right].
\end{equation}
Density fluctuations then propagate as waves rather than diffuse.
Near the transition, where $\alpha_0\rightarrow 0+$, we find
$q_{c-} \to 0$ and $q_{c+}\simeq 2\sqrt{v_0v_1}/K$ and the propagating waves
resemble sound waves, with $\omega_\pm^I\simeq\pm q\sqrt{v_0v_1}$.  These
propagating sound-like density waves are ubiquitous in collections of
self-propelled particles. We stress, however, that if the limit $q\rightarrow 0$
is taken first, density fluctuations always decay diffusively.

\subsubsection{Properties  of the ordered state}
\label{ordered_dry_props}
We now examine the linear stability of the ordered state by considering the linear
dynamics of fluctuations about $\rho_0>\rho_c$ and ${\bf p}_0=p_0{\bf \hat{x}}$.
The detailed analysis below is
for $d=2$, but the general conclusions including the instability of the
uniform ordered state just past threshold hold in all dimensions.
It is useful to write the order parameter in terms of its magnitude and
direction, by letting ${\bf p}=p{\bf \hat{n}}$, where ${\bf \hat{n}}$ is a unit
vector pointing in the direction of local orientational order. Fluctuations in
the polarization can then be written as $\delta{\bf p}={\bf \hat{n}}_0\delta
p+p_0\delta{\bf n}$.  The condition $|{\bf \hat{n}}|^2=1$ requires that to
linear order ${\bf \hat{n}}_0\cdot\delta{\bf n}=0$. For the chosen coordinate
system ${\bf \hat{n}}_0={\bf \hat{x}}$. Then in two dimensions $\delta{\bf
n}=\delta n{\bf \hat{y}}$ and $\delta{\bf p}=\xhat\delta p+\yhat p_0\delta n$.
The  linearized 
equations are
\begin{subequations}
\begin{gather}
\label{rho-lineq-pol}\left(\partial_t+v_0p_0\partial_x\right)\delta\rho=-
v_0\rho_0\bm\nabla\cdot\delta{\bf p}\;,\\
\label{P-lineq-pol}\left(\partial_t+\lambda_1p_0\partial_x\right)\delta{\bf
p}=-2|\alpha_0|
\xhat\delta{p}+a{\bf p}_0\delta\rho+\lambda_2{\bf p}_0(\bm\nabla\cdot\delta{\bf 
p})\notag\\
-(v_1/\rho_0)\bm\nabla\delta\rho+\lambda_3 p_0\bm\nabla\delta p
+K\nabla^2\delta {\bf p}+{\bf f}\;,
\end{gather}
\end{subequations}
where  $a=-\left(\alpha'+\beta'p_0^2\right)=\alpha\beta'/\beta-\alpha'>0$ and
primes denote 
derivatives with respect to density, e.g., 
$\alpha'=\left(\partial_\rho\alpha\right)_{\rho=\rho_0}$.  For the analysis
below it is useful 
to display explicitly the three coupled equations for the fluctuations. For
compactness, we 
introduce a vector with elements $\bm\phi=\{\delta\rho,\delta p,\delta n\}$ and
write them 
in a matrix form as
\beq
\label{lin-eqs}
\partial_t\bm\phi_{\bf q}(t)={\bm M}({\bf q})\cdot\bm\phi_{\bf q}(t)+{\bm
F}_{\bf q}(t)\;
\eeq
where the matrix ${\bm M}$ is given by
\begin{widetext}
\beq
\label{A}
{\bm M}({\bf q})=\left(\begin{array}{ccc}
-iq_xv_0p_0&-iq_xv_0\rho_0&-iq_yv_0\rho_0p_0\\
a p_0-iq_xv_1/\rho_0&~~-2|\alpha_0|-iq_x\bar\lambda
p_0-Kq^2&iq_y\lambda_2p_0^2\\
-iq_yv_1/(\rho_0p_0)&iq_y\lambda_3p_0&-iq_x\lambda_1p_0-Kq^2
\end{array}\right)
\eeq
\end{widetext}
The sign
of the combination  $\bar\lambda=\lambda_1-\lambda_2-\lambda_3$ is important 
in controlling the nature of the
instabilities in the ordered phase. Notice that all microscopic derivations of
the continuous equations for various systems of self-propelled
particles \citep{Baskaran2008,Bertin2009} yield $\bar\lambda>0$. Finally, the
noise vector ${\bm F}_{\bf q}(t)$ has components ${\bm F}_{\bf
q}(t)=\{0,{f}^L_{\bf q}(t),{\bf f}^T_{\bf q}(t)\}$, with $f^L_{\bf q}(t)={\bf
\hat{q}}\cdot{\bf f}_{\bf q}(t)$, ${\bf f}^T_{\bf q}(t)={\bf f}_{\bf
q}(t)-{\bf \hat{q}}{f}^L_{\bf q}(t)$ and ${\bf \hat{q}}={\bf q}/q$.
The dispersion relations of the hydrodynamic modes are the eigenvalues of the
matrix ${\bm 
M}$.  The solution of the full cubic eigenvalue problem is not, however, very
instructive. 
It is more useful to discuss a few simplified
cases~\citep{Bertin2009,Mishra2010}.

\begin{figure}
\centering
\includegraphics[width=0.8\columnwidth]{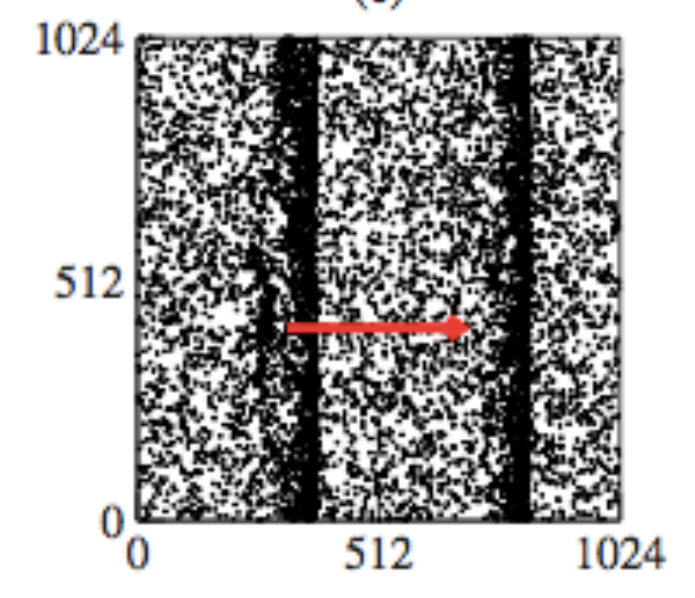}
\caption{(color online) Typical snapshots in the ordered phase. Points represent
the position 
of individual particles and the red arrow points along the global direction of
motion. Note 
that in the simulations sharp bands can be observed only if the system size is
larger than the 
typical width of the bands. Adapted with permission from~\textcite{Chate2008}.
}
\label{fig:stripes}
\end{figure}

\paragraph{Linear instability near the mean-field order-disorder transition.}
We first consider the behavior of fluctuations as the mean-field transition is
approached from the ordered phase, i.e., for $\alpha_0\rightarrow 0^-$ and
$p_0\rightarrow 0^+$. Fluctuations $\delta p$ in the magnitude of polarization
decay at rate $\sim|\alpha_0|$ and become long-lived near the transition. In
fact in this limit one can neglect the coupling to director fluctuations and
only consider the coupled dynamics of $\delta\rho$ and $\delta p$. This
decoupling becomes exact for
wavectors ${\bf q}$ along the direction of broken symmetry, ${\bf q}=q{\bf
\hat{x}}$,
The decay of director fluctuations is described by a stable, propagating mode,
$\omega_n=-q\lambda_1p_0^2-iKq^2$.  For small $q$ the imaginary part of the
frequencies $\omega^{p}_\pm(\mathbf{q})$ of the modes describing the dynamics of
density and order parameter magnitude 
takes the form
\beq
\label{Im-omegap}
Im[\omega_+^p]=-\left[s_2q^2+s_4q^4+{\cal O}(q^6)\right]
\eeq
with
$s_2=\frac{v_0v_1}{2|\alpha_0|}\left[1-\frac{v_0a^{2}\rho_0^2}{
4v_1|\alpha_0|\beta}\right] $, implying an instability at small $q$ when
$s_2<0$, corresponding to $|\alpha_0|\leq\frac{v_0\rho_0^2a^2}{4v_1\beta}$.
Microscopic calculations yield $\alpha_0\sim v_0$ and $v_1\sim
v_0$~\citep{Bertin2009}, indicating that the instability exists for arbitrarily
small $v_0$ and in a narrow region of densities above the mean field
order-disorder transition. The ${\cal O}(q^4)$ term in Eq.~\eqref{Im-omegap}
stabilizes fluctuations at large wavevectors. In detail, near the mean-field
transition ($\alpha_0\rightarrow 0^-$) $s_2\simeq
-{a^{2}}/{8\beta\alpha_0^2}$ and $s_4\simeq
{5a^{4}}/{128\beta^2\alpha_0^5}$ and the mode is unstable for
$q<q_c\simeq 4\sqrt{\beta|\alpha_0|^3/(5a^{2})}\sim(\rho_c-\rho_0)^{3/2}$.
Numerical solution of the nonlinear hydrodynamic equations with and without
noise have shown that in this region the uniform polarized state is replaced by
complex spatio-temporal structures. In the absence of noise, one finds
propagating solitary waves in the form of ordered bands aligned transverse to
the direction of broken symmetry and propagating along $\xhat$ amidst a
disordered background \citep{Bertin2009,Mishra2010}.  As discussed
below, these bands have been observed in numerical simulations of the Vicsek
model
\citep{Chate2008a} and are shown in Fig.~\ref{fig:stripes}.
 
\paragraph{``Sound waves''.}
Deep in the ordered phase, i.e., for large negative
$\alpha_0$,
fluctuations $\delta p$ in the magnitude of the order parameter decay on
non-hydrodynamic time scales. We  assume $\delta p$ has relaxed to zero on the
time scales of interest and let for simplicity $p_0\simeq 1$. In addition, we
include a diffusive current and the associated noise  that were neglected in
the density Eq.~\eqref{rho-eq}. 
The coupled equations for density and polar director
fluctuations, $\delta\rho$ and $\delta n$, take the  form
\begin{subequations}
\begin{gather}
\label{rho-lineq-pol2}\left(\partial_t+v_0\partial_x\right)\delta\rho=-
v_0\rho_0\partial_y\delta n+D\nabla^2\delta\rho+\bm\nabla\cdot{\bf 
f_\rho}\;,\\
\label{n-lineq-pol}
\left(\partial_t+\lambda_1\partial_x\right)\delta{n}=-
v_1\partial_y\left(\frac{\delta\rho}{\rho_0}\right)
+K\nabla^2\delta {n}+f_y\;,
\end{gather}
\end{subequations}
where $D$ is a diffusion constant and ${\bf f}_\rho$ describes white, Gaussian
noise. The 
hydrodynamic modes, obtained from Eqs.~\eqref{rho-lineq-pol2} and 
\eqref{n-lineq-pol}, are 
stable propagating sound-like waves, with dispersion relation 
\beq
\label{sound}\omega_\pm^s=qc_\pm(\theta)-iq^2{\cal K}_\pm(\theta)+{\cal
O}(q^3)\;,
\eeq
with speeds and dampings $c_\pm(\theta)$ and ${\cal K}_\pm(\theta)$, whose
detailed forms are given in the review by \textcite{Toner2005} and depend
on the angle $\theta$ between ${\bf q}$ and the direction $\xhat$ of broken
symmetry.
For $\theta=\pi/2$, i.e., ${\bf q}$ along $y$, we find $c_{\pm}
=\pm \sqrt{v_0v_1}$ and ${\cal K}_\pm = (K+D)/2$, while for $\theta=0$,
corresponding to ${\bf q}$ along $x$, the two modes are decoupled and
$\omega_+^s=qv_0-iDq^2$ and $\omega_-^s=-q\lambda_1-iKq^2$. The difference in
the propagation speeds is due to the lack of Galilean invariance. In a normal
fluid, sound waves propagate because of mass and momentum conservation and
inertia. In contrast, the presence of propagating
long-wavelength density disturbances here is a signature of the spontaneously
broken orientational symmetry which renders $\delta n$ ``massless''
\citep{Ramaswamy2010}.

Going back to the full form of the matrix ${\bm M}$ in (\ref{A}), we notice that to leading order in $q$
the relaxation of $\delta p$ is governed by the equation
\beq
\partial_t\delta p_\qo =-2|\alpha_{0}|\delta p_\qo+a p_0\delta\rho_\qo+{\cal
O}(q)\;.
\eeq
This reminds us to be precise about what it means for a mode to be ``fast'':
fluctuations in the magnitude of the order parameter cannot be neglected, but
rather are slaved to density fluctuations on long timescales: 
$\delta p_\qo\simeq (a p_0/2|\alpha_{0}|) \delta\rho_\qo$. Inserting this in
(\ref{lin-eqs}) yields an effective dynamics for $\delta n$ and
$\rho$ with an
instability of splay fluctuations if $\lambda_3 >
\frac{2v_1\beta}{\rho_0p_0a} > 0$~\citep{Mishra2010}. An analysis of the
eigenvalues of the full matrix ${\bm M}$ confirms this result.

\begin{figure}
\centering
\includegraphics[width=0.45\columnwidth]{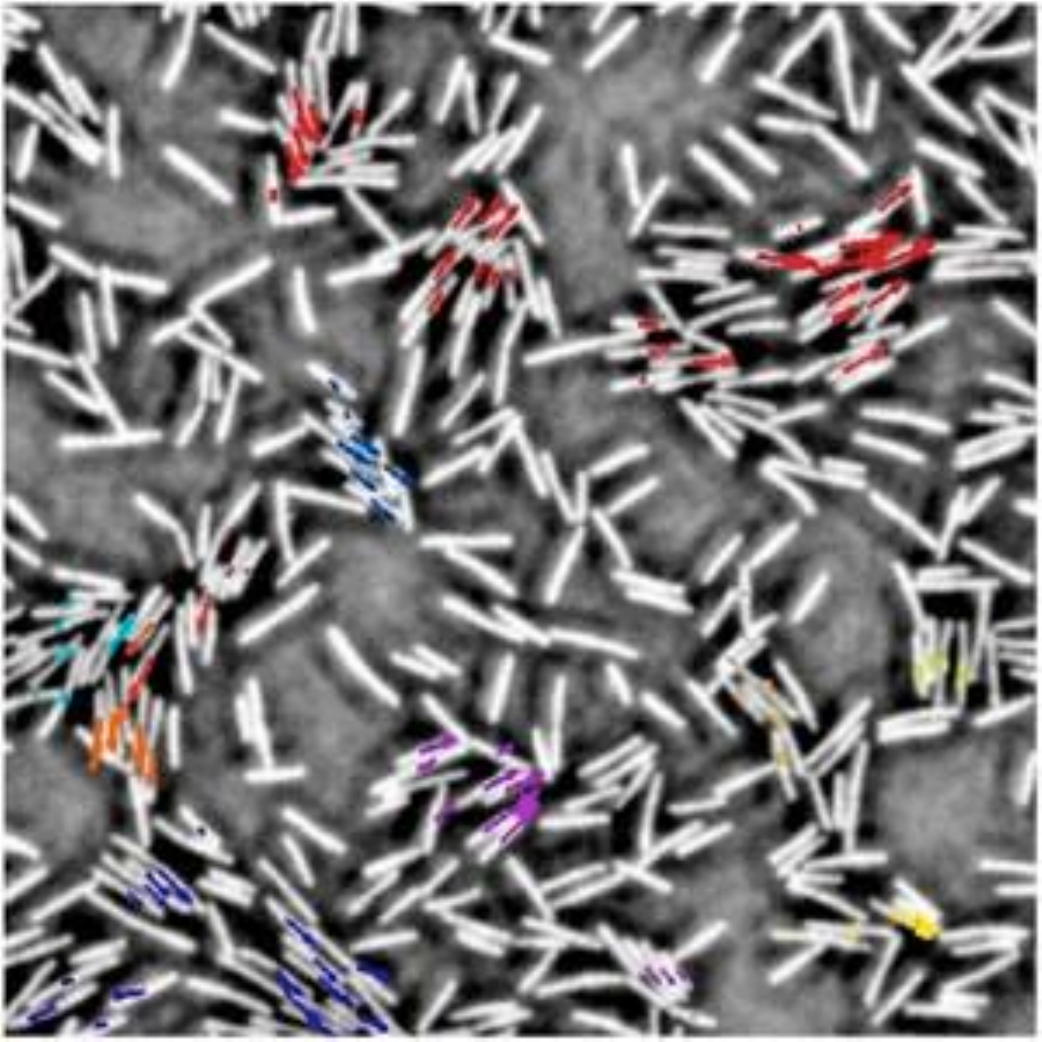}
\includegraphics[width=0.45\columnwidth]{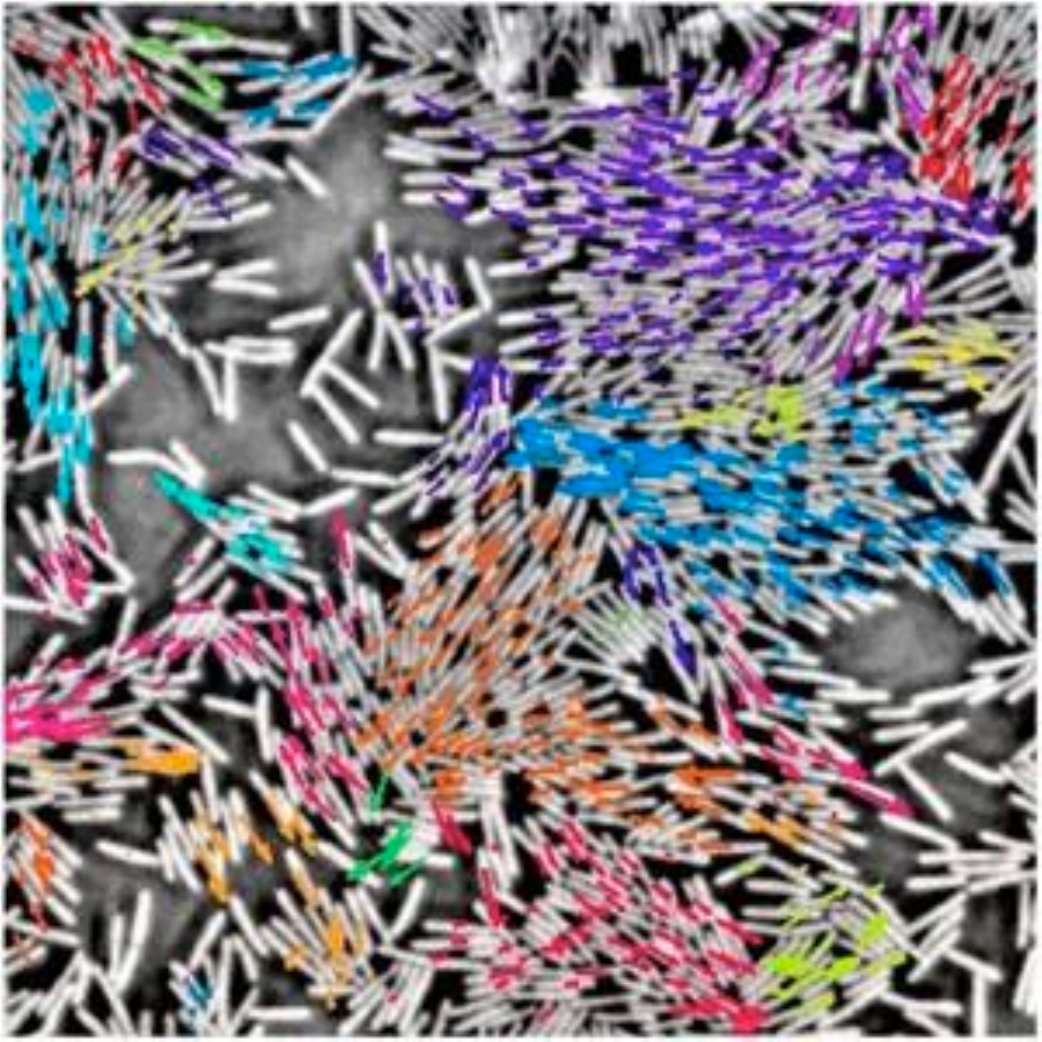}
\caption{(color online) Swimming {\it bacillus subtilis} bacteria exhibit strong
polar order and fluctuations that that are smaller at low density (A) than at
high density (B).  
Nearby bacteria with arrows of the same color belong to the same dynamic
cluster; the arrows indicate a bacterium's speed and direction.  Adapted with
permission from \textcite{Zhang2010}.}
\label{fig:zhang}
\end{figure}

\paragraph{Giant density fluctuations.}
\label{gdf}
The linearized equations with noise can also be used to calculate correlation
functions. Of particular interest is the static structure factor
\beq
\label{Sq}
S(\bfq)=\frac{1}{\rho_0 V}\langle\delta\rho_\bfq(t)\delta\rho_{-\bfq}
(t)\rangle=\int_{-\infty}^{\infty}\frac{d\omega}{2\pi}~S({\bf q},\omega)\;,
\eeq
where 
\beq
\label{Sqom}
S({\bf q},\omega)=\frac{1}{\rho_0 V}\int_0^{\infty} dt  \exp(i \omega t)
\langle \delta\rho_\bfq(0)\delta\rho_{-\bfq}( t)  \rangle
\eeq
 is the
dynamic structure factor. We are interested in calculating density fluctuations
in the region away from the mean-field transition, where the ordered state is
linearly
stable. Eqs.~\eqref{rho-lineq-pol2} and \eqref{n-lineq-pol} yield $S({\bf
q},\omega)$. 
For a reader wishing to work through the calculation, note that to leading order
in the wavevector the noise ${\bf f}_\rho$ in the density equations can be
neglected, except for ${\bf q}$ along the direction of broken symmetry
($\theta=0$). 
Integrating over frequency as in Eq.~\eqref{Sq}, again to leading order in $q$,
one obtains
\begin{widetext}
\beq
\label{Sq-ans}
S({\bf q})=\frac{v_0^2\rho_0\Delta\sin^2\theta}
{(v_0+\lambda_1)^2\cos^2\theta+4v_0v_1\sin^2\theta}\left[\frac{1}{{\cal 
K}_+(\theta)}+\frac{1}{{\cal K}_-(\theta)}\right]~\frac{1}{q^2}\;.
\eeq
\end{widetext}
For $\theta = 0$ the leading $1/q^2$ singularity displayed in (\ref{Sq-ans})
vanishes; to determine the finite value of $S(\mathbf{q})$ in that limit
requires going to higher order in $q$ which we will not do here. 
As we will see below, the divergence of the static structure function at large
wavelengths (as $1/q^2$ in the present linearized treatment) is a remarkable
and robust property of uniaxially ordered active systems, whether polar or
nematic, and follows generally from the orientational order that develops in a
sufficiently dense collection of self-driven particles with anisotropic body
shape~\citep{Toner1995, Toner1998,Simha2002,Simha2002a,Ramaswamy2003,Toner2005}.
The $1/q^2$ is also a dimension-independent property of our linearized theory;
the numerical prefactors do, however, depend on dimensionality $d$.
The divergence of the static structure function for $q\rightarrow 0$  implies an
important violation of the familiar scaling of number fluctuations
in equilibrium. In the limit of vanishing wavevector the
structure factor is simply a measure of number fluctuations,
with
\begin{equation}
\label{sf}
\lim_{q\rightarrow 0}S(q)=\frac{\Delta N^2}{\langle N\rangle}\;,
\end{equation}
where $N$ and $\langle N\rangle$ are the instantaneous and average number of
particles in 
a region of size $V$, respectively,  and  $\Delta N^2=\langle(N-\langle 
N\rangle)^2\rangle$ is the variance of number fluctuations. In an equilibrium
system $\Delta 
N\sim\sqrt{\langle N\rangle}$, so that $\Delta N/\langle
N\rangle\sim1/\sqrt{\langle 
N\rangle}\rightarrow 0$ as $\langle N\rangle\rightarrow \infty$. The $\sim
1/q^2$ 
divergence of $S(q)$ for $q\rightarrow 0$ predicted by Eq.~\eqref{Sq-ans}
implies that in 
an active system $\Delta N$ grows faster than $\sqrt{\langle N\rangle}$.  To
understand this 
we rewrite Eq.~\eqref{Sq-ans} as $\Delta N\sim\sqrt{\langle N\rangle
S(q\rightarrow 0)}$. 
Assuming that the smallest accessible wavevector is of the order of the inverse
of the system 
size $V^{-1/d}$, with $d$ the space dimension, we find $S(q\rightarrow 0)\sim 
V^{2/d}\sim\langle N\rangle^{2/d}$. This gives
\beq
\label{DN}\Delta N\sim \langle N\rangle ^a\;,\hspace{0.2in}
a=\frac12+\frac{1}{d}\;.
\eeq
The linear theory reviewed above predicts a strong enhancement of density
fluctuations in the ordered state, with  an exponent $a= 1$ in two dimensions
\citep{Toner1995,Toner2005}. We will see in section \ref{sec:nemsubstrate} that
a
similar linear theory also predicts the same scaling in the ordered state of
active apolar nematic. In general, we expect the exponent to be substantially
modified as compared to the theoretical value by fluctuation effects arising
from the coupling of modes at different wavenumbers as a result of nonlinear
terms in the equations of motion.  ``Giant density fluctuations" have now
been seen experimentally in both polar~\citep{Deseigne2010} and apolar
\citep{Narayan2007} vibrated granular matter. See section
\ref{subsub:active-nematic} for a more
detailed discussion of the apolar case. The scaling exponents, although
measured 
over a
limited dynamical range, are consistent, for the apolar case, with the
linearized treatment presented here and in~\citep{Ramaswamy2003}, and
with the renormalization-group treatment
\citep{Toner1995,Toner1998,Toner2005} in $d=2$ for polar
flocks. Giant number fluctuations have also been seen in simulations of
Vicsek-type models 
by Chat\'e and collaborators~\citep{Chate2006,Chate2008,Chate2008a}.
 Much remains to be understood about number
fluctuations in flocks, as revealed for instance by the experiments of
\citep{Zhang2010} (see Fig.~\ref{fig:zhang}).
Recent work has also revealed other mechanisms that can yield large
number fluctuations and even true phase separation in active systems  \emph{in
the absence of an ordered state}~\citep{Fily2012,Tailleur2008,Cates2012}. We
refer the reader to section \ref{subsec:current-dry} for further discussion of
this point.

Recall that in thermal equilibrium the thermodynamic sum rule relates $S(q=0)$
to the isothermal compressibility $\chi_T=-\frac{1}{V}\left(\frac{\partial
V}{\partial p}\right)_T$ according to $S(q\rightarrow 0)=\rho_0k_BT\chi_T$, with
$\rho_0=\langle N\rangle/V$ the mean number density. One might then be
tempted to say that orientationally ordered active systems are characterized by
a diverging effective compressibility. This would in general be \textit{wrong}.
Augmenting the free-energy functional (\ref{Fp}) with a linear coupling $\int
U(\mathbf{r},t) \delta \rho(\mathbf{r},t) d^d r$ and calculating the response
function $(\delta \langle \rho \rangle / \delta U)_{\mathbf{q}, \omega}$ can
readily be shown to yield a \textit{finite} result for $\mathbf{q}, \omega \to
0$. The giant density fluctuations seen in the ordered state of active systems
are an excess noise from the invasion
of the density dynamics by the soft Goldstone mode of orientational order, not
an enhanced response to perturbations. 

\paragraph{Long-range order of dry polar active matter in 2d.} 
The ordered polar state is remarkable in that it exhibits long-range-order of a
continuous order parameter in two dimensions. This has been referred to as a
violation of the Mermin-Wagner theorem, although it is important to keep in mind
that the latter only holds for systems in thermal equilibrium. From
Eqs.~\eqref{rho-lineq-pol2} and \eqref{n-lineq-pol} one can immediately obtain
the correlation function of fluctuations in the  direction $\delta n$ of polar
order, with the result $\langle|\delta n_{\bf q}(t)|^2\rangle\sim{1}/{q^2}$.
Taken literally, this result implies  quasi-long-range order in two dimension in
analogy with the equilibrium $XY$ model.
It has been shown, however, that nonlinearities in Eqs.~\eqref{rho-lineq-pol2}
and \eqref{n-lineq-pol} are strongly relevant in $d=2$ and lead to a
singular renormalization of the effective stiffness of the polar director at
small wavenumbers $q$, so that $\langle|\delta n_{\bf q}(t)|^2\rangle$ diverges
more slowly than $1/q^2$ for most directions of ${\bf q}$, thus preserving
long-range order~\citep{Toner1995,Toner1998,Toner2005}. A qualitative
explanation
of how this happens can be given as in~\textcite{Ramaswamy2010}, but a remarkable,
pictorial yet quantitative account can be found in lecture notes by
\textcite{Toner2009}. Exact arguments for scaling exponents in two
dimensions are given by \textcite{Toner2012} in the limit where particle number is
fast, i.e., not conserved locally.

\begin{figure}[h]
\centering
\includegraphics[width=0.47\columnwidth]{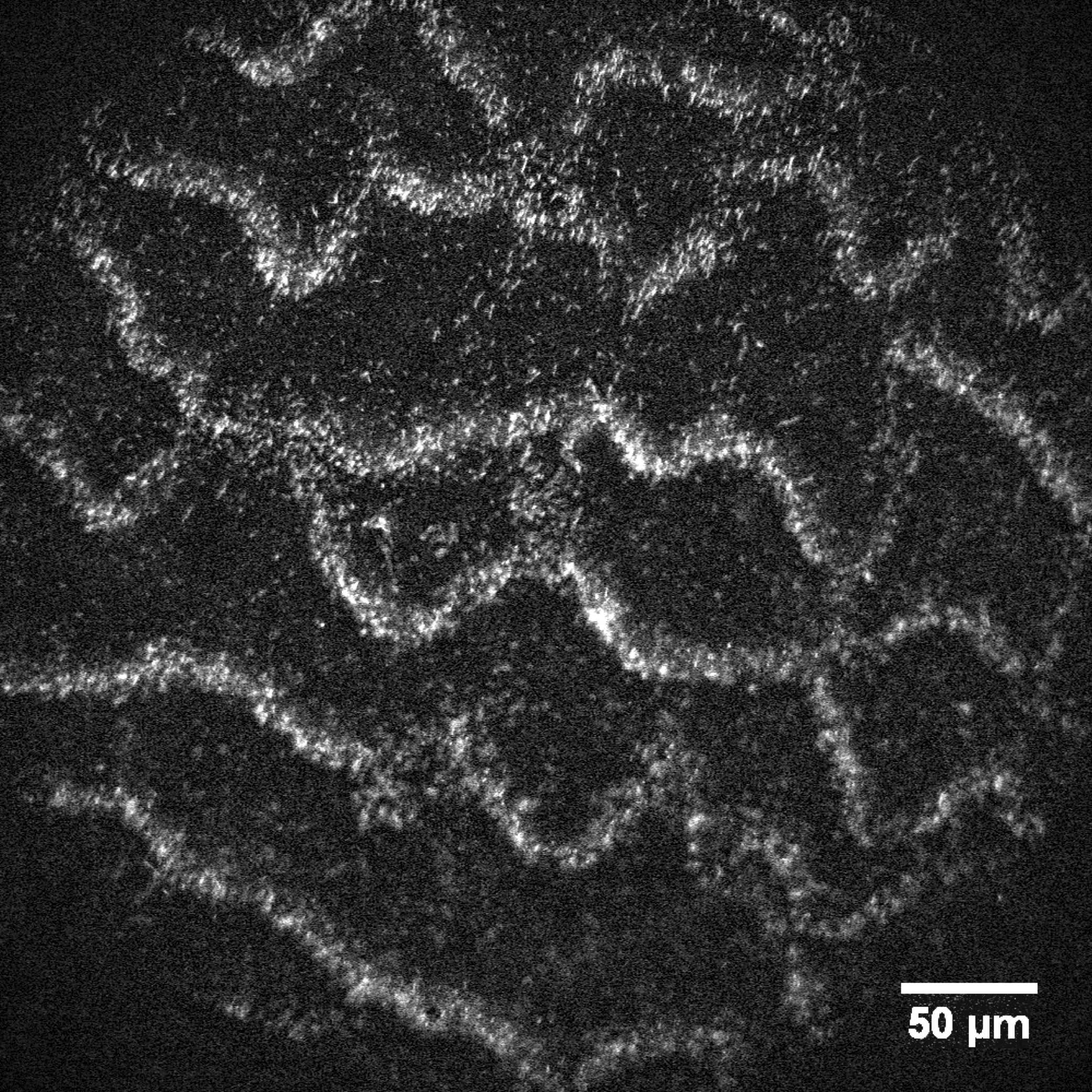}
\includegraphics[width=0.47\columnwidth]{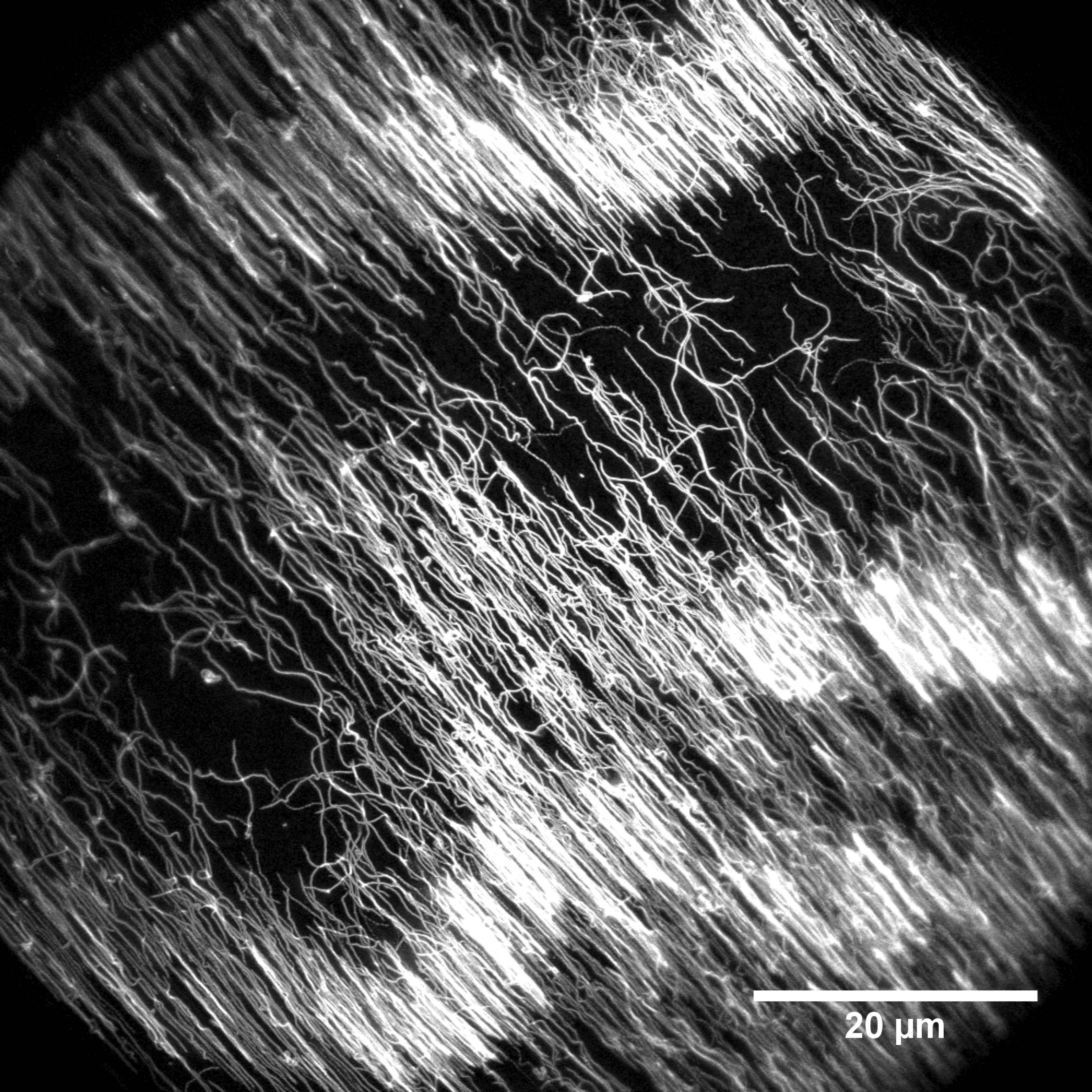}
\caption{Density waves in a dense actin motility assay. The left image is a
snapshot of the traveling wavefronts of actin observed in motility assays for
actin density above about $20~{\rm filaments}/\mu m^2$, adapted with permission
from \textcite{Schaller2010}.   
The right image (courtesy of V. Schaller)
is obtained by overlaying maximal intensity projections of $20$ consecutive
images, corresponding to a total time of $2.5~{\rm seconds}$.
The time overlay allows one to trace the trajectory of the filaments, showing
that filaments in the high-density regions move collectively with high
orientational persistence, while filaments lying outside the bands perform
uncorrelated persistent random walks. The elongated density modulation travels
in the direction of the filaments' long direction at a  speed that remains
approximately constant over the time scale of the experiment. }
\label{Fig:Schaller}
\end{figure}

\paragraph{Numerical simulation of Vicsek-type models.}
Although a review of numerical simulations of agent based models is beyond the
scope of 
this paper, for completeness we summarize the main findings. 
Quantitative agreement has been found between numerical experiments on
microscopic 
models and the predictions of the coarse-grained theory, including long-range
order in 
$d=2$, the form of the propagating modes, anomalous density fluctuations, and 
super-diffusion of tagged particles~\citep{Toner2005,Gregoire2004}.
Early numerical studies~\citep{Vicsek1995} and some later variants
~\citep{Aldana2007,Gonci2008,Baglietto2009} found a behavior
consistent with a continuous onset of flocking, as a mean-field solution  would predict. However,
the currently agreed picture, in systems where particle number is conserved, is
of a discontinuous
transition~\citep{Gregoire2004,Chate2006,Chate2007,Chate2008a}. It was found in
fact 
that in a large domain of parameter space, including the transition region, the
dynamics is 
dominated by propagating solitary structures consisting of well-defined regions
of high density 
and high polar order in a low-density, disordered background. The ordered
regions consist 
of stripes or bands aligned transverse to the direction of mean order and
traveling along the 
direction of mean motion with speed of order $v_0$, as shown in
Fig.~\ref{fig:stripes}. 
Away from the transitions, for weak noise strengths, a homogeneous ordered phase
is found, 
although with anomalously large density fluctuations. Numerical solution of the
nonlinear 
continuum model described by Eq.~\eqref{polar-eqs} has also yielded traveling
wave 
structures~\citep{Bertin2009,Mishra2010,Gopinath2012}.
Finally, traveling density waves have been observed in actin motility assays at
very high actin density~\citep{Schaller2010,Butt2010}, as shown in
Fig.~\ref{Fig:Schaller}.
Although it is tempting to identify these actin density waves with the traveling
bands seen in simulations of the Vicsek model, the connection is 
at best qualitative at the moment. In particular, it remains to be established
whether these actin suspensions can indeed be modeled as dry systems,
or hydrodynamic interactions are important, as suggested
in~\textcite{Schaller2011a}.

\subsection{Active Nematic on a substrate}  \label{sec:nemsubstrate}

Nematic order is the simplest kind of orientational order, with a spontaneously
chosen axis which we shall designate the $\hat{\bf x}$ direction, and no
distinction between $\hat{\bf x}$ and $-\hat{\bf x}$. Lacking a permanent
vectorial asymmetry, the system, although driven, has no opportunity to
translate its lack of time-reversal symmetry into an average nonzero
drift velocity. The reader would be forgiven for thinking that such a non-moving
state cannot possibly display any of the characteristics of a flock.
Remarkably, however, some of the most extreme fluctuation properties of
ordered phases of active particles are predicted \citep{Ramaswamy2003} and
observed in experiments \citep{Narayan2007} and earlier in simulations
\citep{Chate2006,Mishra2006} on this flock that goes nowhere on average.
As with polar active systems, several levels of description are possible for an
active nematic -- minimal agent-based stochastic models \citep{Chate2006} and
Fokker-Planck treatments thereof \citep{Shi2010}, more detailed dynamical
models that generalize rod-like-polymer dynamics to include the physics of
motors
walking on cytoskeletal filaments \citep{Ahmadi2005,Ahmadi2006}, or direct
continuum
approaches based on partial differential equations (PDEs) for the slow variables
\citep{Simha2002}. The microscopic models can in turn be coarse-grained
\citep{Ahmadi2006,Mishra2009,Mishra2011,Shi2010} to yield
the continuum equations in averaged \citep{Ahmadi2005,Shi2010} or stochastic
\citep{Mishra2009,Mishra2011} form.

\subsubsection{Active Nematic}
\label{subsub:active-nematic}
\begin{figure}
\centering
\includegraphics[width=0.89\columnwidth]{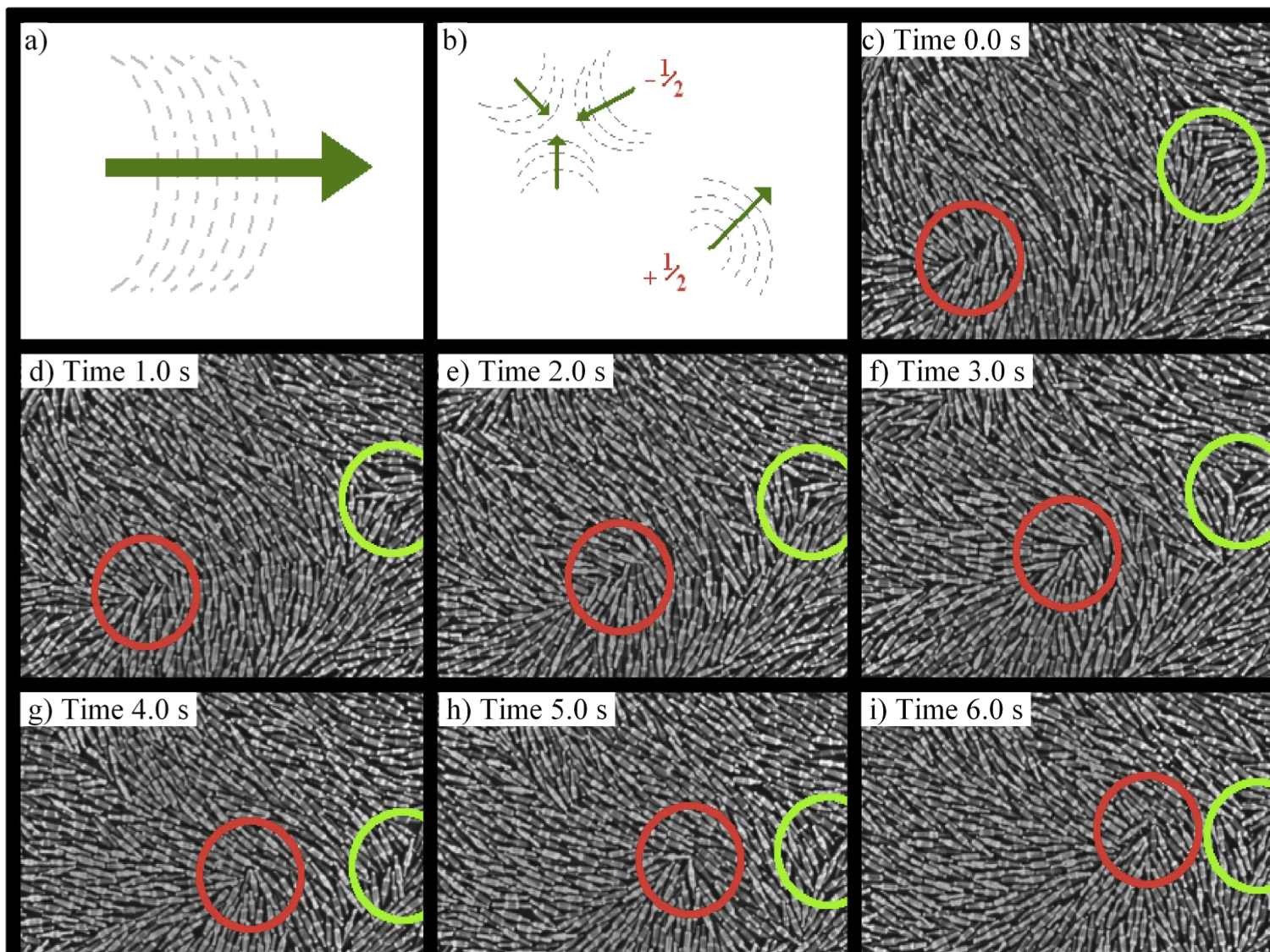}
\caption{(color online) Curvature gives rise to
self-propulsion in an apolar active medium, illustrating Eq.
(\ref{eq:active_curr}). A disclination of strength +1/2 (circled in red) in a
nematic phase of a vibrated layer of rods. The locally polar director structure,
together with the absence of time-reversal symmetry in a driven state leads to
directed motion. The curvatures around a nearby defect of strength -1/2 (circled
in green) lack a preferred direction; that defect does not move. Adapted with
permission from
\textcite{Narayan2007}.}
\label{fig:civ}
\end{figure}

In this section we construct the equations of
motion for active apolar nematics and discover their remarkable properties.
We continue to work in a simplified description in which the medium through or
over which
the active particles move is merely a momentum sink, without its own dynamics.
Effects
associated with fluid flow are deferred to a later section. We consider again a
collection of 
elongated active particles
labeled by $n$, with positions
$\mathbf{r}_{n}(t)$ and orientation described by unit vectors
$\mbox{\boldmath $\hat\nu$}_{n}(t)$. As in the case of polar systems on a
substrate, the only
conservation law is that of number. The slow variables of such a
formulation are thus the number density $\rho(\mathbf{r},t)$ at location
$\mathbf{r}$ and time $t$, as defined in Eq.~\eqref{rho} in section
\ref{sec:polardry}, and the apolar nematic orientational order parameter, a
traceless symmetric tensor $\bsf{Q}$ with components
\beq
\label{Q}
Q_{\alpha\beta}({\bf r},t) = \frac{1}{\rho({\bf
r},t)}
\sum_{n}\left(\hat{\nu}_{n
\alpha}(t)\hat{\nu}_{n
\beta}(t)-\frac{1}{d}\delta_{\alpha\beta}\right)\delta({\bf r}-{\bf
r}_n(t)) 
\eeq
measuring the local degree of mutual alignment of the axes of the
constituent particles, without distinguishing head from tail. In constructing
the equations of motion we keep track of which terms are permitted in a system
at thermal equilibrium and which arise strictly from nonequilibrium activity.
As we are ignoring inertia and fluid flow, the dynamics of the orientation is
governed by
the equation
\beq
\label{eq:nem_gen}
\gamma_Q\partial_t \bsf{Q} = - \frac{\delta F_Q}{ \delta
\bsf{Q}} + \bsf{f}_Q
\eeq
describing the balance between frictional torques governed by a rotational
viscosity 
$\gamma_Q$ and
thermodynamic torques from a free-energy functional
\begin{eqnarray}
\label{eq:nemfree}
F_Q &=&  \int_{\bf r}\,\Big[
\frac{\alpha_Q(\rho)}{2}\bsf{Q}:\bsf{Q}+\frac{\beta_Q}{4}
(\bsf{Q}:\bsf{Q})^2 +\frac{ K_Q}{2}\left(\bm\nabla \bsf{Q}\right)^{2}\nonumber\\
&&+C_Q
\bsf{Q}:\bm\nabla \bm\nabla \frac{\delta \rho}{\rho_0} + \frac{A}{2}
\left(\frac{\delta \rho}{\rho_0}\right)^2
\Big]
\end{eqnarray}
that includes a tendency to nematic order for $\alpha_Q < 0$ and $\beta_Q >
0$, Frank elasticity as well as variations in the \textit{magnitude} of
$\bsf{Q}$ via $K_Q$, bilinear couplings of $\bsf{Q}$ and $\rho$,
and a compression modulus $A$ penalizing fluctuations $\delta
\rho$ in the density about its mean value $\rho_0$. Note that all parameters
in (\ref{eq:nemfree}) can depend on $\rho$, as indicated explicitly for
$\alpha_Q$.  We also stress that this form of the free energy only applies  in
two dimensions. For active nematic in three dimensions
 terms of order ${\bf Q}^3$ also need to be included. We have included in
(\ref{eq:nem_gen}) a spatio-temporally white
statistically isotropic tensor noise $\bsf{f}_Q$ to take into account
fluctuations of thermal or active origin. It is implicit that all terms in
(\ref{eq:nem_gen}) are traceless and symmetric. We have ignored several possible
terms in the equation of motion for $\bsf{Q}$ -- additional couplings to $\rho$
via off-diagonal kinetic coefficients, nonlinear terms not expressible as the
variational derivative of a scalar functional, and terms of sub-leading order in
gradients -- none of which plays an important role in our analysis. The unique
physics of the self-driven nematic state \citep{Ramaswamy2003} enters through
active
contributions to the current $\mathbf{J}$ in the conservation law
\beq
\label{eq:cont_gen}
\partial_t \rho = - \bm\nabla \cdot \mathbf{J}. 
\eeq
\begin{figure}
\centering
\includegraphics[width=0.89\columnwidth]{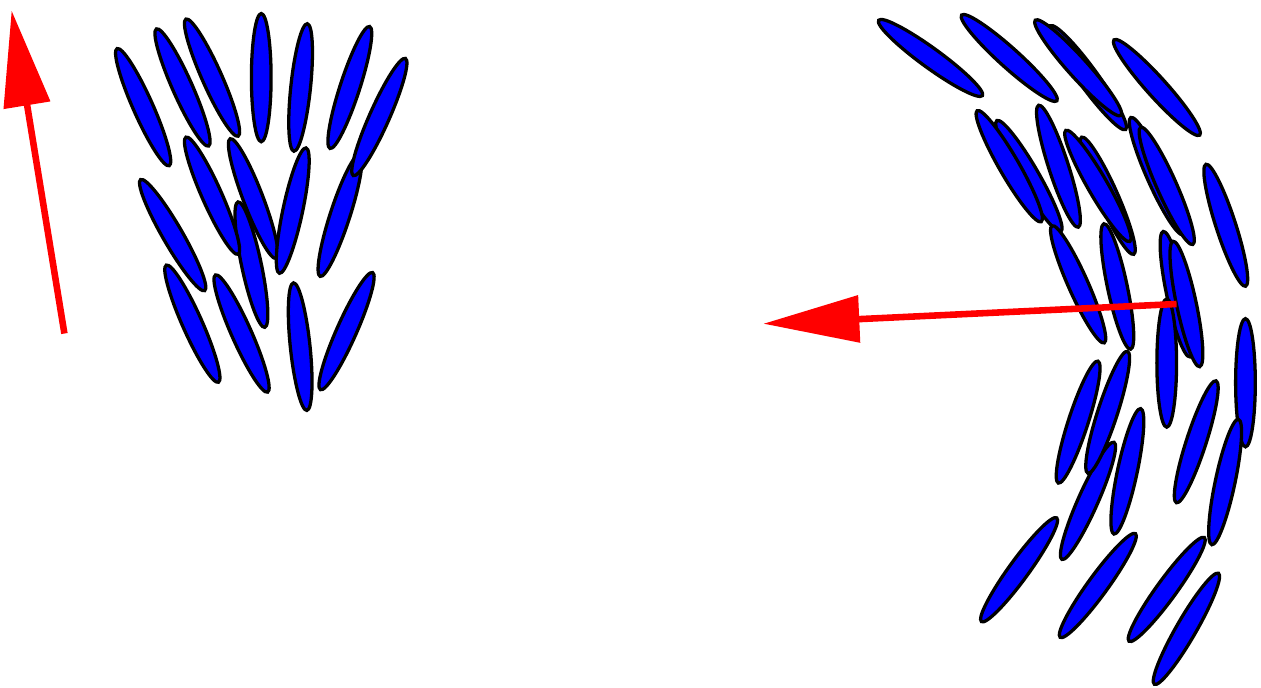}
\caption{(color online) Curvature generates temporary polarity in a
nematic. Activity gives rise to a current in a direction determined by this
polarity.}
\label{fig:curvcurr}
\end{figure}
There are several ways of generating these active terms: one can write them down
on the grounds that nothing rules them out in a nonequilibrium state
\citep{Simha2002,Ramaswamy2003}, derive them from specific microscopic models
involving motors and filaments \citep{Ahmadi2005,Ahmadi2006} or apolar flocking
agents \citep{Chate2006,Mishra2009,Mishra2011}, or frame them in the language of
linear irreversible thermodynamics \citep{Groot1984,Julicher2007}. This last we
will do later, in section \ref{subsub_activecurrents}. For now, Figure
\ref{fig:curvcurr} provides a pictorial argument. In a system with only an
apolar order parameter, curvature $\bm\nabla \cdot \bsf{Q}$ implies a local
polarity and hence in a nonequilibrium situation maintained by a nonzero activity must in general give a current 
\beq
\label{eq:active_curr}
\mathbf{J}_{active} = \zeta_Q \bm\nabla \cdot \bsf{Q}\;,
\eeq
where $\zeta_Q$ is a phenomenological active parameter.
Graphic experimental evidence of curvature-induced current can
be found in Fig. \ref{fig:civ}; for details see \textcite{Narayan2007},
Supporting
Material.
As shown by 
\textcite{Ramaswamy2003} and reiterated below, this $\zeta_Q$ term leads to giant
number fluctuations in active nematics. This is despite the fact that
\eqref{eq:active_curr} has one more gradient than the current in the polar case
in \eqref{polar-eqs}. We will comment below on this subtlety after we summarize
the calculation and the relevant experiments and simulations. 

Using (\ref{eq:nemfree}) in (\ref{eq:nem_gen}), and
including in the current $\mathbf{J}$ in (\ref{eq:cont_gen}) both the active
contribution (\ref{eq:active_curr}) and a purely passive diffusive part $-
\frac{1}{\gamma} \frac{\bm\nabla \delta F_Q }{\delta \rho}$, with a mobility $\gamma^{-1}$, leads to
\beq
\label{rho-eq-N}\partial_t\rho=D\nabla^2\rho + B \nabla^2 \bm\nabla\bm\nabla:\bsf{Q} 
+ \zeta_Q  \bm\nabla \bm\nabla : \bsf{Q}
+\bm\nabla\cdot{\bf
f}^\rho\;,
\eeq
with $D =  A /( \rho_0^2\gamma)$, $B =   C_Q / (\rho_0\gamma)$. We have treated
$\zeta_Q$  as constant, and we have introduced
number-conserving
fluctuations through the random current $\mathbf{f}^{\rho}$ which, again, we
take to be statistically isotropic and delta-correlated in space and time. 

Despite its apparent structural simplicity the apolar active nematic state has
been explored less than its polar counterpart. The ordering transition
appears numerically \citep{Chate2006} to be of the Kosterlitz-Thouless type, but
analytical approaches must confront the difficulties of the coupling of
order parameter and density and have not been attempted so far, beyond
mean-field \citep{Ahmadi2005,Ahmadi2006} which predicts a continuous transition
in two
dimensions. Working with equations of
a form similar to (\ref{eq:nem_gen}) and (\ref{rho-eq-N}), 
\textcite{Shi2010} obtained in a
Boltzmann-equation approach from the model of 
\textcite{Chate2006},with a finite-wavenumber instability of the uniform nematic
state just past onset, and a ``perpetually evolving state''. 
Note that similar behavior has been obtained by Giomi and collaborators for
"wet" nematic fluids~\citep{Giomi2011,Giomi2012a} and has been observed in very
recent experiments on \emph{in vitro} suspensions of microtubule-kinesin
bundles~\citep{Dogic2012}. Finally, there are
preliminary numerical indications \citep{Mishra2009,Mishra2011} of
anomalous growth kinetics of active nematic order following a quench from the
isotropic phase, with a strong clumping of the density. We will not pursue these
issues further in this review. In the remainder of this section we restrict
ourselves to a simple understanding of the statistics of linearized density
fluctuations deep in a well-ordered active nematic. Tests of the resulting
predictions, through experiments and numerical studies, will be discussed
thereafter.

For the purposes of this section it is enough to work in two dimensions denoted
$(x, y)$, so that $\bsf{Q}$ has components $Q_{xx} = - Q_{yy} = Q = S \cos 2
\theta, \, Q_{xy} = Q_{yx} = P = S \sin 2 \theta$. 
We take $\alpha_Q < 0$, $\beta_Q>0$ in (\ref{eq:nemfree}), yielding a phase with
uniform nematic order and density $\rho_0$ in mean-field theory, and choose
our basis so that this reference state has $\theta = 0$.
Expressing (\ref{eq:nem_gen}) and (\ref{rho-eq-N}) to linear order in
perturbations about the reference state, $\rho = \rho_0 + \delta
\rho(\mathbf{r},t)$, $\bsf{Q} = \mbox{diag}(S,-S) + \delta \bsf{Q}$, where
$\delta Q_{xy} = \delta Q_{yx} \simeq 2 S \theta$ and $\delta Q_{xx} = - \delta
Q_{yy} = O(\theta^2)$, and working in terms of the space-time Fourier transforms
$\delta \rho_{\mathbf{q} \omega}, \theta_{\mathbf{q} \omega}$, it is
straightforward to calculate the correlation functions
$S^{\rho}_{\mathbf{q} \omega} = \frac{1}{\rho_0 V} \langle |\delta
\rho_{\mathbf{q} \omega}|^2\rangle$, $S^{\theta}_{\mathbf{q} \omega} =
\frac{1}{V} \langle |\theta_{\mathbf{q} \omega}|^2\rangle$, and
$S^{\rho \theta}_{\mathbf{q} \omega} = \frac{1}{V}\langle \delta
\rho_{\mathbf{q} \omega}\theta^*_{\mathbf{q} \omega}\rangle$. 
Rather than
dwelling on the details of this calculation \citep{Ramaswamy2003}, we suggest
that readers verify for themselves a few essential facts. (a) Parameter ranges
of non-zero measure can be found such that the real parts of the eigenvalues of
the dynamical matrix for this coupled problem are negative, i.e., active
nematics are not generically unstable. (b) Terms involving $C_Q$ from
(\ref{eq:nemfree}) are irrelevant -- sub-dominant in wavenumber in
(\ref{rho-eq-N}) and leading only to a finite renormalization of parameters in
(\ref{eq:nem_gen}). (c) Ignoring $C_Q$, the dynamics of $\bsf{Q}$ through
(\ref{eq:nem_gen}) becomes autonomous, independent of $\rho$. In the nematic
phase it is straightforward to show that $\theta_{\mathbf{q}}$ has equal-time
correlations $S^{\theta}_{\mathbf{q}} = \int (d \omega / 2 \pi)
S^{\theta}_{\mathbf{q} \omega} \propto 1/q^2$, and a lifetime $\propto 1/q^2$.
The term $\bm\nabla \bm\nabla:\bsf{Q}$ through which $\theta_{\mathbf{q} \omega}$
appears in the density equation (\ref{rho-eq-N}) can be re-expressed as
$\partial_x \partial_y \theta$. This autonomy of $\theta$ means that this term
can be viewed as colored noise with variance $\propto q_x^2 q_y^2/q^2$, and
relaxation time $\sim 1/q^2$. The zero-frequency weight of this
noise is thus $\sim q_x^2 q_y^2/q^4$, i.e., of order $q^0$ but anisotropic. By
contrast, the conserving noise $\bm\nabla \cdot \mathbf{f}^{\rho}$ has
zero-frequency weight that vanishes as $q^2$ and is thus irrelevant in the face
of the active contribution $\bm\nabla \bm\nabla : \bsf{Q}$, whose long-time,
long-distance effects are now revealed to be analogous to those of
a number-non-conserving noise. The deterministic part of the dynamics of
$\rho$, on the other hand, is diffusive and hence number-conserving. This
combination implies immediately that the equal-time correlator
$S^{\rho}_{\mathbf{q}} \sim \mbox{noise strength $\times$ relaxation time} \sim
q_x^2 q_y^2/q^6$, i.e., direction dependent but of order $1/q^2$, as argued
in greater detail in section \ref{gdf} for a linearized description of
density fluctuations in a \textit{polar} flock.

There is a puzzle here at first sight. We remarked above that the apolar
contribution to the active current (\ref{eq:active_curr}) is one order higher in
gradients than the polar term. Why then do density fluctuations in the
linearized theory for polar and apolar flocks show equally singular behaviour?
The reason is that in polar flocks density gradients act back substantially on
the polarization $\mathbf{p}$ through the pressure-like term $v_1 \bm\nabla \rho /
\rho_0$ in (\ref{P-eq-2}), reducing the orientational fluctuations that
engendered them. The effect of density gradients on $\theta$ in the apolar case
is far weaker, a quadrupolar aligning torque $\sim \partial_x \partial_z \rho$
which does not substantially iron out the orientational curvature that
engendered the density fluctuations. 

It should be possible to check, on a living system, the rather startling
prediction that flocks that go nowhere have enormous density fluctuations.
Melanocyte suspensions at high concentration show well-ordered nematic phases
when spread on a substrate as seen by \textcite{Gruler1999}. However, we know of no
attempts to measure density fluctuations in these systems. The prediction of
giant number fluctuations has been tested and confirmed first in a computer
simulation of an apolar flocking model by \textcite{Chate2006} and then in an
experiment on a vertically vibrated monolayer of head-tail symmetric
millimeter-sized bits of copper wire by \textcite{Narayan2007}.
Of particular interest is the fact that the experiment observed a
logarithmic time decay of the autocorrelation of the local density, a
stronger test of the theory than the mere occurrence of giant
fluctuations. Issues regarding the origin of the fluctuations in the
experiment were nonetheless raised by \textcite{Aranson2008}, with a
response by \textcite{Narayan2008}. Faced with such
large density fluctuations, it is natural to ask whether the active nematic
state is phase-separated in the sense of \textcite{Das2000}. This idea was tested
by
\textcite{Mishra2006} to a simulation model in which particles were advected by the
nematic curvature  \citep{Lebwohl1973}, and indeed the steady
state showed the characteristics of \textcite{Das2000}'s fluctuation-dominated
phase
separation.

\subsubsection{Self-propelled hard rods: a system of  ``mixed" symmetry?}
\begin{figure}
\centering
\includegraphics[width=0.97\columnwidth]{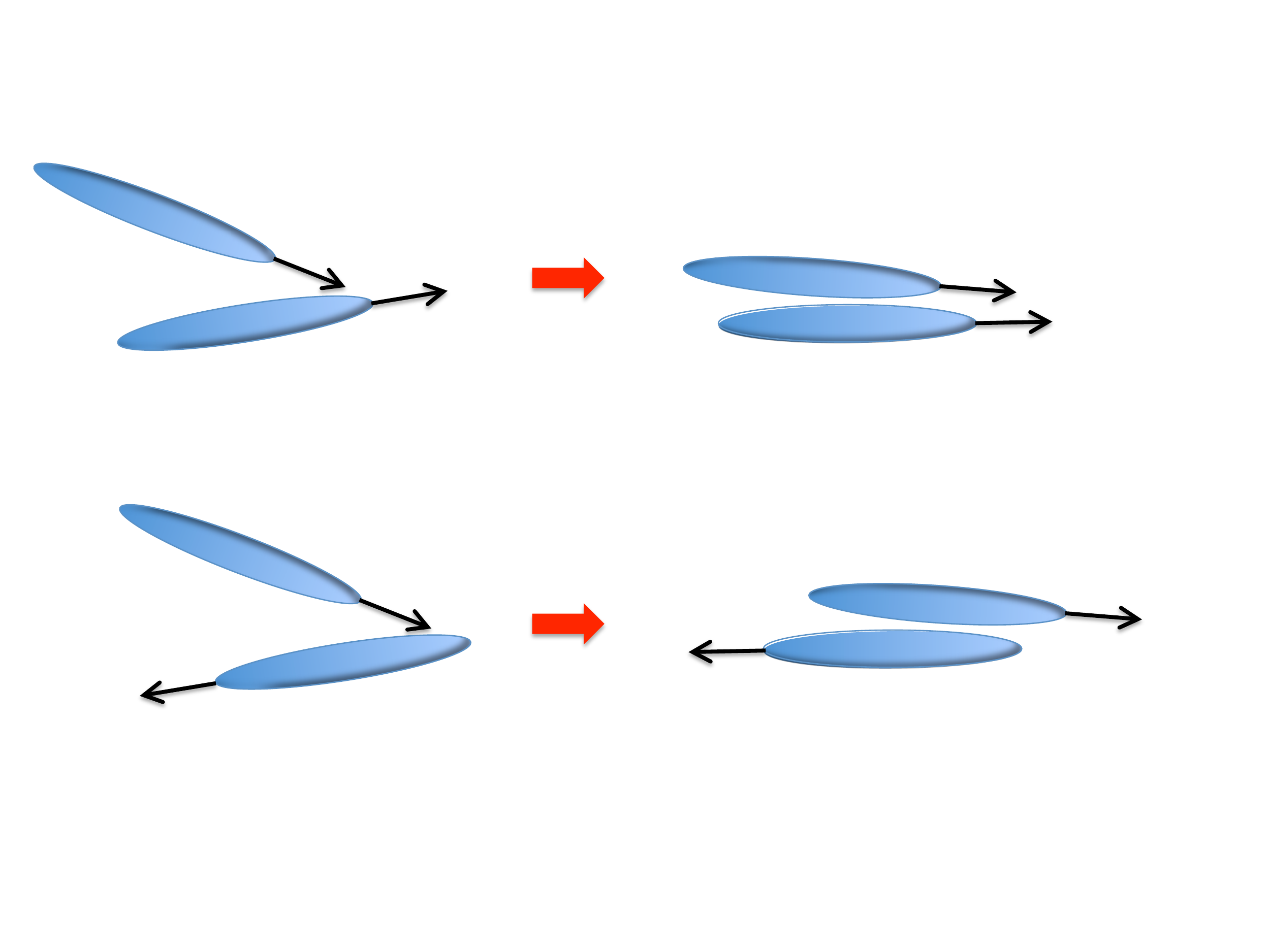}
\caption{(color online) The cartoon shows that the binary collision of
self-propelled hard 
rods provides an \emph{apolar} aligning interaction, in the sense that it aligns
rods 
regardless of their polarity.}
\label{fig:HR-collisiondiag}
\end{figure}

The Vicsek model contains an explicit alignment rule that yields an ordered
polar, moving state. The alignment interaction is in this case explicitly polar,
in the sense that it aligns particles head-to-head and tail-to-tail. A different
model of self-propelled particles where the alignment interaction has nematic,
rather than polar, symmetry has been studied in detail
by \textcite{Baskaran2008,Baskaran2008a,Baskaran2010}. These authors considered the
dynamics of a collection of self-propelled hard rods moving on a
frictional
substrate and interacting via hard core collisions. Numerical simulations of
self-propelled hard rods have also been carried out by
~\textcite{Peruani2006,Kraikivski2006,Peruani2008,Ginelli2010a,Yang2010} and have
revealed
a rich behavior. The model consists of rods interacting via excluded volume
interactions and self-propelled at speed $v_0$ along their  long axis. The
detailed analysis by \textcite{Baskaran2010} of the collisions of two
self-propelled
hard rods (see Fig.~\ref{fig:HR-collisiondiag}) shows that although
self-propulsion modifies the collision by enhancing longitudinal momentum
transfer and effectively tends to align two colliding rods, the alignment is in
this case ``apolar", in the sense that it tends to align particles
without distinguishing head from tail. \textcite{Baskaran2010} used tools of
nonequilibrium statistical mechanics to derive a modified Onsager-type theory
of self-propelled hard rods, obtaining a Smoluchowski equation that is modified
in several ways by self-propulsion. They then obtained hydrodynamic equations by
explicit coarse-graining of the kinetic theory. One result of this work is that
self-propelled hard rods do not order in a polar moving state in bulk. In spite
of the polarity of the individual particles, provided by the self-propulsion,
the symmetry of the system remains nematic at large scales. Self-propulsion
does, however, enhance nematic order, as shown earlier by
\textcite{Kraikivski2006} in a model of actin motility
assays, where actin filaments are modeled as rigid rods moving frictionally on a
in two-dimensional substrate and the propulsion is provided by myosin motors
tethered to the plane.  It is well known (in a mean field approximation
neglecting fluctuations) 
that thermal hard rods of length
$\ell$ undergo a continuous isotropic-nematic transition  in two dimension at a
density   $\rho_{Ons}=3/(2\pi\ell^2)$~\citep{Doi1986}, as first shown by Onsager
in 1949~\citep{Onsager1949}. Self-propelled rods perform a persistent random
walk~\citep{Kareiva1983,Hagen2009,Kudrolli2010} consisting of ballistic flights
randomized by rotational diffusion. At long time the dynamics is diffusive and
isotropic, with effective diffusion constant $D_e=D+v_0^2/2D_r$, where $D$ is
the longitudinal thermal diffusion coefficient and $D_r$ the rotational
diffusion rate. Self-propelled rods
of length $\ell$ then behave as rods with an effective length
$\ell_{eff}\simeq\sqrt{D_e/D_r}=\ell\sqrt{1+v_0^2/(2(DD_r)^{1/2}}$ and undergo
an 
isotropic-nematic transition at a density
 $\rho_{IN}=\rho_{Ons}/(1+\frac{v_0^2}{2\ell^2D_r^2})$, where we have used 
 $D\sim\ell^2 D_r$.  The microscopic derivation
carried out
in~\textcite{Baskaran2008a}
 supports this estimate. This result has  been verified recently via large scale
simulations of 
 self-propelled hard rods~\citep{Yang2010,Ginelli2010a}.

The statistical mechanics of self-propelled hard rods and the
derivation of the modified Smoluchowski equations is described in detail
in~\textcite{Baskaran2010} and will not be repeated here. Instead  the
hydrodynamic equations for this system are introduced phenomenologically below
and their consequences are discussed.

Once again, the only conserved field is the density $\rho({\bf r},t)$ of
self-propelled rods. Theory and simulations have indicated that although
self-propelled
particles with apolar interactions do not order in polar states, their dynamics
is characterized by an important interplay of polar order,
described by the
vector order parameter ${\bf p}$ defined in Eq.~\eqref{p}, and nematic order,
characterized by the alignment tensor $Q_{ij}$ of Eq.~\eqref{Q}. In other words,
although the polarization ${\bf p}$ does not describe a broken symmetry, it
seems necessary to incorporate its dynamics to capture the  rich physics of the
system into a continuum description.  We then consider coupled equations for the
conserved particle density, $\rho$, and the two coupled orientational order
parameters, ${\bf p}$ and ${\bsf{Q}}$.  Once again, the equations can be written
in terms of a ``free energy" as
\begin{subequations}
\begin{gather}
\label{rho-eq-X}\partial_t\rho=-\bm\nabla\cdot\left(v_0\rho
{\bf p}\right)+D\nabla^2\rho+\frac{1}{\gamma}\bm\nabla\bm\nabla{\bf :}
\frac{\delta F_{pQ}}{\delta {\bf Q}}+\bm\nabla\cdot{\bf f}^\rho\;,\\
\label{P-eq-X}\partial_t{\bf p}+\lambda_1({\bf p}\cdot\bm\nabla){\bf
p}-\delta_1{\bf p}\cdot{\bsf{Q}}+\delta_2{\bf Q:Q}{\bf p}=-\frac{1}{\gamma_p}\frac{\delta
F_{pQ}}{\delta{\bf p}}+{\bf f}\;,\\
\label{Q-eq-X} \partial_t \bsf{Q}+\lambda'_1({\bf
p}\cdot\bm\nabla)\bsf{Q}=-\frac{1}{\gamma_Q}\left[\frac{\delta F_{pQ}}{\delta
\bsf{Q}}\right]_{ST}+\bsf{f}^Q\;,
\end{gather}
\end{subequations}
where the subscript $ST$ denotes the symmetric, traceless part of any second
order tensor ${\bf T}$, i.e., $[{\bf T}]_{ST}$ has components
$T^{ST}_{ij}=\frac12\left(T_{ij}+T_{ji}\right)-\frac12\delta_{ij}T_{kk}$, and
\beqa
\label{FpQ}
F_{pQ}&=&F_Q+ \int_{\bf r}\,\Big\{\frac{\alpha_{pQ}}{2}|{\bf p}|^2+
\frac{ K}{2}\left(\partial_\alpha p_\beta\right)\left(\partial_\alpha
p_\beta\right)\notag\\
&&+P(\rho,p,S)\left(\bm\nabla\cdot{\bf
p}\right)-v_2Q_{\alpha\beta}(\partial_\alpha p_\beta)\Big\}\;,
\eeqa
We have grouped several terms (including some that were made explicit for instance in Eq.\eqref{Fp}) 
by writing a local spontaneous splay in terms of a ``pressure"  $P(\rho, p, S)$.  The reader should note that it is essential
to retain the general dependence of this pressure on $\rho$, $p$ and $S$ to generate all terms obtained from the microscopic theory. Here  $p=|{\bf p}|$ and
the magnitude $S$ of the alignment tensor has been defined by noting that in
uniaxial systems
in two dimensions we can write $Q_{\alpha\beta}=2S\left(n_\alpha
n_\beta-\frac12\delta_{\alpha\beta}\right)$, with ${\bf n}$ a unit vector. Then
$Q_{\alpha\gamma}Q_{\gamma\beta}=S^2\delta_{\alpha\beta}$.
The kinetic theory of Baskaran and Marchetti yields $\alpha_{pQ}>0$ at all
densities, indicating that no isotropic-polar transition
occurs in the system. In
contrast, $\alpha_Q(\rho)$ is found to change sign at a
characteristic density
$\rho_{IN}$, signaling the onset of nematic order. The closure of the
moments equations used in \textcite{Baskaran2008,Baskaran2008a} to derive
hydrodynamics only gives terms up to \emph{quadratic} in the fields in the continuum
equations, but is sufficient to establish the absence of a polar state and to
evaluate the renormalization of the isotropic-nematic transition density due to
self-propulsion. Cubic terms, such as the  $\sim \beta_Q{\bf Q}^3$
needed to evaluate the value of the nematic order parameter in the ordered state
and the active $\sim \delta_2{\bf Q:Q}{\bf p}$ can be obtained from a higher order closure and are included here for completeness.
 A higher order closure also confirms the absence of a  term proportional to ${\bf p}^3$
in the polarization equation, which could, if present, yield a bulk polar state. 
We note that a ${\bf p}\cdot{\bsf{Q}}\cdot{\bf p}$ term in the free energy is
permitted
and would yield both the ${\bf
p} \cdot {\bsf{Q}}$ term on the left hand side of  the ${\bf p}$ equation, as
well as a  ${\bf p}{\bf p}$ term in the $Q$ equation, with related
coefficients. The joint presence of these terms can lead to the existence
of a polar ordered state. The hard-rod kinetic theory of Baskaran and
Marchetti produces only the ${\bf p} \cdot {\bsf{Q}}$ term. The fact that the
theory  does
not generate the ${\bf p} \cdot {\bsf{Q}}$ term, which would be obligatory
had the dynamics been governed entirely by a free-energy functional,
underlines the nonequilibrium nature of the dynamics we are
constructing. It would be of great interest to find the minimal
extension of self-propelled hard-rod kinetic theory that could produce a
phase with polar order.

\paragraph{Homogeneous Steady States and their Stability.}

\begin{figure}
\centering
\includegraphics[width=0.9\columnwidth]{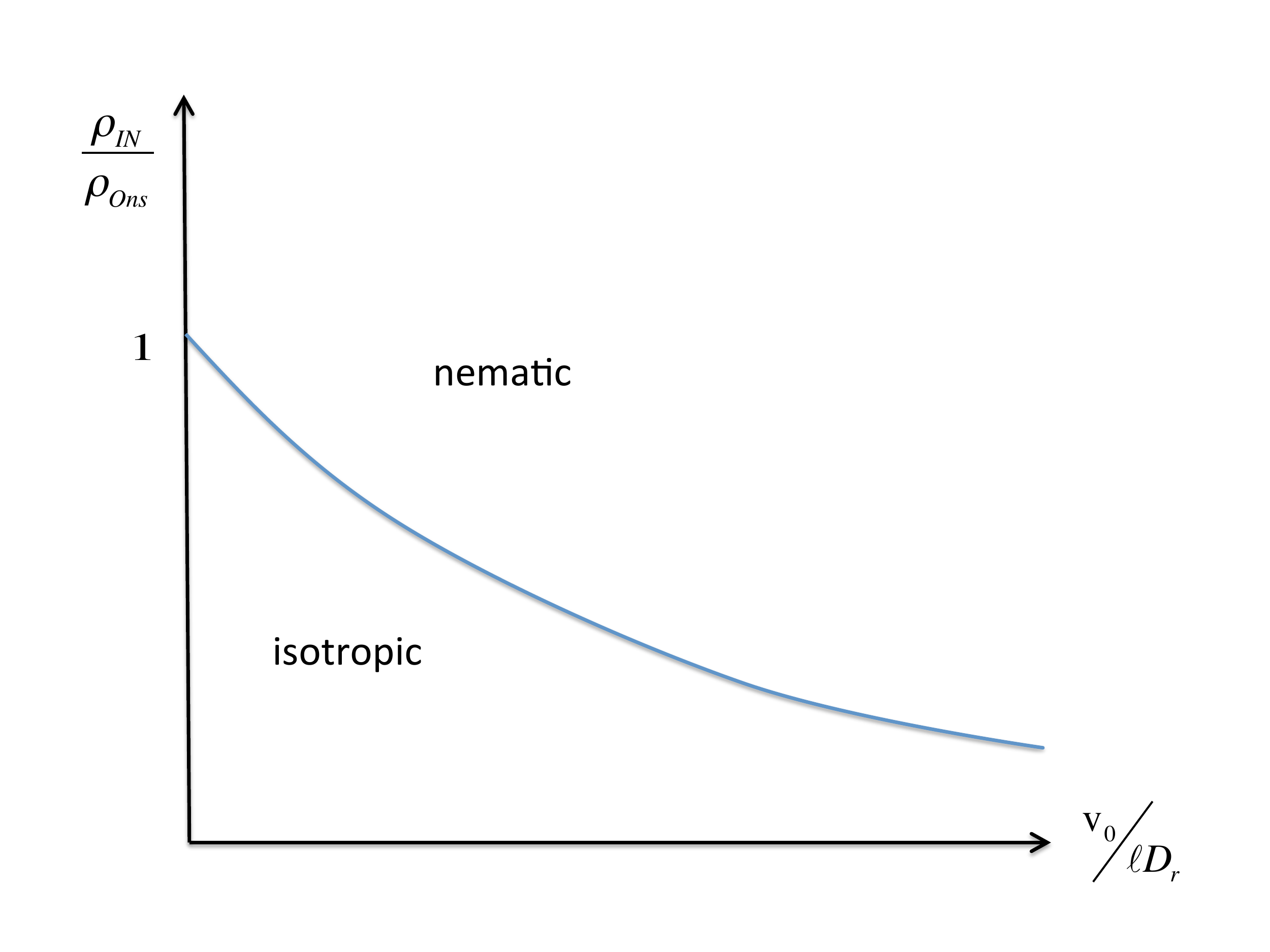}
\caption{(color online) The mean-field isotropic-nematic transition line for
self-propelled 
hard rods. Adapted with permission from ~\textcite{Baskaran2008a}.}
\label{fig:SProds-transition}
\end{figure}

The only homogeneous steady states of the system are an isotropic state with
$\rho=\rho_0$ and ${\bf p}={\bsf{Q}}=0$ and a nematic state where ${\bf p}=0$,
but ${\bsf{Q}}$ is finite. The transition occurs at $\rho_{IN}$ where
$\alpha_{pQ}\sim(\rho_{IN}-\rho) $ changes sign. As in the case of the active
nematic, we work in two dimensions denoted
$x$ and $y$, so that $\bsf{Q}$ has components $Q_{xx} = - Q_{yy} = S \cos 2
\theta, \, Q_{xy} = Q_{yx}  = S \sin 2 \theta$. 
In the nematic state $\alpha_{Q} < 0$, $\beta_Q>0$ in (\ref{Q-eq-X}), yielding a
phase 
with uniform nematic order and $S=\sqrt{-\alpha_{Q}/(2\beta_Q)}$.  We choose
our coordinates so that this reference state has $\theta = 0$. The phase
transition line is 
shown in Fig.~\ref{fig:SProds-transition} as a function of density $\rho_0$ and
self-
propulsion speed $v_0$ for the microscopic hard rod model discussed in 
~\textcite{Baskaran2008,Baskaran2008a}.

As in the case of polar and nematic active systems, the isotropic state is
stable. It can also 
support finite-wavevector propagating sound-like waves~\citep{Baskaran2008}.
The properties of the ordered state, on the other hand, are more subtle, and not
yet fully 
explored. If we consider a \emph{homogeneous } perturbation of the nematic
state, we find 
that fluctuations $\delta p_x$ of the polarization along the direction of
nematic order 
are unstable if $\lambda'_1 S_0>\alpha_{pQ}$. This global instability occurs
outside the 
hydrodynamic regime as it involves fluctuations in the polarization and its
significance 
remains a bit of a puzzle. Deep in the nematic state, neglecting fluctuations of
both 
polarization and magnitude of the nematic order parameter, the homogeneous
nematic state 
is unstable above a critical self propulsion speed. The instability arises from
a subtle 
interplay of splay and bend deformations and diffusion longitudinal and
transverse to the 
direction of nematic order and has been discussed in detail
in~\textcite{Baskaran2008}.

\begin{figure}[h]
\centering
\includegraphics[width=0.47\columnwidth]{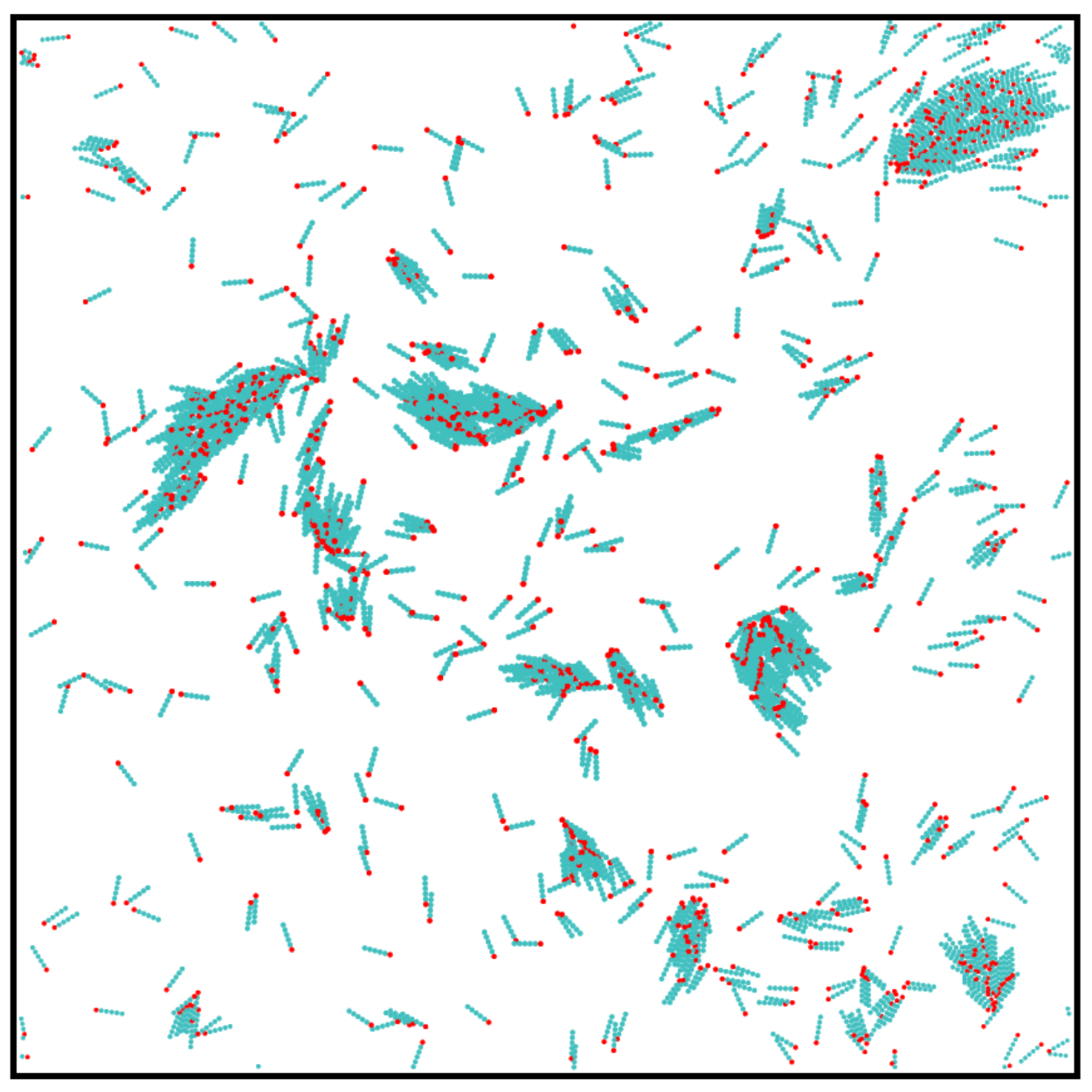}
\includegraphics[width=0.47\columnwidth]{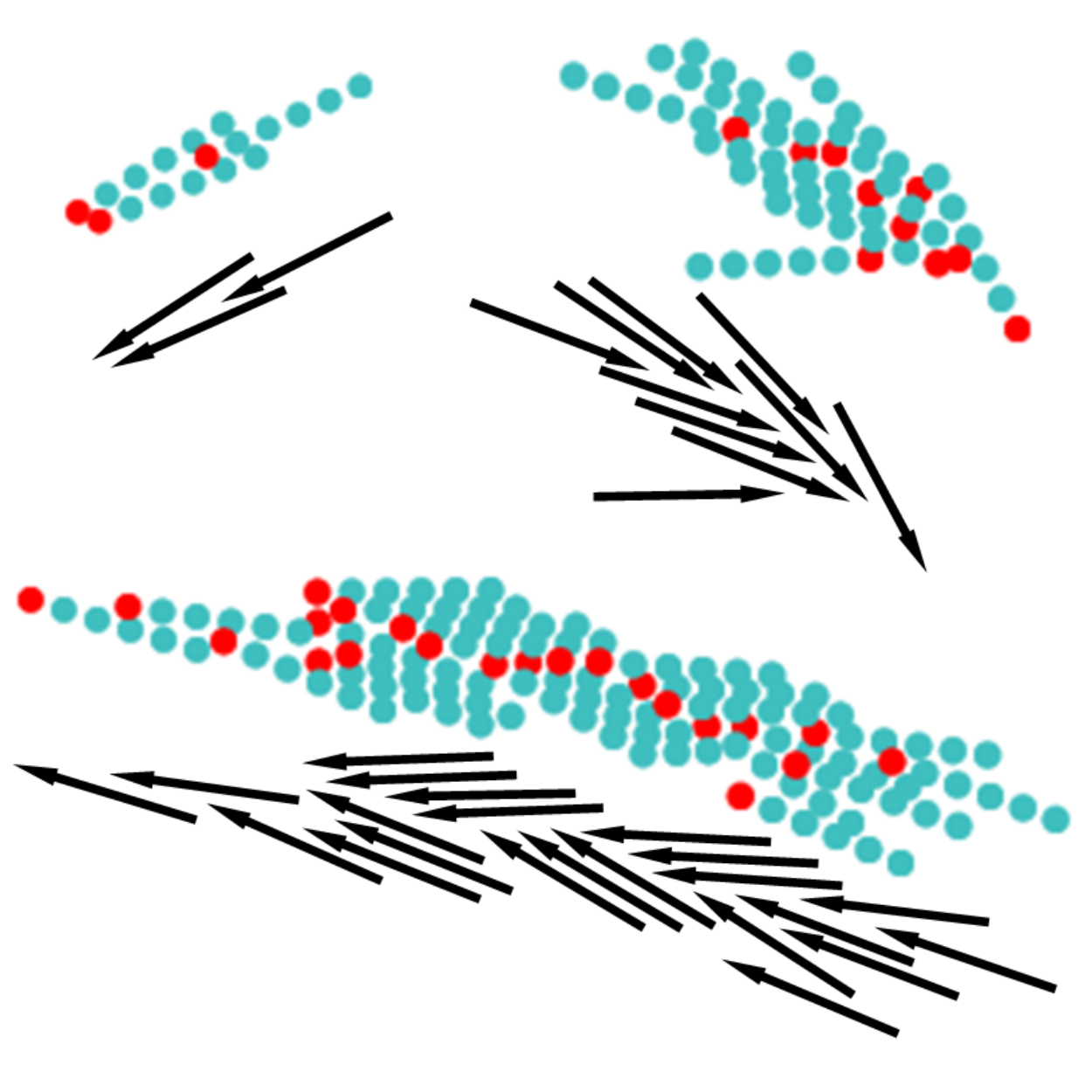}
\caption{(color online) Snapshots from simulations of self-propelled hard rods.
The red dots mark the front end of the rods, i.e., the direction of
self-propulsion. The rods are otherwise head-tail symmetric. The left image is
for a density $\rho_{rod}L^2_{rods}=0.7744$, where $L_{rod}$ is the length of
the rod, and $1/Pe=0.005$, where $Pe=L_{rod}v_0/D_\Vert$ is the Peclet number,
with $v_0$ the self-propulsion speed of each rod and $D_\Vert$ the diffusion
coefficient for motion along the long direction of the rod. The right image
shows a close-up of
clusters of sizes n=3, 10, and 22, highlighting the partially blocked structure
and the ``smectic-like" ordering of rods within a cluster. The clusters are
chosen from a simulation with parameters $1/Pe=0.000 95$ and the same density as
in the left image. Adapted with permission  from ~\textcite{Yang2010}.}
\label{Fig:Gompper}
\end{figure}
Numerical simulations of collections of self-propelled rods with steric
repulsion have 
recently revealed a rich behavior, quite distinct from that of polar Vicsek-type
models. As 
predicted by theory, self-propelled rods with only excluded volume interactions
do not  order 
in a macroscopically polarized state, but only exhibit nematic order, which
appears to be  
long-ranged in two dimensions. Again, simulations have also confirmed that self-
propulsion 
enhances the tendency for nematic ordering as well as aggregation and 
clustering~\citep{Peruani2006,Yang2010,Ginelli2010a}.
After an initial transient, the rods 
form polar clusters that travel in a directed fashion. Large clusters can form
by collisions of 
smaller ones and break up due to collisions with other clusters or due to noise.
\begin{figure}
\centering
\includegraphics[width=0.7\columnwidth]{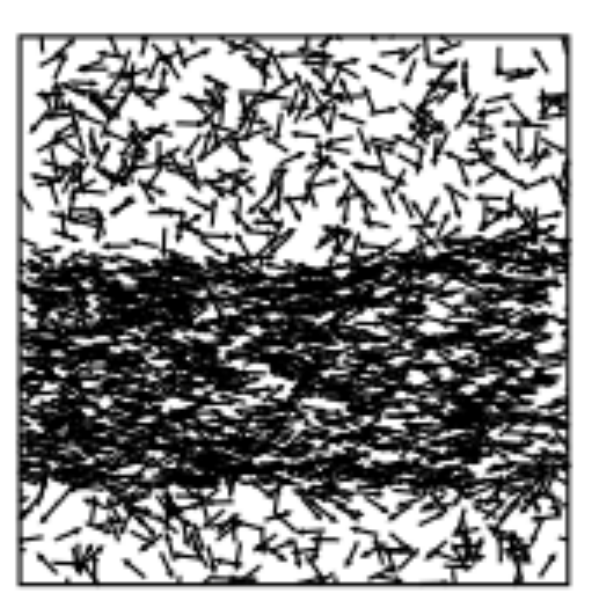}
\caption{Typical snapshots of nematic bands above the order-disorder 
transition.in the ordered phase. Arrows indicate the polar orientation of
particles; only a 
fraction of the particles are shown for clarity reasons.  Adapted with
permission from Ref~\citep{Ginelli2010a}.
 }
\label{fig:nematic-band}
\end{figure}
Eventually the system reaches a stationary state, in which the formation rate of
any cluster size equals its breakup rate, and the rods aggregate in large
stationary clusters. Above the order-disorder transition, 
phase separation manifests itself with the formation of nematic bands,
consisting of high density regions
where the particles are on average aligned with the long direction of the band, but
move in both 
directions, exhibiting no polar order~\citep{Ginelli2010a}, as shown in
Fig.~\ref{fig:nematic-band}. 
Self-propulsion has also been shown to increase the segregation tendency in
mixture of self-propelled rods with distinctly
different motilities~\citep{McCandlish2012}.
In closing this section we remark that mixed-symmetry models of the sort just
presented may well be the best description of the collective crawling of
rod-like bacteria such as those studied by \textcite{Wu2011}.
Cluster formation qualitatively similar to that observed in simulations of SP
rods has been observed in recent experiments in myxobacteria  \textcite{Peruani2012}.
\begin{figure}[h]
\centering
\includegraphics[width=0.45\columnwidth]{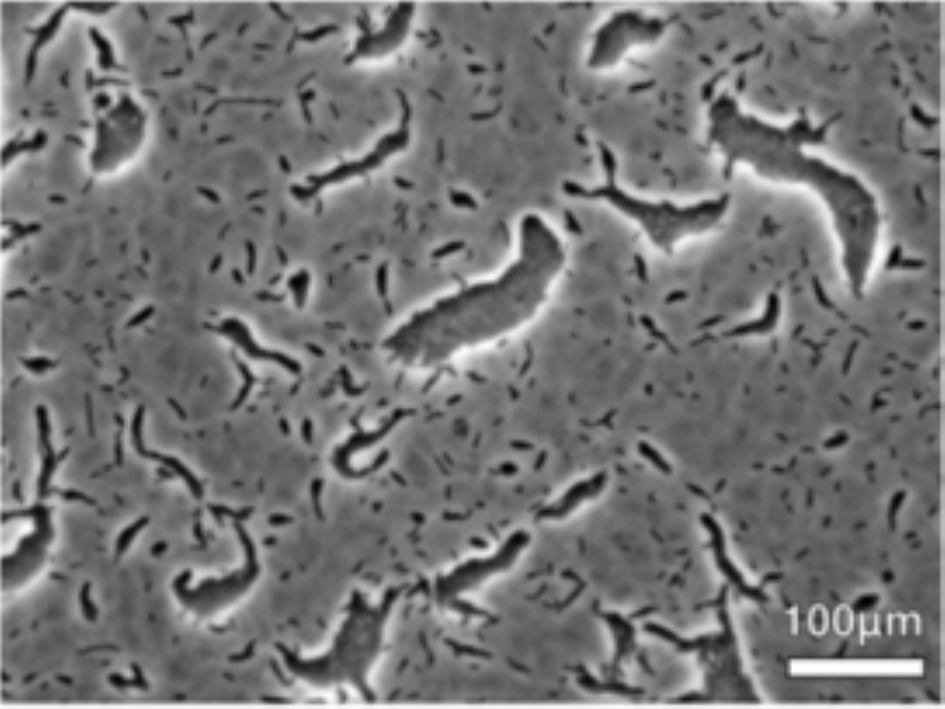}
\includegraphics[width=0.45\columnwidth]{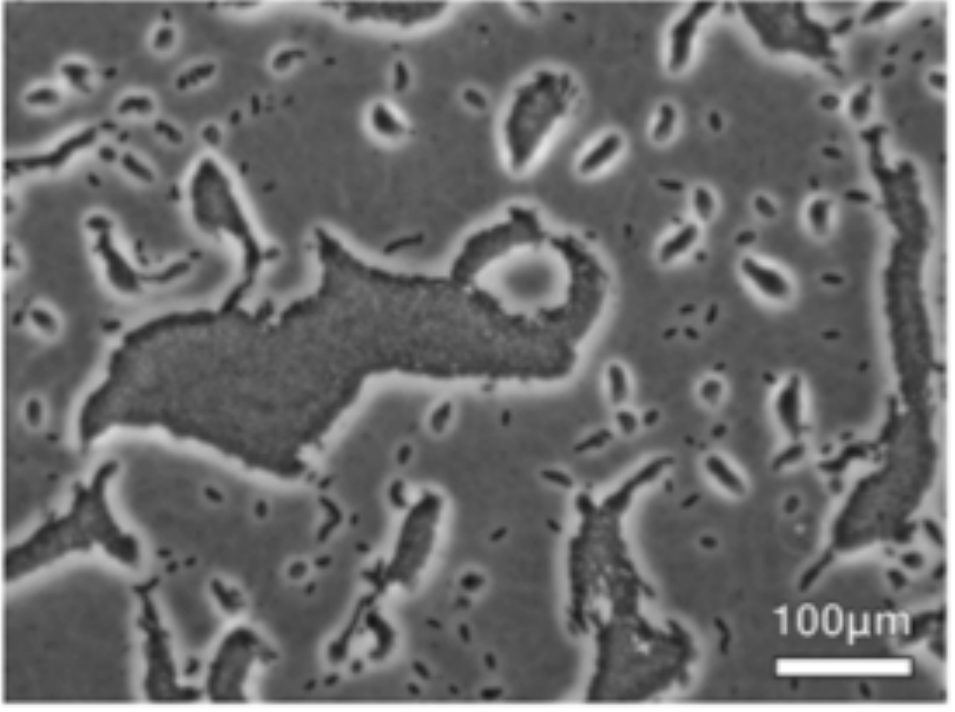}
\caption{ Cluster formation in myxobacteria mutants (SA2407) that cannot reverse
their gliding direction. At long times the dynamical clustering process reaches
a steady state that depends on cell density. The images show the clusters
obtained at two different packing fractions $\eta=\rho a$, with $\rho$ the
two-dimensional cell density and $a=4.4\mu m^2$ the average area covered by a
bacterium, with $\eta=0.16$ (left) and $\eta=0.24$ (right). Adapted with
permission from \textcite{Peruani2012}. }
\label{Fig:Peruani}
\end{figure}

\subsection{Current Status of Dry Active Matter}
\label{subsec:current-dry}
Although much progress has been made in the understanding and classification of
dry active matter, a number of open questions remain.

Clustering and phase separation, with associated giant number fluctuations, are
ubiquitous in active systems. It was first suggested that giant number
fluctuations may be a distinct property of the ordered state (nematic and polar)
of active systems, intimately related with the existence of a spontaneously
broken orientational symmetry~\citep{Toner1998,Simha2002,Ramaswamy2003}. In
ordered states such large fluctuations can indeed be understood as arising from
curvature-driven active currents unique to the ordered state of active
systems~\citep{Narayan2007}. More recently giant number fluctuations consistent
with a standard deviation growing linearly with the mean number of particles,
rather than with the $1/2$  exponent expected in systems where the central limit
theorem applies, 
have been reported in active systems of symmetric disks with no alignment rule
that do not exhibit any orientational broken symmetry~\citep{Fily2012}. In this
case large density fluctuations associated with strong clustering and phase
separation seem to arise from the general mechanism proposed in
\textcite{Tailleur2008,Cates2010} and reviewed recently in \textcite{Cates2012}
associated with the breaking down of detailed balance in systems that are driven
out of equilibrium by a local input of energy on each constituent, as in
self-propelled systems or bacterial suspensions. Strong density inhomogeneities
leading to persistent clustering have also been seen  in layers of vibrated
granular spheres~\citep{Aranson2008,Prevost2004}. Again, although in this case
other mechanisms such as the inelasticity of collisions may play a role, more
work is needed to elucidate the generic aspect of this ubiquitous phenomenon.

Although some controversy still remains~\citep{Vicsek1995,Aldana2007,Gonci2008},
there is now strong numerical evidence~\citep{Chate2007} that the
order-disorder
transition in dry active systems is discontinuous, with an associated wealth of
coexistence phenomena. One unusual aspect of the transition is that density
fluctuations destabilize the ordered state right at the mean-field order
transition. This  behavior seems to be associated with a phenomenon that has
been referred to as ``dynamical self-regulation" 
\citep{Gopinath2012,Baskaran2012} associated with the fact that in this class of
active systems the parameter that controls the transition, namely the density of
active particles, is not tuned from the outside, as in familiar
equilibrium
phase transitions, but rather it is dynamically convected or diffused by non
equilibrium active currents controlled by the  order parameter itself. It would
of course be very interesting to attempt to use formal field theory and
renormalization group methods to shed some light on this unusual non equilibrium
transition.

The behavior of collections of self-propelled or active particles is strongly
affected by the presence of boundaries and
obstacles~\citep{Wensink2008,Elgeti2009}. Ratchet effects have been demonstrated
experimentally and theoretically for active particles interacting with
asymmetric obstacles or on asymmetric substrates, without any external
forcing~\citep{Angelani2009,DiLeonardo2010}. For bacteria undergoing run and
tumble dynamics or self-propelled particles performing persistent random walks,
the guiding effect of asymmetric boundaries can yield a variety of rectification
effects~\citep{Galadja2007,Wan2008}, that can even power submillimiter gears, as
demonstrated in recent experiments~\citep{Sokolov2010,DiLeonardo2010}.
Understanding the interaction of bacteria with obstacles and with passive
particles is crucial for harnessing the collective power of these living systems
and developing micron-scale mechanical machines powered by microorganisms.
Whether these phenomena can be described in the dry limit discussed in this
section
or require a proper treatment of flow as introduced in the next section
and even hydrodynamic interactions  is still an open question.

Finally, there has been a surge of recent interest in the properties of dense
active matter and the living crystalline or glassy states that may be formed in
these systems. 
The nonequilibrium freezing of active particles may be directly relevant to the
behavior of suspensions of self-propelled Janus colloids or other artificial
microswimmers~\citep{Palacci2010}. Recent work has demonstrated that active
particles do form crystalline states, but the freezing and melting is in this
case a true non equilibrium phenomenon that cannot be described simply in terms
of an effective temperature for the system~\citep{Bialke2012}. In vitro
experiment on confluent monolayers of epithelial cells suggest that the
displacement field and stress distribution in these living systems  strongly
resemble both the dynamical heterogeneities of glasses and the soft modes of
jammed
packings~\citep{Angelini2011,Angelini2010,Trepat2009,Poujade2007,Petitjean2010}.
This observation has led to new interest in the study of active jammed and glassy
states~\citep{Henkes2011} obtained by packing self-propelled particles at high
density in confined regions or by adding attractive interactions. In these
models the interaction with the substrate has so far been described simply as
frictional damping. Although a rich and novel dynamical behavior has emerged, it
has become clear that a more realistic description of stress transfer with the
substrate will be needed to reproduce the active stress distribution observed
experimentally~\citep{Trepat2009} in  living cellular material.

\section{Active gels: self-driven polar
and apolar filaments in a fluid}
\label{wet}

In this section we describe an alternative method for constructing hydrodynamic
theories for a class of active materials. This approach
involves  a
systematic derivation of the hydrodynamic equations based on a generalized
hydrodynamic approach close to equilibrium following closely
the work of Martin Parodi, and Pershan for nemato-hydrodynamics
\citep{Martin1972,Gennes1993}.
As in the previous sections, the theory is mostly based on symmetries and does
not involve significant microscopic considerations; it is thus
applicable to a whole range of systems, which share the appropriate polar or
nematic symmetries and which are liquid at long times. 
 We focus here on an  `active gel', defined as  a fluid or
suspension of orientable objects endowed
with active stresses, with momentum damping coming from the viscosity
of the bulk fluid medium, rather than from friction with a
substrate or a
porous
medium \citep{Simha2002,Simha2002a,Kruse2004,Julicher2007}. 
The equations that
emerge are those proposed in \textcite{Simha2002} for self-propelling organisms,
but the development in \textcite{Kruse2004,Julicher2007} was carried out in the
context of the cytoskeleton of living cells, a network of polar actin filaments,
made active by molecular motors that consume ATP.

In  section \ref{nematicgels} we consider an active system
with a polarization field $\mathbf{p}$ but we suppose that the physics of this
system is invariant under a change of  $\mathbf{p}$ to $-\mathbf{p}$. We
therefore
consider an active \emph{nematic} gel. The polarization vector can in this case
be
considered as the director field for the nematic order.  An alternative approach
is to describe the ordering not by a director field but by a nematic alignment
tensor, ${\bf Q}$.
This approach, also part of the phenomenological treatment in \textcite{Simha2002},
is presented in \textcite{Salbreux2009}.  A brief discussion of \emph{polar}
active gels and of the effects
of polarity on the dynamics is given in  section \ref{polargel}.

 \subsection{Hydrodynamic equations of active gels}
\label{activegels}
For the sake of  simplicity in this section we only discuss a one component
active polar gel considering therefore that the complex composition of active
materials such as the
cytoskeleton can 
be described by an effective single component. The original formulation of
\textcite{Simha2002} for the hydrodynamics of self-propelled orientable suspensions
was already an explicitly multicomponent formulation of active liquid-crystal
hydrodynamics, but without a formal link to the nonequilibrium thermodynamic
approach to active systems of \textcite{Julicher2007}. Generalizations of the
latter approach to multicomponent systems are possible and have been recently
proposed by \textcite{Callan-Jones2011,Joanny2007}. The multicomponent theory takes
properly into account the relative permeation between the various components
which are neglected in the simpler one-component description. As in the previous
sections, we retain as slow variables, the number density $\rho$, the
polarization $\bf p$ and the momentum density ${\bf g} = \rho m{ \bf v}$ where $\bf
v$ is the local velocity of the gel and  $m$ the (effective)  mass of the 
molecules.

\subsubsection{Entropy production}
\label{entropy}

The derivation of generalized hydrodynamic equations is based on
the
identification 
of fluxes and forces from 
the entropy production rate $\dot {\cal S}$ \textit{\`{a} la} \textcite{Groot1984}.
We do not wish to consider heat
exchange here; we will assume that the active gel has a constant
temperature, meaning that it is in contact with a reservoir at a finite
temperature $T$. In this case, the entropy production rate is  related to the
rate of change in the free energy of the active gel
$T\dot {\cal S}=-\frac{dF}{dt} $.

For a passive gel at rest, the free energy density $f$ is a function of the 
two intensive variables, the density $\rho$ and the polarization $\bf p$ and its
differential is 
$df=\mu d\rho -h_{\alpha} dp_{\alpha}$. The field conjugate to the density is
the 
chemical potential $\mu$ and the field conjugate to the polarization is the
orientational 
field $\bf h$. For a passive system moving at a velocity $\bf v$, the density of
kinetic 
energy $\frac{1}{2} \rho m {\bf v}^2$  must be added to the free energy. 

For an active gel one must also take into account the fact that energy is 
constantly injected into the gel locally. A simple intuitive way to introduce
the energy 
injection is to assume that as in the case of the cytoskeleton, this is due to
a 
nonequilibrium chemical reaction such as the consumption of ATP. If the
energy gain per ATP molecule is denoted by $\Delta \mu$, and the rate of
advancement 
of the reaction (the number of ATP molecules consumed per unit time and unit
volume) 
is denoted by $r$, then the associated rate of change of the free energy per
unit volume is 
$-r \Delta \mu$. 
Taking into account all contributions, we find the entropy 
production rate of an active gel at a constant temperature $T$
\citep{Kruse2004,Kruse2005}
\begin{equation}
 T \dot{\cal S}=\int d{\bf r} \{ -\frac{\partial}{\partial t}(\frac{1}{2}\rho m
{\bf  v}^2) - \mu \frac{\partial \rho}{\partial t}+h_{\alpha}p_{\alpha} + r
\Delta \mu \}\;.
\end{equation}

\subsubsection{Conservation laws}
\label{conserv-laws}
The two conserved quantities in an active gel are the density and the momentum. 
The density conservation law reads
\begin{equation}
 \frac{\partial \rho}{\partial t} +{\bm\nabla}\cdot(\rho {\bf v})=0\;.
 \label{cons-rho}
\end{equation}
The momentum conservation law can be written as
\begin{equation}
 \frac{\partial g_{\alpha}}{\partial t}+\partial_{\beta}\Pi_{\alpha \beta}=0\;,
 \label{cons-g}
\end{equation}
where the momentum flux in the system is  $\Pi_{\alpha \beta}=
\rho m v_{\alpha}v_{\beta}
-\sigma^t_{\alpha \beta}$, where the first term is associated with
the so called
Reynolds stress
and  $\sigma^t_{\alpha \beta}$ is the total stress in the system. For most
active gels, in 
particular for biological systems, the Reynolds number is very small and we
ignore in the 
following the Reynolds stress contribution.

\subsubsection{Thermodynamics of polar systems}
\label{thermo}
For generality in this section we consider active gels in three dimensions.
The polarization free energy of an active polar material is a functional of the
three components 
of the polarization vector $\bf p$. However, if the system is not
in the vicinity
of 
a critical point there are only two soft modes associated with rotations of 
the polarization. The modulus of the polarization is not a 
hydrodynamic variable and is taken to be constant. Without any loss of
generality, we assume that the 
polarization is a unit vector, and that the effect of its modulus is integrated
in 
the phenomenological transport coefficients. 
The polarization free energy of the active gel is then the classical Frank free 
energy of a nematic liquid crystal \citep{Gennes1993}
\begin{align}
\label{frank}
 {F_p}=\int_{\bf r}\Big[&\frac{K_1}{2} ({\bm\nabla}\cdot {\bf p})^2 + 
 \frac{K_2}{2} ({\bf p}\cdot ({\bm\nabla}\times {\bf p}))^2 \nonumber\\
 &+ \frac{K_3}{2}
({\bf p}\times ({\bm\nabla}\times {\bf p}))^2 + v  {\bm\nabla}\cdot {\bf p}- 
\frac{1}{2}h^0_{\parallel} {\bf p}^2 \Big]\;.
\end{align}
The first three  terms correspond to the free energies of splay, twist and
bend deformations. The three Frank constants $K_i$ are positive. We have 
also added in this free energy a Lagrange multiplier $h^0_{\parallel}$ to insure
that 
the polarization is a unit vector. This free energy is very similar to the free
energy of 
Eq. \eqref{Fp}, although in Eq.~\eqref{Fp} we had made the additional approximation that 
the Frank constants are equal. 
In the case where the polarization
is a 
critical variable, one would also need to add to Eq.~\eqref{frank} a Landau
expansion 
in powers of the polarization modulus as done in Eq.~\eqref{Fp}.

The orientational field is obtained by differentiation of the free
energy
\eqref{frank}. 
It is useful to decompose it into a component parallel to the polarization
$h_{\parallel}$ 
and a component perpendicular to the polarization $h_{\perp}$ 

In the simple case where the system is two-dimensional, the polarization can be 
characterized by its polar angle $\theta$. In the approximation where the Frank 
constants are equal the perpendicular molecular field is $h_{\perp}=K\nabla^2
\theta$.

In a non-isotropic medium the stress is not symmetric. There is an
antisymmetric 
component of the stress associated to torques in the medium. As for nematic
liquid 
crystals, this anti-symmetric component can be calculated from the conservation 
of momentum in the fluid \citep{Gennes1993} and is equal to $\sigma^A_{\alpha
\beta}=\frac{1}{2}
(h_{\alpha}p_ {\beta}-p_{\alpha}h_{\beta}) \sim h_{\perp}$.

\subsubsection{Fluxes, forces and time reversal}
\label{fluxes-forces}

Using the conservation laws and performing integrations by parts, the entropy
production 
can be written as 
\begin{equation}
\label{entropy}
  T \dot{\cal S}=\int d{\bf r} \{ \sigma_{\alpha \beta} 
  v_{\alpha \beta}+P_{\alpha}h_{\alpha}+ r\Delta \mu     \}\;,
\end{equation}
where $\sigma_{\alpha \beta}$ is the symmetric deviatoric stress tensor defined 
by 
\beq
\sigma^t_{\alpha \beta}=\sigma_{\alpha \beta}+\sigma^A_{\alpha
\beta}-\delta_{\alpha\beta} P\;,
\label{sigma-dev}
\eeq
 with $\sigma_{\alpha\beta}^A$ the antisymmetric
part of the stress tensor and $P$  the pressure. We have also introduced the strain rate tensor  $v_{\alpha\beta}$ and the
antisymmetric part of the velocity gradients  associated to
the 
vorticity $\omega_{\alpha\beta}$, defined as
\begin{subequations}
\begin{gather}
\label{strain-rate}
v_{\alpha \beta}=\frac 1 2
(\partial_{\alpha} v_{\beta} + \partial_{\beta} v_{\alpha})\;,\\
\label{vorticity}
\omega_{\alpha \beta}=\frac 1 2 (\partial_{\alpha} v_{\beta} - 
\partial_{\beta} v_{\alpha})\;.
\end{gather}
\label{strain-vorticity}
\end{subequations}
Finally,  $P_{\alpha}=\frac{D p_\alpha}{Dt}$, with
\beq
\frac{D p_\alpha}{Dt}=\frac{\partial p_{\alpha}}{\partial
t} + 
v_{\beta} \partial_{\beta} p_{\alpha}+ \omega_{\alpha \beta}p_{\beta}\;,
\label{convected-der}
\eeq
 is the 
convected derivative of 
the polarization.
Note that if the system is chiral, kinetic momentum conservation must be taken 
into account properly and the local rotation is no longer given by 
$ \omega_{\alpha \beta}$ \citep{Furthauer2012}.

This form of the entropy production allows for the identification of three
forces: 
$v_{\alpha \beta}$ which has a signature $-1$ under time reversal, $h_{\alpha}$ 
which has a signature 
$+1$ under time reversal and $\Delta \mu$ which also has a signature $+1$ under 
time reversal. The conjugated fluxes are respectively 
$\sigma_{\alpha \beta}$, $P_{\alpha}$ and $r$.
The fluxes and associated driving forces that control the hydrodynamics of
active gels are summarized in Table ~\ref{Table:flux-force}.
\begin{table}
\centering
\begin{tabular}{|p{1cm}|p{1cm}|}
\hline\hline
flux & force\\ \hline
$\sigma_{\alpha\beta}$  & $v_{\alpha\beta}$\\
$P_\alpha$ &  $h_\alpha$\\
$r$ & $\Delta\mu$\\
\hline\hline
\end{tabular}
\caption{Fluxes and associated forces controlling the entropy production in  a
one-component nematic active gel.}
\label{Table:flux-force}
\end{table}

In a linear generalized hydrodynamic theory, the constitutive equations of the
active 
gel are obtained by writing the most general linear relation between fluxes and 
forces respecting the symmetries of the problem, such as translational and 
rotational 
symmetries with one vector $p_{\alpha}$ and one tensor 
$q_{\alpha \beta}=p_{\alpha}p_{\beta} - \frac {1} {3} \delta _{\alpha \beta}$ in
$3d$.
Particular care must be taken in considering the time reversal symmetry. The
fluxes 
must all be separated into a reactive component with a signature opposite to
that 
of the conjugate force and a dissipative component with the same signature as
the 
conjugate force. As an example, the reactive component of the stress is the
elastic 
stress and the dissipative 
component is the viscous stress.  Only the dissipative component of each flux 
contributes to the entropy production.

\subsection{Active nematic gels}
\label{nematicgels}

\subsubsection{Constitutive equations}
\label{const-eqs}
We first consider an active polar liquid for which the relationship between
fluxes and 
forces is local in time. 

It is convenient to split all tensors into  diagonal and  traceless parts
part: 
$\sigma_{\alpha \beta}=\sigma \delta _{\alpha \beta} + {\tilde \sigma}_{\alpha
\beta}$, 
with $\sigma=(1/3)\sigma_{\alpha\alpha}$, ${\tilde \sigma}_{\alpha \alpha}=0$, and $d$ the dimensionality. 
 Similarly, we let $v_{\alpha \beta}= \frac u 3 
\delta _{\alpha \beta} +{\tilde v}_{\alpha \beta}$, where $u =\partial_{\gamma}
v_{\gamma}$ is the divergence of the velocity field. Finally, all fluxes are written as the sums of reactive and dissipative parts,
\begin{subequations}
\begin{gather}
\sigma_{\alpha\beta}=\sigma_{\alpha\beta}^r+\sigma_{\alpha\beta}^d\;,\\
P_\alpha=P_\alpha^r+P_\alpha^d\;,\\
r=r^r+r^d\;.
\end{gather}
\end{subequations}

\paragraph{Dissipative fluxes.}
 Only fluxes and forces with the same time signature are coupled and 
 $\sigma_{\alpha \beta}$ is only coupled to $v_{\alpha \beta}$. This leads 
 to the constitutive equations
 \begin{subequations}
 \begin{gather}
  \sigma^d = {\bar \eta} u \\
  {\tilde \sigma}^d_{\alpha \beta} = 2\eta {\tilde v}_{\alpha \beta}\;.
  \end{gather}
   \label{eq:diss-fluxes}
 \end{subequations}
We ignore here the tensorial character of the viscosity and assume
 only two viscosities, as for an 
isotropic fluid: the shear viscosity $\eta$ and
the 
longitudinal viscosity $\bar \eta$. 
 The two other fluxes are coupled and the corresponding constitutive equations
read \citep{Kruse2004,Kruse2005}
 \begin{eqnarray}
  P^d_{\alpha} &=& \frac{h_{\alpha}}{\gamma_1} + \epsilon p_{\alpha} \Delta
\mu\;, \\
  r^d &=& \Lambda \Delta \mu + \epsilon p_{\alpha} h_{\alpha}\;.
 \end{eqnarray}
$\gamma_1$ is the rotational viscosity and we use here the Onsager symmetry
relation which imposes that the ``dissipative'' Onsager matrix is symmetric.

\paragraph{Reactive fluxes}

The reactive Onsager matrix is antisymmetric and couples fluxes and forces of
opposite 
time reversal signatures. 
\begin{subequations}
\begin{gather}\label{eq:reac-stress-trace}
 \sigma^r = -{\bar \zeta} \Delta \mu +{\bar \nu}_1 p_{\alpha} h_{\alpha}\;,\\
 \label{eq:reac-stress-dev}
 {\tilde \sigma}^r_{\alpha \beta} = -\zeta \Delta \mu q_{\alpha\beta}  +
 \frac{\nu_1}{2}(p_{\alpha}h_ {\beta}+p_{\beta}h_ {\alpha}-\frac{2}{3}
 p_{\gamma}h_ {\gamma} \delta_{\alpha \beta})\;, \\
  \label{eq:reac-flux-P}
 P^r_{\alpha} = -{\bar \nu}_1 p_{\alpha}\frac u 3 -\nu_1 p_{\beta}
 {\tilde v}_{\alpha \beta}\;,\\
  \label{eq:reac-flux-r}
 r^r = {\bar \zeta} \frac{u}{3}+ \zeta q_{\alpha \beta}{\tilde v}_{\alpha
\beta}\;.
\end{gather}
\label{eq:reac-stress}
\end{subequations}
In most of the following we consider incompressible fluids so that $u=
\bm\nabla\cdot{\bf v}=0$. 
In this case the diagonal component of the stress can be included in the
pressure which 
is a Lagrange multiplier ensuring incompressibility and one can set ${\bar
\zeta} = 
{\bar \nu}_1= {\bar \eta}=0$.

To summarize, the hydrodynamic equations for an incompressible
one-component active fluid of nematic symmetry are given by
\begin{subequations}
\label{eq:one-fluid_nem_hydro}
\begin{gather}
\label{eq:NS}
m\rho\left(\partial_t+{\bf v}\cdot\bm\nabla\right){\bf v}=-\bm\nabla
P+\bm\nabla\cdot\bm\sigma\;,\\
\label{eq:polarization}
\left(\partial_t+{\bf
v}\cdot\bm\nabla\right)p_\alpha+\omega_{\alpha\beta}p_\beta=-\nu_1
v_{\alpha\beta}p_\beta+\frac{1}{\gamma_1}h_\alpha+\epsilon\Delta\mu p_\alpha\;,
\end{gather}
\end{subequations}
to be supplemented with the incompressibility condition $\bm\nabla\cdot{\bf
v}=0$. Assuming the Frank constants are all equal to $K$, the molecular field
$h_\alpha$ is given by
\begin{equation}
\label{eq:oneFrank_molfield}
h_\alpha=K\nabla^2 p_\alpha+h_\Vert^0 p_\alpha\;,
\end{equation}
with $h_\Vert^0$ a Lagrange multiplier to be determined by the condition $|{\bf
p}|=1$. Finally, it is convenient for the following to write the deviatoric stress tensor given by the sum of trace and
deviatoric parts of the dissipative and reactive components by separating out passive and active parts as
\beq
\sigma_{\alpha\beta}=\sigma^p_{\alpha\beta}+\sigma^a_{\alpha\beta}\;,
\label{eq:totalstress}
\eeq
with passive and active contributions given by
\begin{subequations}
\begin{gather}
\label{sigma-passive}
\sigma^p_{\alpha\beta}= 2\eta {\tilde v}_{\alpha \beta} +
 \frac{\nu_1}{2}(p_{\alpha}h_ {\beta}+p_{\beta}h_ {\alpha}-\frac{2}{3}
 p_{\gamma}h_ {\gamma} \delta_{\alpha \beta})\;,\\
 \label{sigma-active}
\sigma^a_{\alpha\beta}=  -\zeta \Delta \mu q_{\alpha\beta} \;.
\end{gather}
 \end{subequations}
In many biological applications inertial effects are negligible and the
Navier-Stokes equation \eqref{eq:NS} can be replaced by the Stokes equation
obtained by simply neglecting all inertial terms on the left hand side of
Eq.~\eqref{eq:NS} and corresponding to a force balance equation, given by
\begin{equation}
\label{eq:forcebalance_general}
-\bm\nabla P+\bm\nabla\cdot\bm\sigma=0\;.
\end{equation}

\subsubsection{Microscopic interpretation of the transport coefficients}
\label{micro-transport}

The Onsager approach that we have outlined introduces several transport
coefficients. 
Some of these coefficients exist for passive nematic liquid crystals such as the
viscosities, 
the rotational viscosity $\gamma_1$ or the flow -coupling coefficient $\nu_1$. 
The two important new coefficients are the transport coefficients associated 
with the activity of the system, $\epsilon$ and $\zeta$.
The active stress in the system is $\sigma_a= -\zeta \Delta \mu$. 
In the case of the cytoskeleton 
this can be viewed as the stress due to the molecular motors, which tend to
contract
the gel. The sign of the activity coefficient  $\zeta$ is not imposed by
theory. 
A negative value corresponds to
a contractile stress as in the  actin cytoskeleton. A positive value of $\zeta$
corresponds 
to an extensile stress as observed in certain bacterial suspensions.
Conceptually, active stresses in living matter were first discussed by
\textcite{Finlayson1969}. These authors, however, specifically steer clear of
uniaxial stresses
arising from motor-protein contractility. The first incorporation of
self-propelling stresses into the generalized hydrodynamics of orientable
fluids was by \textcite{Simha2002}, although it has long been understood that the
minimal description of a single force-free swimmer is a force dipole
\citep{Brennen1977,Pedley1992}.

The other active coefficient, $\epsilon$, is an active orientational field that
tends 
to align the polarization 
when it is positive. In the limit where the modulus of the polarization is $p=1$
one 
can always consider that $\epsilon=0$ and introduce an effective activity
coefficient 
$ \zeta +\epsilon \gamma_1 \nu_1$. In the following we therefore choose $\epsilon
=0$ 
and use this effective value of the activity coefficient $\zeta$.

\subsubsection{Active currents in nematic and polar systems from forces and
fluxes}
\label{subsub_activecurrents}
Let us take the opportunity to show that the active currents in nematic
(section \ref{sec:nemsubstrate}) and polar (section \ref{sec:polardry}) systems
follow naturally from the forces-and-fluxes framework. In this short treatment
we ignore fluid flow. Consider a mesoscopic region in our active medium, in a
chemical potential gradient $\nabla \Phi$ corresponding to the concentration
$\rho$, and subjected to a nonzero chemical potential difference $\Delta \mu$
between a fuel (ATP) and its reaction products (ADP and inorganic phosphate).
For small departures
from equilibrium, the fluxes $r$ (the rate of consumption of ATP molecules) and
$\mathbf{J}$ must be linearly related to
$\Delta \mu$ and $\nabla \Phi$. The presence of local orientational order in the
form of $\mathbf{p}$ and $\bsf{Q}$ allows one to construct scalars $\mathbf{p}
\cdot \nabla \Phi$ and $\nabla \cdot \bsf{Q} \cdot \nabla \Phi$; thus $r$ in
general gets a contribution $(\zeta_p \mathbf{p} + \bar\zeta_Q \nabla \cdot \bsf{Q})
\cdot \nabla \Phi$, where $\zeta_p$ and $\bar\zeta_Q$ are kinetic coefficients
depending in general on $\rho$ and other scalar quantities. The symmetry of
dissipative Onsager coefficients then implies a contribution
\beq
\label{eq:active_curr_both}
\mathbf{J}_{active} = (\zeta_p \mathbf{p} + \bar\zeta_Q \nabla \cdot \bsf{Q})
\Delta \mu
\eeq
to the current. In the presence of a maintained constant value of $\Delta
\mu$ the current (\ref{eq:active_curr_both}) rationalizes, through the
$\zeta_p$ and $\bar\zeta_Q$ terms, the form of (\ref{rho-eq}) (with $\zeta_p\Delta\mu v_0\rho$) and
(\ref{eq:active_curr}) (with $\bar\zeta_Q\Delta\mu\rightarrow\zeta_Q$). In particular, it underlines the fact that
active currents do not require an explicit polar order parameter. Even without
$\mathbf{p}$, the term in $\bar\zeta_Q$ in (\ref{eq:active_curr_both}) tells us that
curvature in the spatial arrangement of active filaments gives rise to particle
motion. In retrospect this is not shocking: a splayed or bent configuration of a
nematic phase has a vectorial asymmetry, as argued through Fig.
\ref{fig:curvcurr}. In a system out of equilibrium this asymmetry should reflect
itself in a current. As shown by \textcite{Ramaswamy2003} and discussed in section
\ref{sec:nemsubstrate}, $\bar\zeta_Q$ leads to giant number fluctuations in active
nematics.

\subsubsection{Viscoelastic active gel}
\label{viscoel}
An active gel is not in general a simple liquid but rather a viscoelastic medium
with 
a finite viscoelastic relaxation time, which is only liquid at  long time
scales. 
In a passive visco-elastic medium the constitutive relation between stress
and
strain 
is non local in time. The simplest description of a visco-elastic
medium is the 
so-called Maxwell model where the system only has one relaxation time
\citep{Larson1988}.
Within this model the constitutive equation is 
\begin{equation}
 \frac{D{\tilde \sigma}_{\alpha \beta}}{Dt}+\frac{1}{\tau}{\tilde
\sigma}_{\alpha \beta}
 = 2E {\tilde v}_{\alpha \beta}\;.
\end{equation}
The Maxwell model involves two material constants, the viscoelastic relaxation
time 
$\tau$ and the shear modulus $E$. The long time viscosity of the medium is then 
$\eta =E \tau$. In order to respect  rotational and
translational invariance,
we use 
here a convected Maxwell model with a convected time derivative of the stress
tensor
$\frac{D {\tilde \sigma}_{\alpha \beta}}{D t}=
\frac{\partial{\tilde \sigma}_{\alpha \beta} }{\partial t}+ v_{\gamma}
\partial_{\gamma}{\tilde \sigma}_{\alpha \beta}+\omega_{\alpha \gamma}{\tilde
\sigma}_{\gamma \beta}+{\tilde \sigma}_{\gamma \alpha}
\omega_{\beta\gamma}$. 
Note that  there are several ways of defining the convective derivative of tensors. For simplicity we use the same notation $\frac{D}{Dt}$ to denote convected derivatives of vectors and tensors.

The generalization of the Onsager hydrodynamic approach of the previous section 
to a viscous elastic polar passive medium lead to the constitutive equations for
an 
active gel \citep{Julicher2007}.
\begin{subequations}
\label{constitutive}
\begin{gather}
 2\eta v_{\alpha \beta}=(1 + \tau\frac{D}{Dt})\left({\tilde \sigma}_{\alpha
\beta} + 
 \zeta \Delta \mu q_{\alpha \beta} - \frac{\nu_1}{2}
 (p_{\alpha}h_{\beta}+p_{\beta}h_{\alpha})\right)\;,\\
 \frac{D p_{\alpha}}{Dt} = \frac{1}{ \gamma_1}
 (1 + \tau\frac{D}{Dt})h_{\alpha}-\nu_1 v_{\alpha \beta}p_{\beta}\;,
\end{gather}
\end{subequations}
where for simplicity we have only considered an incompressible active gel where
the modulus of the polarization is unity.
Note that the memory of the system not only plays a role for the stress but also
for 
the dynamics of the orientation and that we have supposed that the two
corresponding
relaxation times are equal.

It is important to note that the dynamical equation for the polarization is very
similar to 
Eq.~ \eqref{P-eq} used in section \ref{dry} which has been obtained using the
same
symmetry arguments; Eq.~\eqref{P-eq} ignores memory effects and therefore the
visco-elasticity of the polarization response; all the extra terms in
Eq.~\eqref{P-eq} do not appear here because the Onsager approach that we use
only
derives the linear hydrodynamic theory.   

\subsection{Active polar gels}
\label{polargel}

For the sake of simplicity, we have only presented here the derivation of the 
hydrodynamic theory of active gels in the simplest case where the system is a
single component fluid and has nematic 
symmetry, 
${\bf p}$ being the director,  
 and ignore any type of noise. Several extensions 
of this theory have been proposed. 

For a polar system, there is an extra polar term in the free energy
\eqref{frank}, ${F}_p=
\int d{\bf r}~ v(\rho) \left({\bf \nabla}\cdot {\bf p}\right)$. This spontaneous
splay term is a surface term 
if the coefficient  $v$ is a constant. If $v$ depends on the local density, this
term, that was included in Eq.~\eqref{Fp},
yields a ``pressure-gradient" proportional to $\bm\nabla\rho$ in the equation
for the polarization.  Other
 non linear polar 
terms that can be 
added on the right hand side of  the dynamical equation for the polarization 
\eqref{constitutive}
are proportional to 
${\bf p}\cdot \nabla {\bf p}$, $\nabla {\bf p^2}$ and $ {\bf p} \nabla\cdot {\bf
p}$. These
terms have already been considered in Eq.\eqref{P-eq}. The first of these terms
cannot 
be derived from 
a free energy and is therefore an active term proportional to $\Delta \mu$. Its
effect in 
a dynamics linearized about an
ordered state was considered for active liquid-crystalline suspensions by
\textcite{Simha2002}. The two
other terms can be derived from a free energy and have both active and 
passive contributions.   Within the Onsager linear hydrodynamics scheme,  other
polar terms 
in the equations show up only at nonlinear order or at sub-leading order in a
gradient expansion.  Active stresses proportional to $\Delta\mu(\partial_i p_j +
\partial_j p_i)$ unique to polar fluids
are obtained from the microscopic theory~\citep{Marchetti2007}, but are
considered nonlinear in the driving forces in the context of the Onsager
approach. The effect of these polar terms has been studied in
detail
by Giomi and collaborators \citep{Giomi2008}. In general these polar terms are
important when describing an
active suspension as opposed to the one-component system considered here. In
this case these terms yield
spatial inhomogeneities in the concentration of active
particles~\citep{Tjhung2011,Giomi2012} that are not obtained in active
suspensions with nematic symmetry~\citep{Giomi2011,Giomi2012a}.

The effect of noise in active polar gels can be studied by introducing random 
Langevin forces in the constitutive equations. For treatments including thermal
and active non-thermal noise sources in a systematic way, see
\textcite{Lau2003,Lau2009,Hatwalne2004,Chen2007,Basu2008,Sarkar2011}. The
precise description of the statistics of active noise requires a microscopic
description of the gel which is not generic and goes beyond the scope of this
review \citep{Basu2008}. \textcite{Hatwalne2004} introduce the active noise
in the isotropic phase of
an active gel through the fluctuating active stress in the Navier-Stokes
equation, and use scaling arguments to estimate the apparent temperature it
would generate in a tracer diffusion measurement. 

Most active systems described in the earlier sections of this review are
multicomponent systems containing a solvent and active objects. In many
instances these
systems can be described by an effective one-component theory as described here.
However, in particular when considering viscoelastic effects the one-component
theory ignores the permeation of the solvent through the active gel. A detailed
two component theory of active gels that takes into account properly permeation
effects is given by~\textcite{Callan-Jones2011}.

\subsection{Active Defects}
Ordered phases of active matter, like their counterparts at thermal 
equilibrium, should exhibit topological defect configurations, generated
either through specific boundary conditions or spontaneously in the bulk. As in
equilibrium systems, the nature of these defects should depend on whether the
system has polar  or apolar symmetry
 \citep{Palffy-Muhoray1988,Gennes1993,Kung2006}.
The selection criterion for the defect strength in active systems is not
obvious, as one cannot a priori invoke free-energy minimization as in
equilibrium systems. However, it does appear that experiments
see strength $+1$ and strength $+1/2$ defects, respectively, in polar
\citep{Nedelec1997}
and apolar \citep{Narayan2007} active systems. 
Activity confers a particularly
interesting property on defects, namely rotational \citep{Nedelec1997,Kruse2006}
or translational \citep{Narayan2007,Dogic2012} movement, with sense or direction
determined
by the chirality or polarity associated with the defect.

The first quantitative experiments which explicitly demonstrated how active
mixtures of long rods (microtubules) and motors (kinesin) could spontaneously
form defects such as asters and spirals were reported in
\textcite{Nedelec1997,Surrey2001}; see Fig. \ref{fig:Surrey}. These patterns showed
a remarkable resemblance to the microtubule-based spindle patterns in the cell,
thus suggesting that the gross features of spindle patterning could be
understood as arising from a self-organization of simple elements. Several
qualitative features of these experiments including the defect patterns could be
simply understood using continuum models describing the polar orientation of the
rigid filaments and the density of processive motors
\citep{Lee2001,Sankararaman2004}.

A detailed study of the nature of defects and their phase transitions
within the framework of the active gel theory was done by 
\textcite{Kruse2004,Kruse2006}
who showed that flows arising from active stresses leads to
systematic rotation of chiral defects. We summarize here the calculation,
working as in \textcite{Kruse2004} with the ordered state described by a vectorial
order parameter $\mathbf{p}$ in two dimensions. Polarity enters nowhere in
the analysis, except as justification for working with a strength $+1$ defect.
We parametrize a two-dimensional defect configuration of the polarization field
${\bf p}$ (of unit magnitude) with topological charge $\pm 1$ using polar
coordinates $(r,\theta)$. Thus charge $+ 1$ defects such as {\it asters}, {\it
vortices} and {\it spirals}, may be represented by an angle $\psi$, with $p_r =
\cos \psi$ and $p_{\theta}= \sin \psi$, such that $\psi=0$ (or $\pi$) is an
aster, $\psi=\pm \pi/2$ a vortex and $\psi=\psi_0$ (any other constant) a
spiral. At equilibrium, a situation corresponding to defect configurations in a
ferroelectric nematic liquid crystal,
the optimal value of $\psi$ is obtained by minimizing the Frank free-energy
functional ${ {F}}$ (\ref{frank})
with $K_2=0$ (no twist as we are working in two dimensions). With appropriate
boundary conditions, it easy to see that the stable defect configurations are
(i) asters when $K_1 < K_3$, so that splay is favored, (ii) vortices
when $K_1 > K_3$, i.e., bend is favored, and (iii) spirals when $K_1=K_3$.
In an active system, however, the stability of defect configurations is obtained
by solving the dynamical equations for the polarization ${\bf p}$
together with the conditions of force balance and overall incompressibility, 
as in section \ref{nematicgels}
%

\begin{figure}
\centering
\includegraphics[width=0.70\columnwidth]{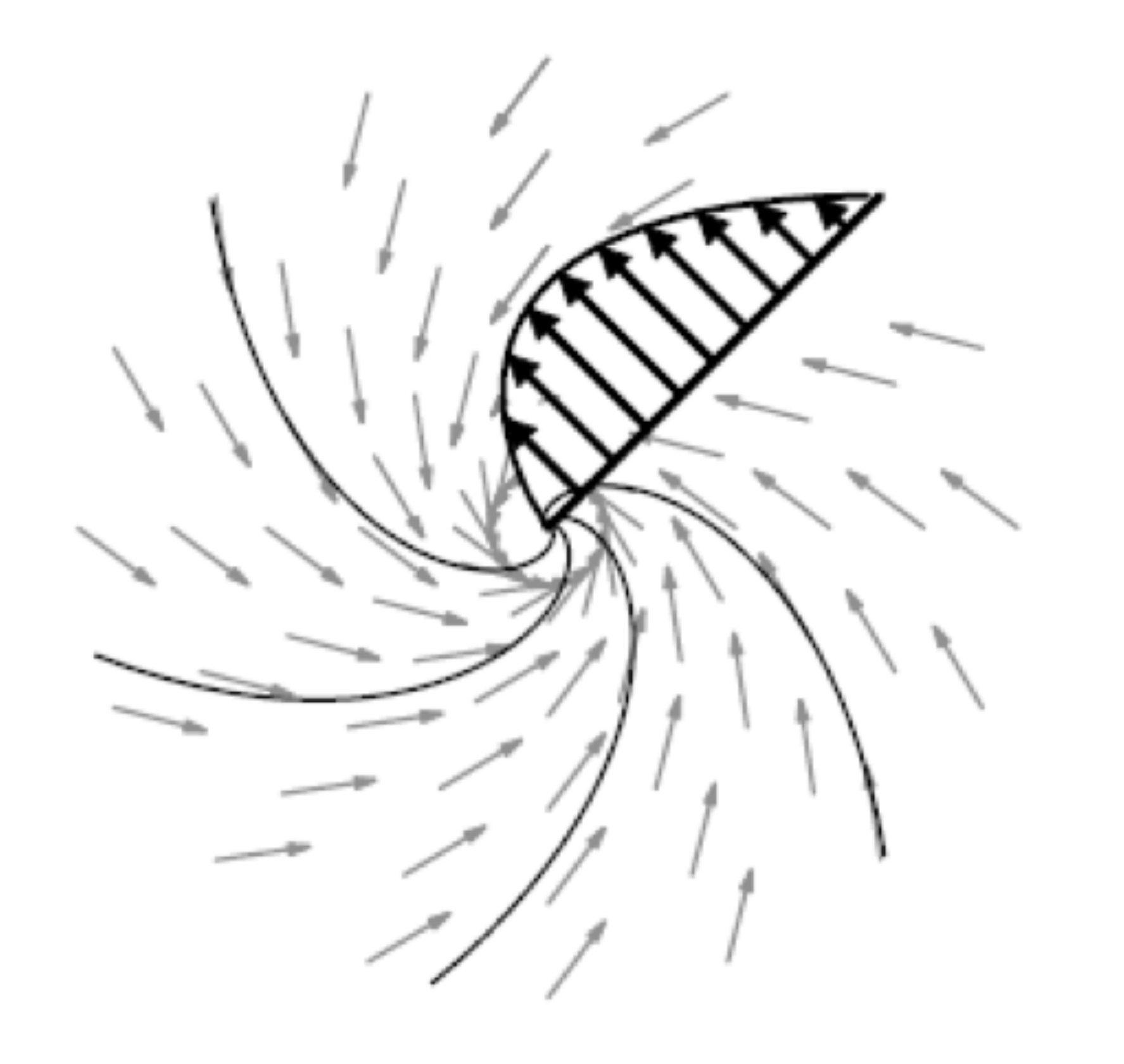}
\caption{(color online) Sketch of a rotating spiral defect in an active
nematic fluid with vanishing elastic anisotropy $\delta K = 0$. The 
gray arrows mark the director, and the solid lines show its spiral
structure. The hydrodynamic velocity field is in the azimuthal direction, and
its profile is indicated by the dark arrows. Adapted with permission from
\textcite{Kruse2004}.}
\label{spiralinstab}
\end{figure}

\begin{figure}
\centering
\includegraphics[width=0.89\columnwidth]{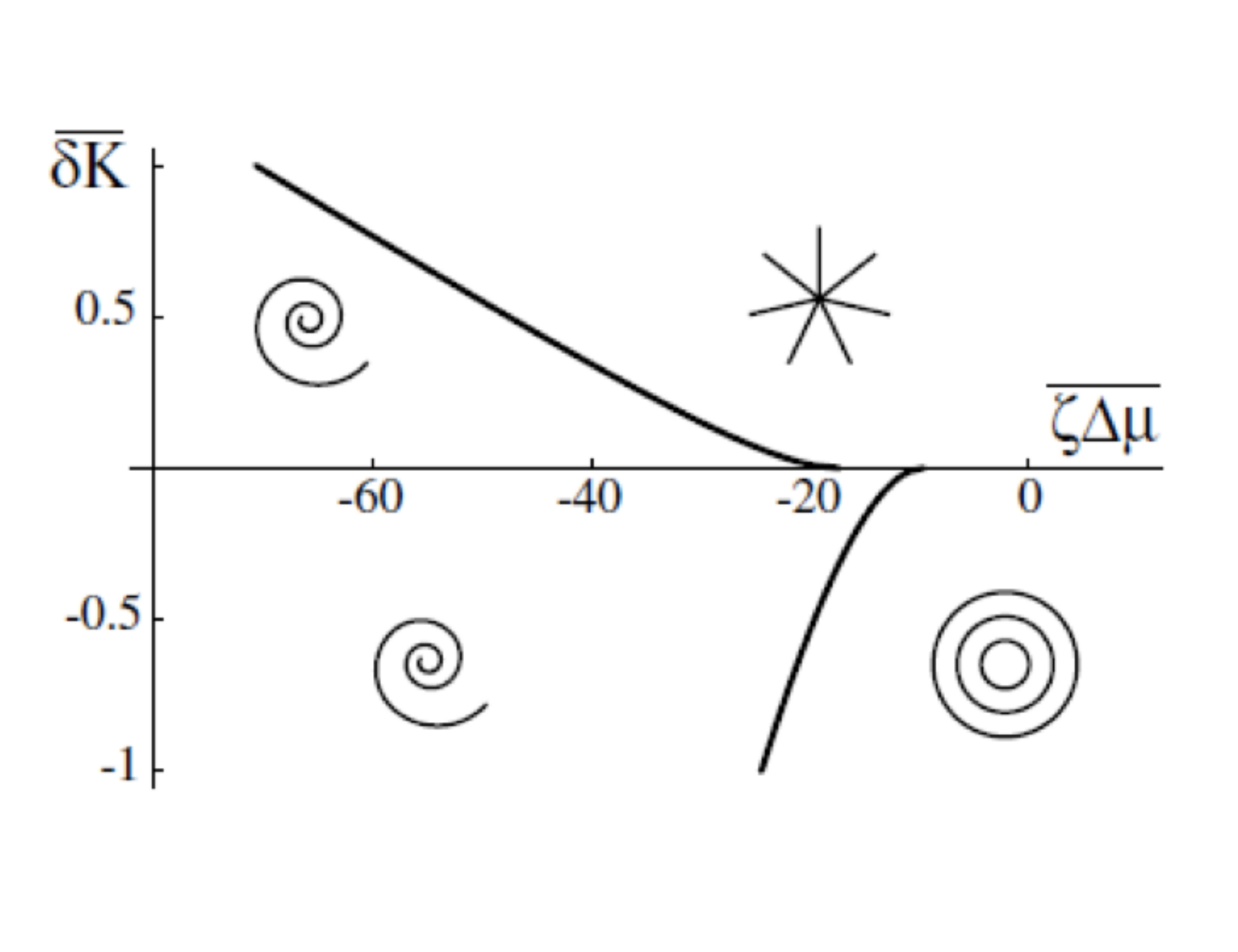}
\caption{(color online) Stability diagram of topological defects in an active
nematic fluid. For small values of the activity $\zeta \Delta \mu$ one gets
asters and vortices, respectively for $\delta K = K_1 - K_3 < 0$ and $>0$. When
$\zeta \Delta \mu$ of sufficiently large magnitude both asters and
spirals are unstable to the formation of a rotating spiral. Adapted with
permission from
\textcite{Kruse2004}.}
\label{defectphasediag}
\end{figure}

Suppose $K_1 < K_3$, so that the aster is stable in the absence of activity,
$\Delta \mu=0$. Now introduce activity. Linear stability analysis shows
that at sufficiently large contractile active stresses $\zeta \Delta \mu < 0$,
the aster gets destabilized giving rise to a spiral with an angle $\psi_0$ set
by the flow alignment parameter $\nu_1$ (assuming stable flow alignment).
An entirely similar analysis, starting from a stable vortex for $K_1>K_3$, with
$\Delta \mu = 0$, shows again an instability for large enough $\Delta \mu$.
The reason this happens here and does not happen for systems without activity
is that the active stresses associated with the perturbed director configuration
give rise to flows whose effect on the director is to reinforce the
perturbation. Figs. \ref{spiralinstab} and \ref{defectphasediag}
illustrate
the flow associated with the spiral instability and the
stability domains of the defects, respectively. The active spiral has a sense of
direction and
will therefore rotate. The angular speed can be obtained by solving the steady
state equations for $\psi$ and $v_{\theta}$, leading to 
  \begin{eqnarray}
v_{\theta}(r) & = & \omega_0 r \log\left(\frac{r}{r_0}\right),
\label{angvel}
\end{eqnarray}
where, as is inevitable on dimensional grounds, $\omega_0$
scales as the ratio of the active stress to a viscosity, with a detailed form
that includes a dependence on the director kinetic coefficient and the
flow-alignment parameter.
In a finite system of size $R$, imposing a vanishing velocity at the outer
boundary due to the presence of a wall, we can set
the length scale $r_0=R$.

In the above analysis we ignored the dynamics of the concentration
field $\rho$. This is probably acceptable when the active units are long and
rigid
or when their concentration is so high that excluded volume
considerations do not permit significant density inhomogeneities. 
We can include $\rho$ \citep{Gowrishankar2012a} through an extra term $v_1 \int d^3
r \rho \nabla \cdot {\bf p}$ in the free energy (\ref{frank}), with the form of
a spontaneous splay that depends on the local concentration, and restore the
concentration equation, $\partial_t \rho = - \nabla \cdot {\bf J}$, where the
filament current ${\bf J} = v_0 \rho {\bf p} - D \nabla \rho$, has an
active advective and a diffusive contribution. Now for large enough  $A$,
the defect configurations are generated by the internal dynamics (and
insensitive to the boundary for large systems) and so have finite size. When
advection is negligible,
the defect size is set by the ratio of the spontaneous splay coupling strength
$A$ to a Frank modulus, $K$, as it would be in an equilibrium polar liquid
crystal. On the other hand,  when advection is appreciable so as to have
density clumping in regions where $\mathbf{p}$ points inwards towards a
common center, the defect size is set by the ratio of  
diffusion to active advection, $D/v_0$. We note that since the advection current ${\bf
J} \propto \rho {\bf p}$, a small perturbation of the vortex configuration
renders it unstable and the only stable defects are inward pointing asters and
spirals \citep{Gowrishankar2012a}.

Since defects now have a finite size, it is possible to have an array of defects
which interact with each other. Such studies have shown that under certain
conditions, one obtains a stable lattice of asters accompanied by a variety of
phase transitions \citep{Ziebert2005,Gowrishankar2012a,Voituriez2006}.
A detailed study of defect-defect interactions and the dynamics of defects
and their merger in this active context are open problems for the future.

So far we have discussed charge $1$ defects generated in  polar active media.
Apolar active media, described by a local orientational tensor ${\bf Q}$,
exhibit $\pm1/2$ strength disclinations, topologically identical to those
obtained in equilibrium nematic liquid crystals (Fig.\,7). The orientation field
around a defect of strength $+1/2$ has a polarity, whereas that around a $-1/2$
has a $3$-fold symmetric appearance. On general grounds, the $+1/2$ defect
should move spontaneously whereas the $-1/2$ defect should show no such
tendency. Precisely this behavior seems to be observed in
the active nematic phase in a vibrated granular-rod monolayer
\citep{Narayan2007}.

\subsection{Current status on active gels}
The nonequilibrium thermodynamic description of active systems is a systematic
approach based on symmetries and in particular on invariance against time
inversion. 
It is however based on a linear expansion of fluxes in terms of forces and can
in principle only 
describe systems close to equilibrium where $\Delta \mu$ tends to zero.
One of the main application though is to biological systems that are 
mostly far from equilibrium systems. There is no systematic extension of 
the theory to systems far from equilibrium. One must rely either  on a
microscopic 
description that generates non-linear contributions by coarse-graining to large 
length scales and long time scales or on experimental results that emphasize 
specific non-linear aspects, which can then be introduced in the theory.
Microscopic 
or mesoscopic descriptions of molecular motors are a good example of the first
case 
and lead to motor forces or velocities that are not linear in $\Delta \mu$
\citep{Julicher1997, Liverpool2009}.
Several experiments suggest that the treadmilling associated to the
polymerization 
and depolymerization of actin in a cell depends on the force applied on the filaments or
the 
local stress in a non-linear way \citep{Mogilner1999, Prost2007}.  The force-dependent
treadmilling is essential for many cellular processes such as cell migration or
cell adhesion and a 
full description  including these effects in the active gel theory has not been
proposed yet 
\citep{Keren2008}.

Another intrinsic limitation of the current active gel theory is the assumption
of linear rheology and the use of the Maxwell 
model with a single relaxation time. A large body of experimental work shows
that for 
many types of cells as well as for actomyosin gels there is a broad distribution
of relaxation 
times leading to a complex modulus decreasing as a power law of frequency with a
small 
exponent between $0.1$ and $0.25$ \citep{Fabry2001}. An ad hoc power law
distribution of
relaxation times could be introduced in the active gel hydrodynamic theory but
despite some 
efforts, this type of law is not understood on general grounds
\citep{Balland2006}. The non-linear rheology of actin networks has also been studied in detail and actin
networks in the 
absence of molecular motors have been found to strain thicken
\citep{Gardel2004}. This
behavior can at least in part be explained by the inextensibility of the actin
filaments. In the 
presence of myosin motors, the active stress induced by the molecular motors
itself stiffens the 
active gel \citep{Koenderink2009}.

The hydrodynamic description of active polar or nematic gels is very close to
that of nematic 
liquid crystals. Nevertheless the existence of an active stress leads to several
non-intuitive  and 
spectacular phenomena.  The most spectacular is the flow instability described
below that 
leads to spontaneously flowing states. Other spectacular effects are associated
with active noise in 
these systems. In all active systems the noise has a thermal component 
and a non thermal active component. The properties of the active noise cannot be
inferred 
from the macroscopic hydrodynamic theory   and must be derived in each case from
a 
specific microscopic theory.  The study of tracer diffusion in a thin active
film for example 
\citep{Basu2011} leads to  an anomalous form of the diffusion constant that does
not depend on the size of the particle but only on the thickness of the film.
We believe that 
there are still many unusual properties of active gels to be discovered and that
in many cases, 
this will require detailed numerical studies of the active gel hydrodynamic
equations such as 
the one performed in \textcite{Marenduzzo2007}.

Active gel models have also been used to describe cross-linked motor-filament
systems that behave as soft solids at large
scales~\citep{MacKintosh2008,Levine2009,Yoshinaga2010,Banerjee2011} and also to understand
the origin of sarcomeric oscillations, both from microscopic as well as from a
continuum viewpoint~\citep{Gunther2007,Banerjee2011}. Further, active gel
models have been shown to successfully account for the experimentally observed
traction stresses exerted by cells and cell sheets on compliant substrates
\citep{Banerjee2011,Mertz2012,Edwards2011}.

Finally, a large part of the theoretical activity on active gels aims at a
quantitative description of 
biological phenomena at the scale of the cell and of cellular processes
involving the 
cytoskeleton. Some success has already been obtained in discussing lamellipodium
motion 
\citep{Kruse2006} or the formation of contractile rings during cell division
\citep{Salbreux2009}. An important step is the connection of the parameters of
the hydrodynamic theory with the more microscopic parameters that can be
monitored experimentally which requires an explicit coarse graining of the
microscopic theories as discussed in section \ref{microscopic}. At larger scale
one can build a hydrodynamic theory of tissues that shares many features with
the active gel theory described here \citep{Ranft2010}.

\section{Hydrodynamic Consequences of Activity}
\label{applications}

In this section we describe a number of remarkable hydrodynamic phenomena
induced by activity. Most of the section is devoted to the description of
materials properties of active gels, such as thin film instabilities and
rheology. In section~\ref{subsec:cells} we also highlight  some of the
remarkable successes of the hydrodynamic theory of
active gels  in describing phenomena observed in living cells. A more complete
review of these latter class of phenomena can be found in \textcite{Joanny2010}.

\subsection{Instabilities of thin liquid active films}
\label{thinfilmsection}

\subsubsection{Spontaneous flow of active liquid films}
\label{spontflowsubsec}
One generic property of active orientable liquids, whether polar or
apolar, is the instability   of any homogeneous non-flowing steady state toward
a
inhomogeneous spontaneously flowing state, as shown by \textcite{Simha2002}. We
illustrate this
instability in the simple geometry of a thin active nematic liquid film
\citep{Voituriez2005}.

\begin{figure}
\centering
\includegraphics[width=0.8\columnwidth]{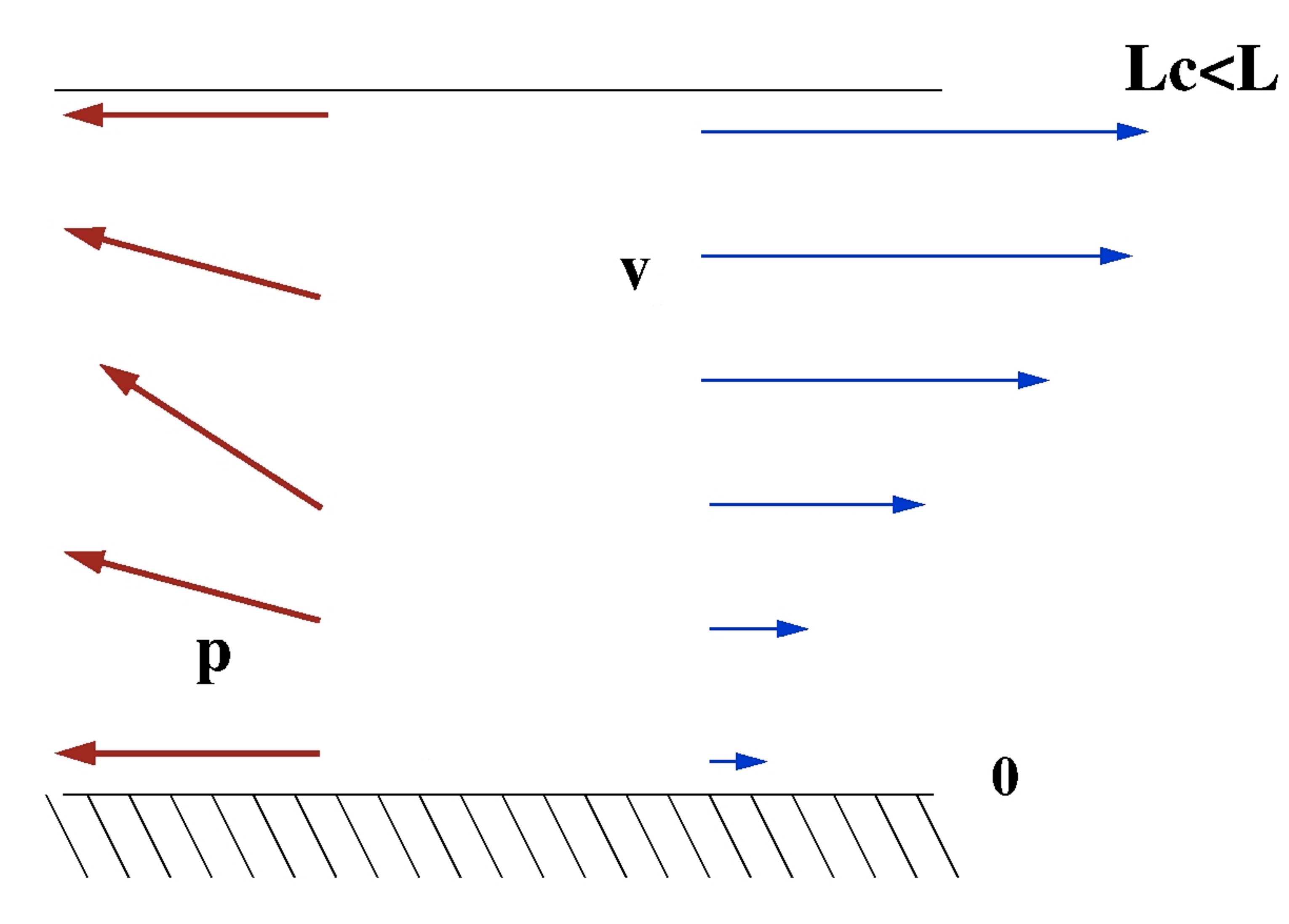}
\caption{ (color online) Spontaneously flowing film of active liquid. }
\label{film}
\end{figure}

We study a thin film of thickness $h$ supported by a solid substrate. For
simplicity we only consider the two-dimensional geometry sketched in Fig.
\ref{film} where the substrate is along $x$ and the normal of the film is along
$y$. We choose anchoring conditions parallel to the film surface so that for
$y=0,h$ the polarization is along $x$, $p_x=1,\  p_y=0$. An obvious solution for
the equations of motion of an active liquid is  a non flowing state $\bf v=\bf
0$ with a constant polarization parallel to $x$. We now discuss the stability of
this steady state. 
We look for a state of the system which is invariant by translation along $x$ so
that all
derivatives with respect to $x$ vanish and with a polarization that is not
parallel to the 
film surfaces $p_x= \cos \theta (y), p_y=\sin \theta (y)$. The velocity is along
the $x$
direction and the shear rate tensor only has one non vanishing component 
$v_{xy}\equiv w=\frac 1 2 \partial_y v_x$.

The total stress can be written as $\sigma^t_{\alpha\beta}=
-P \delta_{\alpha\beta}+\sigma^A_{\alpha\beta}+ \sigma_{\alpha\beta}$ where the 
last term is the deviatoric stress given by the constitutive equations
\eqref{constitutive} and 
the previous term is the antisymmetric component of the stress. Force balance in
the film 
is written as $\partial_y \sigma^t_{yx}=0$. Taking into account the fact that on
the free 
surface the shear stress vanishes, we obtain $\sigma^t_{yx}=0$. Using the
constitutive
equation, this leads to 
\begin{equation}
\label{sigmayx}
 -h_{\perp}=4\eta w-\zeta \Delta \mu \sin 2\theta +\nu_1 (h_{\parallel} \sin
2\theta 
 + h_{\perp}  \cos 2\theta)\;,
\end{equation}
where we have introduced the parallel and perpendicular components of the
orientational 
field $h_{\parallel}=h_x \cos \theta + h_y \sin \theta$ and $h_{\perp}=h_y \cos
\theta -
h_x \sin \theta$

The constitutive equations for the polarization of Eq.~ \eqref{constitutive}
give
\begin{eqnarray}
\label{polar}
 -w \sin \theta &=& \frac{h_x}{\gamma_1}-\nu_1 w\sin \theta\;,   \nonumber\\
 w \cos \theta &=& \frac{h_y}{\gamma_1}-\nu_1w \cos \theta\;.
\end{eqnarray}

Combining Eqs.~ \eqref{sigmayx} and \eqref{polar}, we obtain the perpendicular
field 
and the velocity gradient
\begin{eqnarray}
\label{fred}
 h_{\perp}&=&\frac{\zeta \Delta \mu \sin 2\theta(1 + \nu_1 \cos 2 \theta)}
 {\frac{4 \eta}{\gamma_1} + 1 + \nu_1 ^2 + 2 \nu_1 \cos 2 \theta}\;, \nonumber
\\
 w &=& \frac{\zeta \Delta \mu \sin 2\theta}{\frac{4 \eta}{\gamma_1} + 1 + \nu_1
^2 +
 2 \nu_1 \cos 2 \theta}\;.
\end{eqnarray}
In the approximation where the Frank constants are equal, the perpendicular 
molecular field is $h_{\perp}=-\frac{\delta {F}}{\delta \theta}=K 
\frac{\partial^2 \theta}{\partial y}$. If the angle $\theta$ is small, an
expansion  of Eq.\eqref{fred} for 
the perpendicular 
field to
lowest order in $\theta$ gives $\nabla^2 \theta +\frac{\theta}{L^2}=0$ where the characteristic 
length $L$ is defined by 
\begin{equation}
\label{L2}
 \frac{1}{L^2}= \frac{-2 \zeta \Delta \mu (1+ \nu_1)}{K[\frac{4\eta}{\gamma_1}+
 (1+ \nu_1)^2)]}\;.
\end{equation}
Assuming $1+\nu_1>0$, if the activity coefficient $\zeta$ is negative
corresponding to a contractile stress, 
$L^2>0$ and the length $L$ is indeed real.
The polarization angle varies then as $\theta= \theta_0 \sin
\left(\frac{y}{L}\right)$. This satisfies the 
anchoring condition on the solid surface $y=0$ but the anchoring condition on
the 
free surface $y=h$ can only be satisfied if $h=\pi L$.
If the film is thin $h\leq \pi L=L_c$ there is no solution with a finite
$\theta$ and the non-flowing steady state is stable. For a film thicker than the
critical value $h= \pi L$ 
a solution with a finite value of the polarization angle $\theta$ exists and 
the non-flowing steady 
state is unstable.  If $h> L_c$ there is spontaneous 
symmetry breaking and two possible solutions with amplitudes $\pm \theta_0$. 
The amplitude $\theta_0$ can be obtained by expansion at higher order 
of Eq.~\eqref{fred} and vanishes if $h= \pi L$. In this case 
the second of Eq.~\eqref{fred} gives a finite velocity gradient and the film
spontaneously 
flows with a finite flux. Note that in general the onset of spontaneous flow is
controlled by the sign of the combination
$\zeta(1+\nu_1)$. The flow coupling coefficient (also known as flow alignment
parameter) $\nu_1$ can in  general have both positive and negative
values~\citep{Gennes1993}. It is controlled by the shape of the active units and
the degree of nematic order. Deep in the nematic state, $\nu_1<-1$ corresponds
to elongated rod-like particles, while $\nu_1>1$ corresponds to disk-like
particles. The onset of spontaneous flow is therefore controlled by the
interplay of the contractile/tensile nature of the active forces and the shape
of the active particles. A detailed description of this can be found in
\textcite{Giomi2008,Edwards2009}.

This transition is very similar to the Frederiks transition of nematic
liquid crystals 
in an external electric or magnetic field~\citep{Gennes1993}. The active stress
$\zeta \Delta \mu$ plays
here the role of the external field. If the thickness is larger that the
critical value $L_c$,
a distortion of the polarization appears. Any distortion of the polarization
creates a 
gradient of active stress that must be balanced by a viscous stress which
implies 
the appearance of a finite flow. 

The Frederiks transition could also be considered at a constant film thickness
varying 
the active stress $\zeta \Delta \mu$. The film spontaneously flows if the active
stress 
is large enough (in absolute value). 

Finally, to properly describe the spontaneous flow transition for polar active
films one needs to
consider a two-fluid model that allows for variations in the concentration of
active particles. 
In this case spontaneous flow is accompanied by spatial inhomogeneities in the
concentration or ``banding" not seen in active nematics
~\citep{Giomi2008}. In addition for stronger values of activity in polar films
steady spontaneous flow is replaced by oscillatory flow that become increasingly
complicated and even chaotic for strong polarity.

\subsubsection{Instabilities of thin films} \label{thinfilmfree}
The Fredericks active film instability discussed in section
\ref{spontflowsubsec} occurs in situations where the geometry of the film 
is fixed and its surface cannot deform, whereas many beautiful phenomena in
thin film flow \citep{Oron1997,Sarkar2010} involve
distortions of the free surface. In the context of cell biology, a study of
a film of active fluid with a dynamic free surface is the natural starting
point for a complete description of a moving lamellipodium, which is a thin,
flat,
fluid projection, full of oriented actin, that forms the leading edge of a
crawling cell \citep{Verkhovsky1999,Small2002}.
The spreading of bacterial suspensions \citep{bees2000,bees2002} is another
biological instance in which the processes discussed in this section could
intervene. However we discuss here simpler situations where mechanisms 
such as cell division in the
bacterial case and actin treadmilling 
in the lamellipodium case are not included and that in a first step could only 
be compared to model biomimetic experiments. 
Despite these limitations, the problems of an active drop or an
active film are interesting as novel variants of classic fluid mechanics
problems and as settings for phenomena of relevance to biology. 

We review in this section the hydrodynamics of thin films of
fluid containing orientable degrees of freedom and endowed with locally uniaxial
active stresses, bounded on one side by a solid surface,
and on the other side by a surface free to undulate in response to flows in the
film. 
We present in
some detail the case of an unbounded film \citep{Sankararaman2009}, to highlight
the exotic physical effects that arise from \textit{polar} orientational
order in an active system.
In this example, we consider an active system comprising a solvent
fluid and 
dissolved active particles. In the language of the previous section, we thus
consider 
a multicomponent active fluid.  We assume that the polar material velocity with
respect 
to the background fluid (related to the relative current between the two
components) 
is strictly slaved to the polar order  and equals $v_0 \mathbf{p}$. 
This assumption is meaningful in the case of bacterial colonies
but would have to be reconsidered in the discussion of a cell lamellipodium.
We also sketch results for the case of a
finite, partially wetting drop with small equilibrium contact angle and
\textit{apolar} orientational order \citep{Joanny2012}, where topological
defects
play a role. In both cases, we consider only  planar 
alignment \citep{Gennes1993}
where
the local orientation field is anchored parallel to the free surface and
the rigid base, and free to point in any direction in the plane of anchoring.
Our focus is
of course on the effects specifically due to the active stresses and currents. 

\begin{figure}
\centering
\includegraphics[width=0.650\columnwidth]{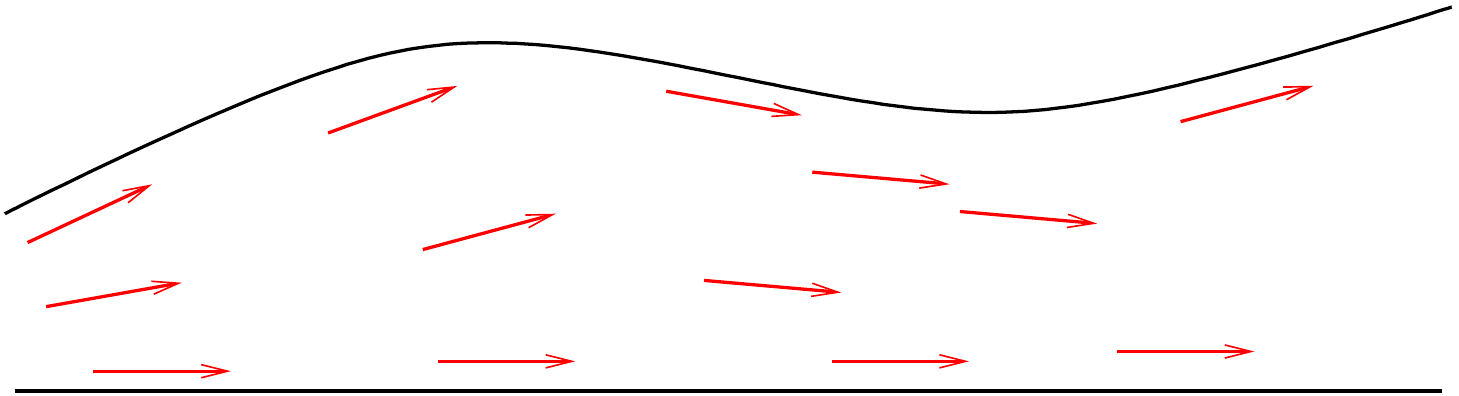}
\caption{(color online) A
film of ordered polar active suspension, illustrating the anchoring boundary
condition on the filaments, parallel to the free surface and the base.}
\label{fig:thinfilmfig}
\end{figure}

Following \textcite{Sankararaman2009}, we consider a fluid film (Fig.
\ref{fig:thinfilmfig}) containing active particles with number density
$\rho(\mathbf{r},t)$ and orientation field $\mathbf{p}(\mathbf{r},t)$ as a
function of time $t$ and three-dimensional position $\mathbf{r} =
(\mathbf{r}_{\perp},z)$, where $z$ denotes the coordinate normal to the
horizontal coordinates $\mathbf{r}_{\perp} = (x,y)$, and the solid surface lies
at $z=0$. In the case of the drop, where we consider only apolar order, we
identify the vector $\mathbf{p}$ with the nematic director field, with
$\mathbf{p} \to - \mathbf{p}$ symmetry. The free surface is located at
$z=h(\mathbf{r}_{\perp},t)$. The flow of the film is characterized by the
$3$-dimensional
incompressible velocity field $\mathbf{v} (\mathbf{r},t)$. We
focus on the case where the film or drop has macroscopic order, i.e., the
mean $\langle \mathbf{p} \rangle$ is nonzero, which means that variations in the
direction, not the magnitude, of $\mathbf{p}$
plays the central role. Our
aim is to obtain an effective equation of motion for the thickness $h$ and the
$z$-averaged
density and polarization. This
requires solving the Stokes equation in the presence of
stresses generated by the active particles. At the end of the section, 
we offer  a qualitative physical explanation as well.

The dynamics of the height $h$ is related to the velocity  through 
the kinematic condition $\dot h = v_z - \mathbf{v_\perp}\cdot { \bm\nabla_\perp}
h$
\citep{Stone2005}.
The incompressibility of the suspension implies volume conservation, so
that the height dynamics becomes a local conservation law \citep{Stone2005}
\begin{equation}
\label{hincompeqn}
\partial_t  h + \bm{\nabla}_\perp \cdot ( h \bar{\mathbf{v}}_{\perp}) = 0
\end{equation}
in the $\perp$ plane, where $\bar{\mathbf{v}}_{\perp}$ is the in-plane
velocity field averaged over the thickness of the film. With our assumptions,
the flux 
of active particles in the laboratory reference frame is ${\bf j}=\rho
({\mathbf{v}} 
+ v_0 \mathbf{p})$ 
so that the concentration $\rho$ obeys the
continuity equation
\beq
\label{conceqfilm}
\partial_t \rho = -\bm\nabla \cdot[\rho (\mathbf{v} + v_0 \mathbf{p})]\;,
\eeq
which simply generalizes (\ref{rho-eq}) to the case where the particles are
moving not through an inert background but through a suspension with
velocity field $\mathbf{v}$. 
The velocity field $\mathbf{v}$ obeys the Stokes equation
\beq
\label{active_stokes}
\eta \nabla^2 \mathbf{v} = \bm\nabla P - \bm\nabla \cdot (\bm\sigma^a +\bm
\sigma^p)\;,
\eeq
with viscosity $\eta$, pressure $P$, and stresses
$\bm\sigma^a = -\zeta \Delta \mu \rho \mathbf{p} \mathbf{p}$
and $\bm\sigma^p$ arising from activity with strength
$\zeta \Delta \mu$ and nematic elasticity \citep{Gennes1993}, respectively, as given in Eqs.~\eqref{sigma-active} and \eqref{sigma-passive}. Note that we have imposed here an explicit linear  dependence of the active stress  $\bm\sigma^a$ on the local concentration $\rho$   of active particles as expected in a bacterial
suspension  
whereas for filament-motor systems the 
active stress increases faster than linear with the filament density. The polar
order parameter
$\mathbf{p}$ obeys (adapting \textcite{Simha2002} to the case where
$\mathbf{p}$ is coupled to the free-surface tilt)
\beq
\label{P-eq-2-wet}
\frac{D p_\alpha}{Dt} +\nu_1 v_{\alpha\beta}p_\beta +\lambda_1({\bf
p}\cdot\bm\nabla)p_\alpha= -\frac{\delta F_p }{ \delta p_\alpha}+\frac{C}{h}
\partial^{\perp}_{\alpha} h
+f_\alpha
\eeq
where $\frac{D}{Dt}
$ is the time-derivative in a frame comoving and corotating with the
fluid, defined in Eq.~\eqref{convected-der}, $\nu_1$ is the flow alignment parameter introduced in Eq.~\eqref{eq:reac-stress-dev}
and familiar from liquid-crystal hydrodynamics
\citep{Gennes1993},
$\lambda_1$ is the coefficient of the advective nonlinearity in Eq.~\eqref{P-eq},
and $v_{\alpha\beta}$ is the strain rate tensor defined in Eq.~\eqref{strain-rate}.
 This is the
liquid limit ($\tau=0$) of 
Eq. \eqref{constitutive} where the polar active term proportional to $\lambda_1$ 
has been included.  
Finally,  $F_p$ is the free-energy
functional for $\mathbf{p}$ and $\rho$ given in Eq.~\eqref{Fp}, whose content will be
discussed further below. Equation (\ref{P-eq-2-wet}) 
generalizes Eq.
(\ref{P-eq-2}) to the case where a fluid medium is present. 

It has been argued in
\textcite{Sankararaman2009} that in a geometry of finite thickness in the
$z$ direction, symmetry could not rule out a term of the form 
$(C / h) \bm\nabla_{\perp} h$, in the $z-$averaged equation of motion for
$\mathbf{p}$. Such a term arises through the interaction of $\mathbf{p}$ with
the free surface, and the coefficient $C$ encodes the preference of
$\mathbf{p}$ to point uphill or downhill with respect to a tilt of the free
surface. Possible microscopic mechanisms and estimates of magnitudes for such a
term are discussed in \textcite{Sankararaman2009}.
A key additional remark needs to be
made: such an effect, while allowed by symmetry, must explicitly involve
properties of the free surface and the base; it cannot emerge simply from a
$z-$averaging of the bulk $3d$ hydrodynamics. We therefore add such a term 
to \eqref{P-eq-2-wet} .

We now proceed to
solve the Stokes equation  (\ref{active_stokes}) for the velocity in terms of
$\rho$,
$\mathbf{p}$ and $h$, restricting ourselves to the lubrication approximation
\textcite{Oron1997, Batchelor2000} $v_z = 0, \, |\partial_z
\mathbf{v}| \gg |\nabla_{\perp}\mathbf{v}|$.

We study perturbations about a reference configuration of the film with
uniform concentration $\rho_0$ and height $h_0$, spontaneously ordered into a
state with nonzero mean polarization $\langle \mathbf{p}\rangle = p_0
\hat{\mathbf{x}}$. As we have chosen the active-particle current relative to the
medium in (\ref{conceqfilm}) to be entirely along $\mathbf{p}$, with no explicit
diffusive contribution, and as the particles cannot escape the fluid film, 
the normal components of $\mathbf{p}$ must vanish at the bounding surfaces
at $z=0$ and $z=h$ in agreement with the planar alignment
condition that we impose. In a perturbed state with a non-uniform film
thickness, 
this means that $p_z(z=h) \simeq \partial_xh$ to linear
order in $\bm\nabla h$. The $z$ direction being the smallest dimension in the
problem, it is consistent to assume that the variation of $\mathbf{p}$ with
respect
to $z$ is at mechanical equilibrium via nematic elasticity. The
instantaneous value of $p_z$ is then related to the thickness profile by 
$p_z = (z/h)\partial_xh$. In the
$z$-averaged description that we are aiming for, $p_z = (1/2)\partial_x h$ and
$\partial_z p_z \simeq h^{-1} \partial_x h$. 

\begin{figure}
\centering
\includegraphics[width=0.6\columnwidth]{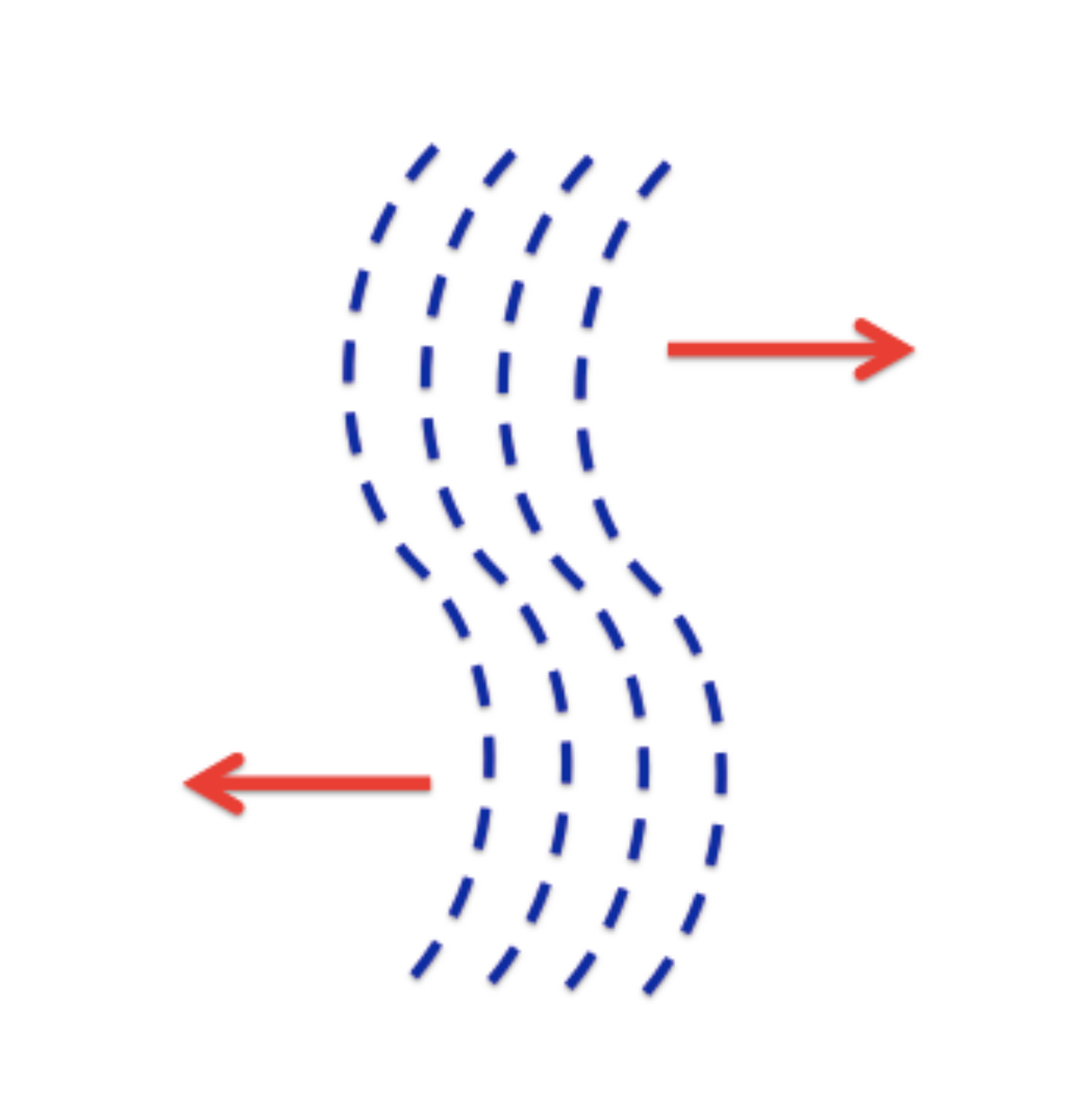}
\caption{ (color online) Curvature of the director of an active liquid
crystalline phase
gives rise to flows. The arrows indicate direction of flow if the active
stresses are contractile. }
\label{fig:flowcurr}
\end{figure}

Let us represent the perturbed state as
$\mathbf{p}_\perp ={\bf  \hat{x}} + \theta{\bf  \hat{y}}, \, \theta \ll 1$
The divergence of the active stress has components 
$\partial_\alpha \sigma_{\alpha x}^a = \zeta \Delta \mu (\partial_y \theta +
\partial_x \rho/\rho_0 + h^{-1}\partial_x h)$, $\partial_\alpha \sigma_{\alpha
y}^a = \zeta
\Delta \mu \partial_x \theta$ and $\partial_\alpha \sigma_{\alpha z}^a = \zeta
\Delta \mu
\partial_x^2 h/2$ to linear order. We apply the thin film approximations of
\citep{Stone2005,Oron1997} to first calculate  the pressure and then obtain
the linearized expression of the averaged in-plane velocity
\begin{equation}
\label{uofz} \mathbf{v}_\perp(z) = \frac{hz - z^2/2}{\eta} \left(\gamma
\bm\nabla_\perp \nabla_\perp^2 h
- \frac{1}{2} \zeta \Delta \mu h \partial_x^2 \bm\nabla_\perp h-
\mathbf{f}_\perp\right),
\end{equation}
where $\mathbf{f}_\perp = \zeta \Delta \mu
[(\partial_y \theta + \partial_x \rho/\rho_0 +h^{-1} \partial_x h){\bf \hat{x}
}+
\partial_x \theta {\bf \hat{y}}]$ contains the dominant contributions of
activity.
Inserting  this result in the incompressibility condition \eqref{hincompeqn} and
linearizing $h = h_0 +
\delta h$, $\rho = \rho_0 + \delta \rho$, leads to the dynamical equations
\begin{eqnarray}
\label{finalhteqn}
\partial_t \delta h_\mathbf{q} &=& -\frac{\zeta \Delta \mu  h_0^2}{3\eta}
[2h_0q_x q_y \theta_\mathbf{q} + h_0 q_x^2\frac{\delta \rho_\mathbf{q}
}{ \rho_0} \nonumber \\
&+& (1 - \frac{1}{2} h_0^2 q^2)q_x^2\delta h_\mathbf{q}]
-\frac{\gamma h_0^3}{3 \mu} q^4 \delta h_\mathbf{q}
\end{eqnarray}
for the in-plane spatial Fourier transforms $\delta h_\mathbf{q}(t)$, $\delta
\rho_\mathbf{q}(t)$, $\theta_{\mathbf{q}}(t)$ of the perturbations in height,
concentration and orientation. The group of terms multiplied by $\zeta
\Delta \mu$ on the right-hand side of (\ref{finalhteqn}) displays four
consequences of activity, \textit{viz.}, from right to left: (i) pumping by curvature (see
Fig. \ref{fig:flowcurr}); (ii) anisotropic osmotic flow, through the interplay
of a uniform orientation field and a concentration inhomogeneity; (iii)
splay-induced flow from tilting the free surface (the spontaneous flow
instability discussed in the previous section, but adapted to the case of a deformable
surface) and (iv) an active anisotropic contribution to the effective tension.
Following these active terms is conventional surface tension. Terms (iii) and
(iv) are, respectively, destabilizing and stabilizing for contractile stresses,
and the other way around for extensile stresses; the reader will see that this
is physically reasonable. 
\begin{figure}
\centering
\includegraphics[width=0.90\columnwidth]{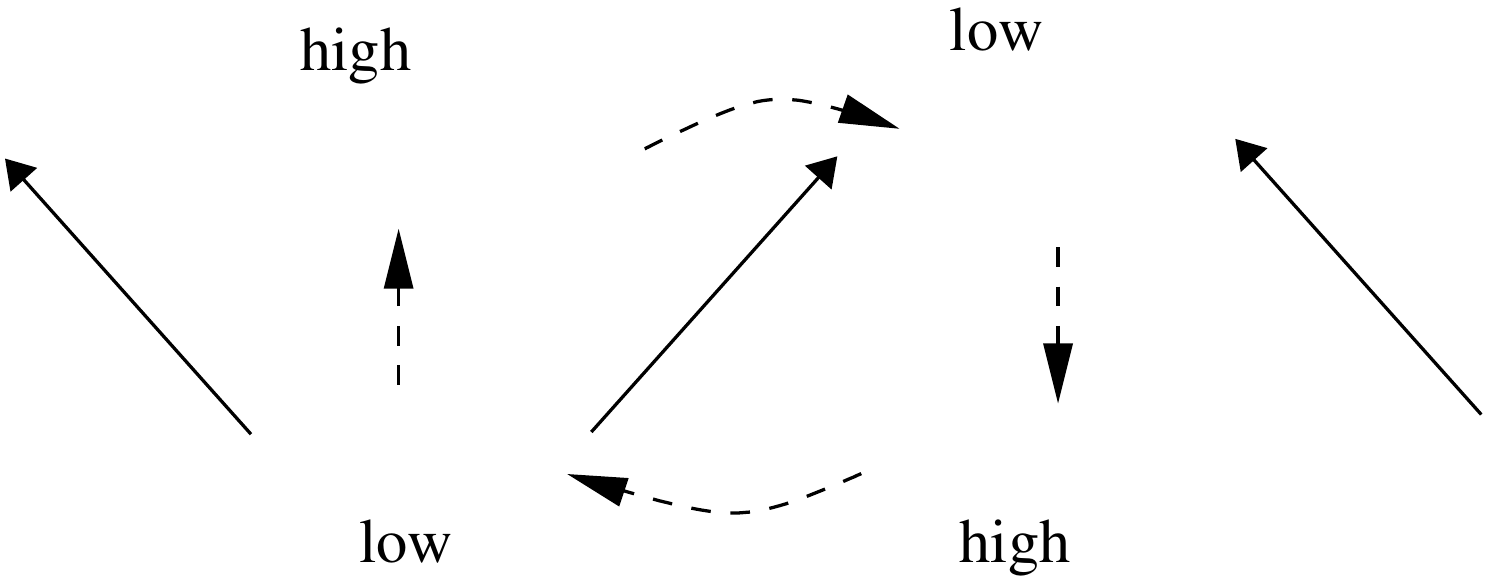}
\caption{ (color online) Tilt of the free surface leads to active flow; the
arrows directions
are for the contractile case.}
\label{fig:tiltpump}
\end{figure}
We now turn to the equation of motion
for the polar orientation $\mathbf{p}$. 
After linearizing (\ref{P-eq-2-wet}), and
averaging over $z$, in the hydrodynamic limit, we obtain 
\begin{eqnarray}
\label{thetaeqn}
\partial_t \theta_{\mathbf{q}} &=&  +\frac{i C}{h_0} q_y
\delta h_{\mathbf{q}}
- \left(D_+ q_x^2 + D_- q_y^2 + i \lambda_1 q_x \right) \theta_{\mathbf{q}}
\nonumber \\
&-& (i \zeta q_y - \Phi q_x q_y) \delta \rho_{\mathbf{q}},
\end{eqnarray}
where $ D_{\pm} = D - (\lambda_1 \pm 1) h_0^2\zeta \Delta \mu / 4 \eta  $, $D
\sim K / \eta$ being a director diffusivity with $K$ the Frank constant
(assuming 
that the Frank constants are all equal),  
and $\Phi =  (\nu_1 -1) h_0^2 \zeta \Delta \mu / 4\rho_0 \eta  $.

Linearizing the active particle conservation law (\ref{conceqfilm}) about the 
ordered uniform state gives to leading order in wavenumber:
\begin{equation}
\label{conceom}
\partial_t \delta \rho_{\mathbf{q}} =
-i \rho_0 v_0 q_y \theta_{\mathbf{q}} 
 - i v_0 q_x \delta \rho_{\mathbf{q}} 
+O(q_x^2 \delta \rho_\mathbf{q}, q_x^2 \delta h_\mathbf{q}).
\end{equation}

The most accessible and interesting instability, arising from the combination
of activity and the tilt coupling $C$, can be understood by
ignoring the concentration and motility ($v_0 = 0$), but retaining contractile
or
extensile active stresses ($\zeta \Delta \mu \neq 0$). The dispersion relation
has the complex form: 
\beq
\label{newinstab}
\omega = \pm\frac{ {1+i \,\mbox{sgn}(q_x C \zeta \Delta \mu)}}{\sqrt{2}}
\left(\frac{h_0^2}{
 3 \eta}\right)^{1/2}|C \zeta \Delta \mu q_x|^{1/2} |q_y|
\eeq
The relative signs of $\zeta \Delta \mu$ and $C$ determine the direction
($\pm \hat{\mathbf{x}}$) of propagation of the unstable mode. Fig.
\ref{fig:tiltpump} 
attempts to explain the mechanism of this intriguing
instability. The effects of concentration fluctuations, and the suppression of
the instability as the motility $v_0$ is increased, are discussed in
\textcite{Sankararaman2009}.

\begin{figure}
\centering
\includegraphics[width=0.50\columnwidth]{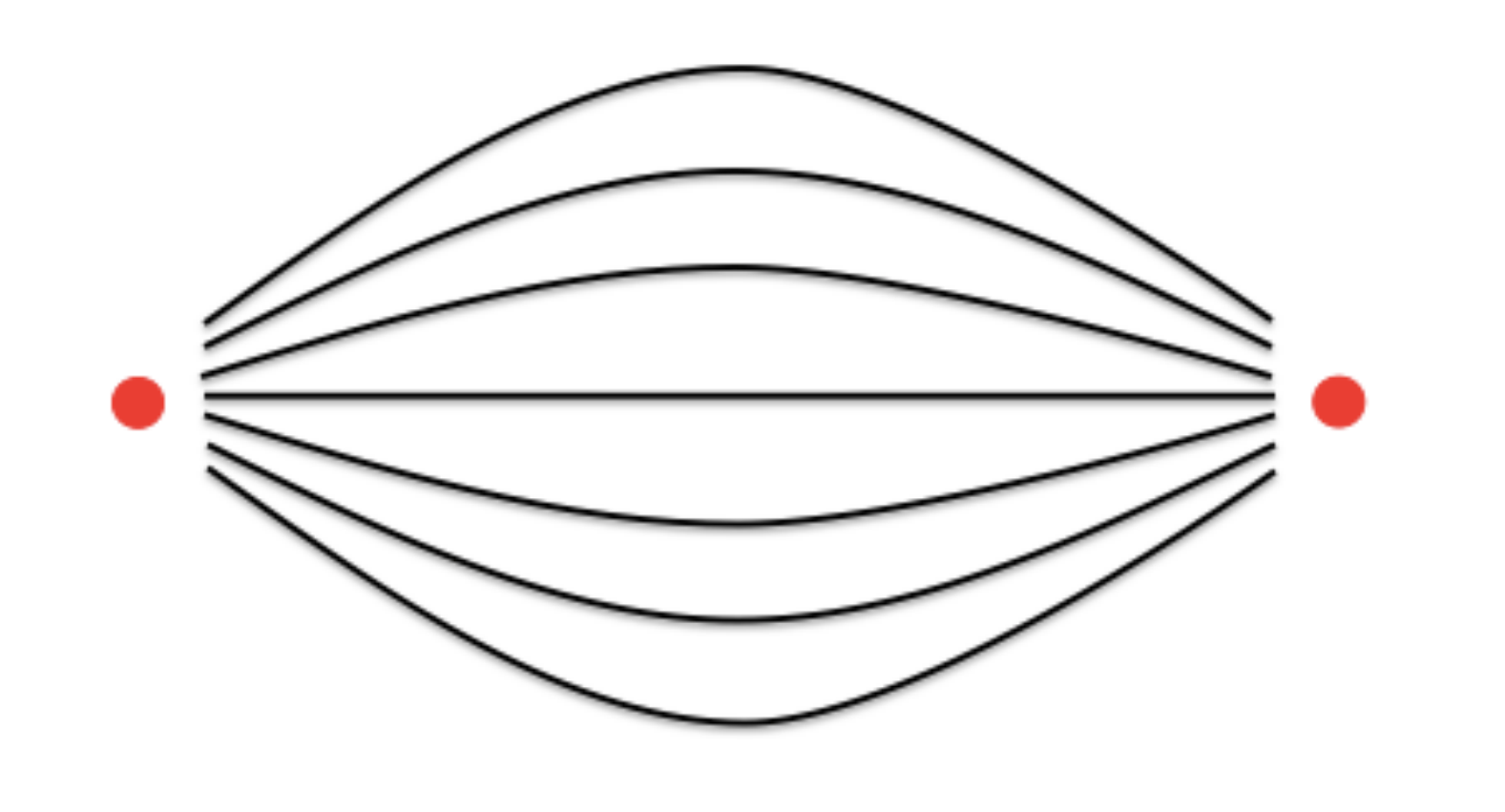}\includegraphics[width=0.450\columnwidth]{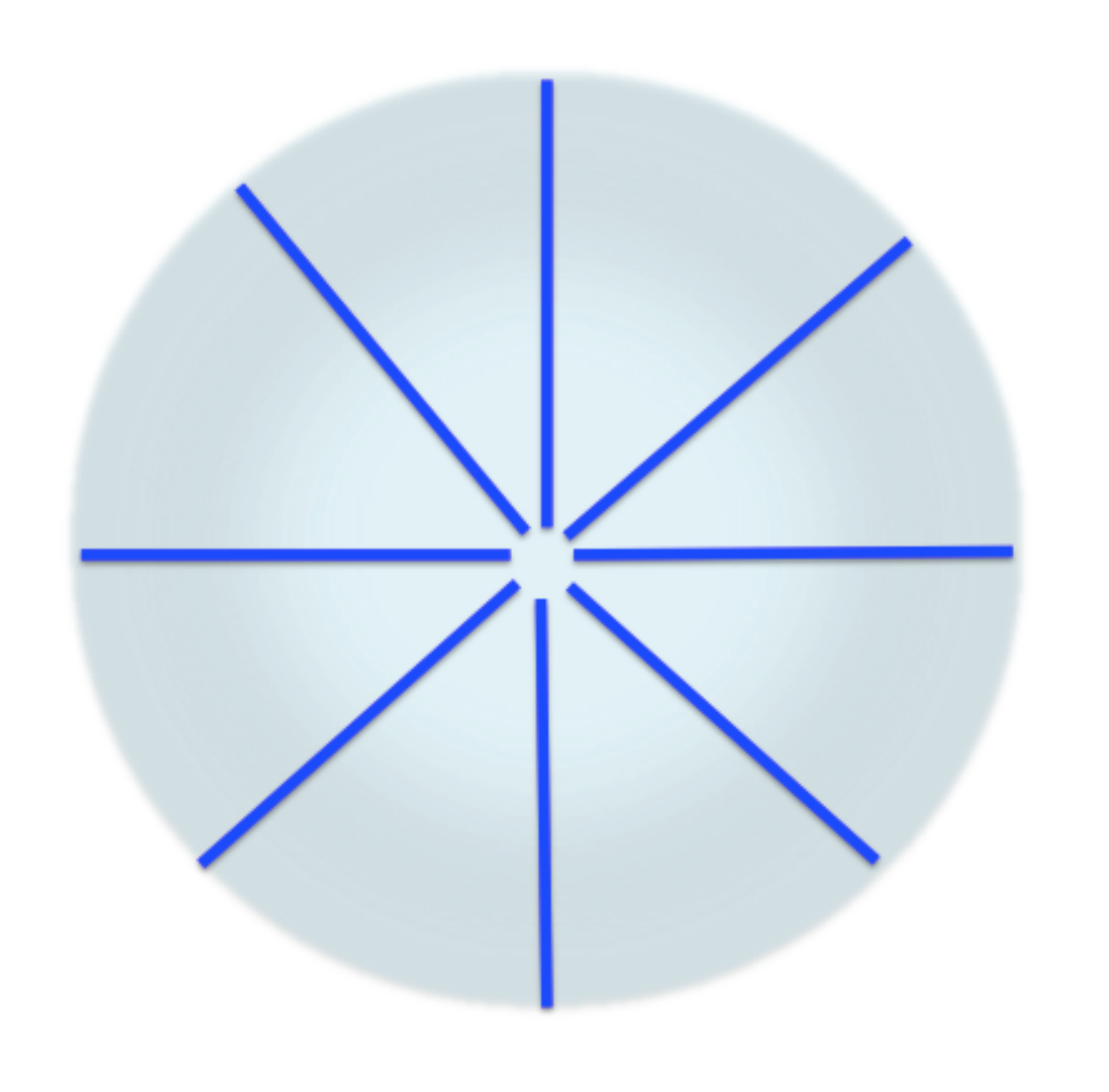}
\caption{ Left: two half- strength defects (ÒboojumsÓ) at the ends of a drop. Right:  a radial strength-1 aster
defect. Both defects are viewed from above the plane in which the drop lies. The
lines are the director orientation.}
\label{watermelon}
\end{figure}

So far we have assumed an unbounded film. Let us now consider 
briefly the spreading kinetics of a finite drop, as discussed by
\textcite{Joanny2012}, containing apolar active filaments in a
state of nematic
order. We choose a planar anchoring condition as in the case of the unbounded
film, i.e.,
no component normal to free surface or base. The geometry of the drop forces a
topological defect in the interior that itself induces a deformation of the
drop. 
The simplest case to consider is a
defect consisting of two ``boojums'' with orientation pattern as in Fig.
\ref{watermelon} (left), with the assumption that the drops spreads uniaxially in the
$x$ direction. Two possible defect structures are two ``boojums'' with
orientation pattern as in Fig. \ref{watermelon} (left) and an aster as in Fig.
\ref{watermelon} (right).
%
%
The active stress amounts to a peculiar kind of disjoining pressure $P_{act} =
\zeta \Delta \mu \ln (h/h_0)$, which can be rationalized on dimensional grounds
by
noting that the activity strength itself has units of stress so that
the dependence on thickness has to be logarithmic. The resulting static shape
of a drop depends on the sign of the active stress: the drop is flat if the active stress
 is extensile, and elevated, if it is contractile. For a linear structure like
Fig. \ref{watermelon} (left), if we assume spreading only along the long axis 
 at fixed width $w$ the result, for a fixed volume $\Omega$ and viscosity
$\eta$, is a drop of linear dimension 
\begin{equation}
\label{1+1d}
 R(t) \sim \left(\frac{\zeta \Delta \mu \Omega^2 t}{w^2\eta}\right)^{1/4}.
\end{equation}
A drop of volume $\Omega$ with an aster defect as in Fig. \ref{watermelon} (right) spreads
isotropically with
\begin{equation}
\label{1d}
 R(t) \sim \left(\frac{\zeta \Delta \mu \Omega^2 t}{\eta}\right)^{1/6}\;.
\end{equation}
Further details including the behavior of other
defect configurations are discussed in \textcite{Joanny2012}.

\subsection{Polar active suspensions with inertia}
In section \ref{wet} we summarized the continuum flocking model of
\textcite{Toner1995,Toner1998} for polar
self-propelling particles moving on a frictional substrate. Deep in the ordered
phase, the theory predicted novel propagating sound-like waves, coupling
orientation and density fluctuations, with direction-dependent wave speeds and
giant density fluctuations. The discussion thereafter, when the ambient fluid
medium was introduced, focused on \textit{destabilizing} effects arising
from the interplay of activity and the hydrodynamic interaction for bulk
suspensions in the Stokesian regime where viscosity dominates.  
However, we claimed in section \ref{sec:introduction} that our
hydrodynamic approach applied to flocks from sub-cellular to oceanic scales.
Bacterial suspensions, cell aggregates, and the cytoskeleton or its extracts
(see section \ref{wet}) are well approximated in the Stokesian limit where
inertia is altogether ignored. For collections of large swimmers, such as fish,
where inertial effects are important and the role of viscosity can be ignored an
alternative emphasis is appropriate. \textcite{Simha2002} originally formulated
their general theory of ordered states and fluctuations for active particles
suspended in a fluid with a view to describing both viscosity- and
inertia-dominated flows. 
Let us briefly review the discussion of \textcite{Simha2002} in the case where
inertia is taken into account through the \emph{acceleration} term in the
momentum equation. We ignore the advective nonlinearity $\mathbf{v} \cdot
\nabla \mathbf{v}$, so the treatment amounts to the unsteady Stokes equation,
but this is precisely the level of description in which the propagating modes
of translationally ordered crystalline or liquid-crystalline phases are
discussed by \textcite{Martin1972}.

The slow variables in a polar active suspension are the number
density, the broken symmetry variable ${\bf p}$ for polar order (defined in
section-I A. ) and the momentum density of the suspension $\bf{g}$ coming from
momentum conservation. We proceed initially as in section \ref{wet} A.2, through
the momentum equation \eqref{cons-g}
$\partial_t g_{\alpha} = - \partial_{\beta} \Pi_{\alpha \beta}$,
 with the stress tensor  $\Pi_{\alpha \beta}=
-\sigma^t_{\alpha \beta}\,+m\rho v_\alpha v_\beta$, where $\sigma^t_{\alpha \beta}$
contains the various contributions listed in Eqs.~\eqref{eq:diss-fluxes}, 
\eqref{eq:reac-stress-trace},\eqref{eq:reac-stress-dev} and
\eqref{eq:totalstress}, including the crucial active stress.
In systems where inertia is important, it is essential to retain
the acceleration $ \partial_t {\bf g}$ even while ignoring the inertial
contributions $m\rho v_\alpha v_\beta$ to the stress which are non-linear in ${\bf g}$.
The equation for the vector order parameter $\bf{p}$ is as in section
\ref{wet}.
The number density is governed by Eq. (\ref{conceqfilm}).

The hydrodynamic modes implied by these equations of motion are obtained by
linearizing and Fourier transforming in space and time \citep{Simha2002}.
Imposing overall incompressibilty with the condition 
$\nabvec \cdot {\bf g} =0 $, the number of modes is five.  We briefly state here
the main results of this analysis \citep{Simha2002}.  First, when viscosity is
ignored and acceleration is included, polar ordered suspensions are not
{\em in general} linearly unstable, that is, a parameter range of nonzero
measure exists for which stable behavior is found. The dynamic response 
displays a whole new range of propagating waves as a result of the interplay of
hydrodynamic flow with fluctuations in orientation and concentration: a pair of
bend-twist waves and three waves, generalizations of those in \textcite{Toner1998},
coupling splay, concentration and drift each with exceedingly complicated
direction-speed wave relations \citep{Simha2002}. The bend/twist waves result from
the
 interplay of $\bm\nabla\times{\bf p}_\bot$ and $\bm\nabla\times{\bf v}_\bot$,
which provides a qualitative difference with respect to dry flocks.
 These propagating modes can be
observed on length scales where it is reasonable to ignore their damping -- due
to viscosity for the total momentum and velocity fluctuations, and diffusion for
concentration fluctuations. This gives a large range of length scales
for large fast swimmers like fish. Possibly experiments like those of
\textcite{Makris2006}, which do speak of fish density waves, could test
the existence of these modes in detail. 
Secondly, within this linearized treatment, equal-time density correlations of the
density show the same features in the ordered phase as we already saw for
dry systems in section \ref{gdf}, with the number variance, scaled by the mean
$N$, predicted to diverge as $N^{1/2 + 1/d}$ in $d$ dimensions as in Eq.
(\ref{DN}). Presumably nonlinearities, which power-counting readily shows to be
relevant below 4 dimensions, but which are even more painful to analyze here
than in dry flocks, will change the exponents but not the essential fact of
supernormal fluctuations. It is not clear that the power-law static structure
factor seen by \textcite{Makris2006} in shoals is evidence for these giant
fluctuations; a fair test of flocking theories requires a school rather than a
shoal.

\subsection{Rheology}
\label{rheologysection}

The active hydrodynamic framework of section \ref{wet} not only
allows us to predict spontaneous-flow instabilities as in sections
\ref{spontflowsubsec} and \ref{thinfilmfree} but also the response of an active
fluid to an \textit{imposed} flow, i.e., the \textit{rheology} of active soft
matter \citep{Hatwalne2004,Liverpool2006,Giomi2010,Haines2009,Saintillan2010a}.

For concreteness, we associate the purely coarse-grained
description of section \ref{wet} with a microscopic picture of a suspension
containing active particles of linear size $\ell$, at concentration $\rho$,
each particle carrying a force dipole of strength $f \ell$, where
$f$ is the propulsive thrust, with the activity of an individual particle
correlated over a time $\tau_0$, and collective fluctuations in the activity
correlated over length scales $\xi$ and timescales $\tau$. 

We apply the approach here to the isotropic phase of active
particles in a fluid, to extract linear rheological properties
\citep{Hatwalne2004,Liverpool2006} and the autocorrelation of spontaneous stress
fluctuations \citep{Hatwalne2004,Chen2007} when noise is included.

We consider an apolar system described by orientation order in terms of the tensor $\bsf{Q}$, defined in Eq.~\eqref{Q}..
We obtain predictions for the rheology of active suspensions through the
coupled dynamics of $\bsf{Q}$ and the hydrodynamic velocity field $\mathbf{v} =
\mathbf{g}/\rho_{tot}$, where $\mathbf{g}$ and $\rho_{tot}$ are the
total densities of momentum and mass of the suspension. Neglecting inertial contribution to the stress tensor, conservation of total
momentum of particles $+$ fluid is expressed by $\partial_t {\bf g} =  \bm\nabla
\cdot \bm\sigma^t$, where the stress tensor $\bm\sigma^t$ is as in section
\ref{const-eqs}. 
In addition, we must allow for noise sources of thermal and non-thermal origin.
The former, though mandated in the equilibrium limit by the
fluctuation-dissipation relation connecting them to the solvent viscosity
in~\eqref{eq:diss-fluxes}, are quantitatively negligible in comparison to
effects arising
from activity. Such active fluctuating stresses can further arise in two ways:
directly, as additive white-noise contributions to the stress in
~\eqref{eq:diss-fluxes}-\eqref{eq:totalstress}, which we shall ignore as they
add no new physics, or indirectly, via the stochastic dynamics of the
orientation field, which is the case of interest. 

Consistent with Section \ref{wet}, we write the active stress in terms of the alignment tensor in the form
\beq
\label{sigproptoq}
\sigma_{\alpha\beta}^{a} = -\zeta \Delta \mu Q_{\alpha \beta}.
\eeq
We note for later reference that we will eventually consider the
concentration dependence of active stresses, $\zeta \propto \rho$, which is
physically reasonable and also follows from an explicit realization in terms of
dipolar force densities associated with the active particles
\citep{Hatwalne2004,Simha2002,Baskaran2009}. We remind the reader that we are dealing with
force-free, neutrally buoyant, self-propelling particles. Thus there are no {\em
external} forces on the system, so that the simplest active particle, on long
timescales, is a permanent force dipole \citep{Brennen1977,Pedley1992}.
Although we have already discussed the magnitude and sign of
$\zeta \Delta \mu$ in section \ref{wet}, it is useful to note here that for a
system with concentration $\rho$ the quantity $\zeta \Delta \mu / \rho
\sim \ell f $ characterizes the strength of the elementary force dipoles
associated with, for example, individual swimming organisms. Negative and
positive $\zeta \Delta \mu$ refer respectively to contractile swimmers, or
``pullers'', and extensile swimmers, or ``pushers'', whose distinct rheological
behavior we outline below.

The rheology is to be obtained from the momentum equation
together with the equations of motion for the order parameter field and
particle concentration. We begin with a description of the linear rheology of
active matter in the isotropic and orientationally ordered phases, and follow
it up with a brief survey of the nonlinear rheology of active matter.

\begin{figure}
\begin{center}
\includegraphics[width=9cm,height=6cm]{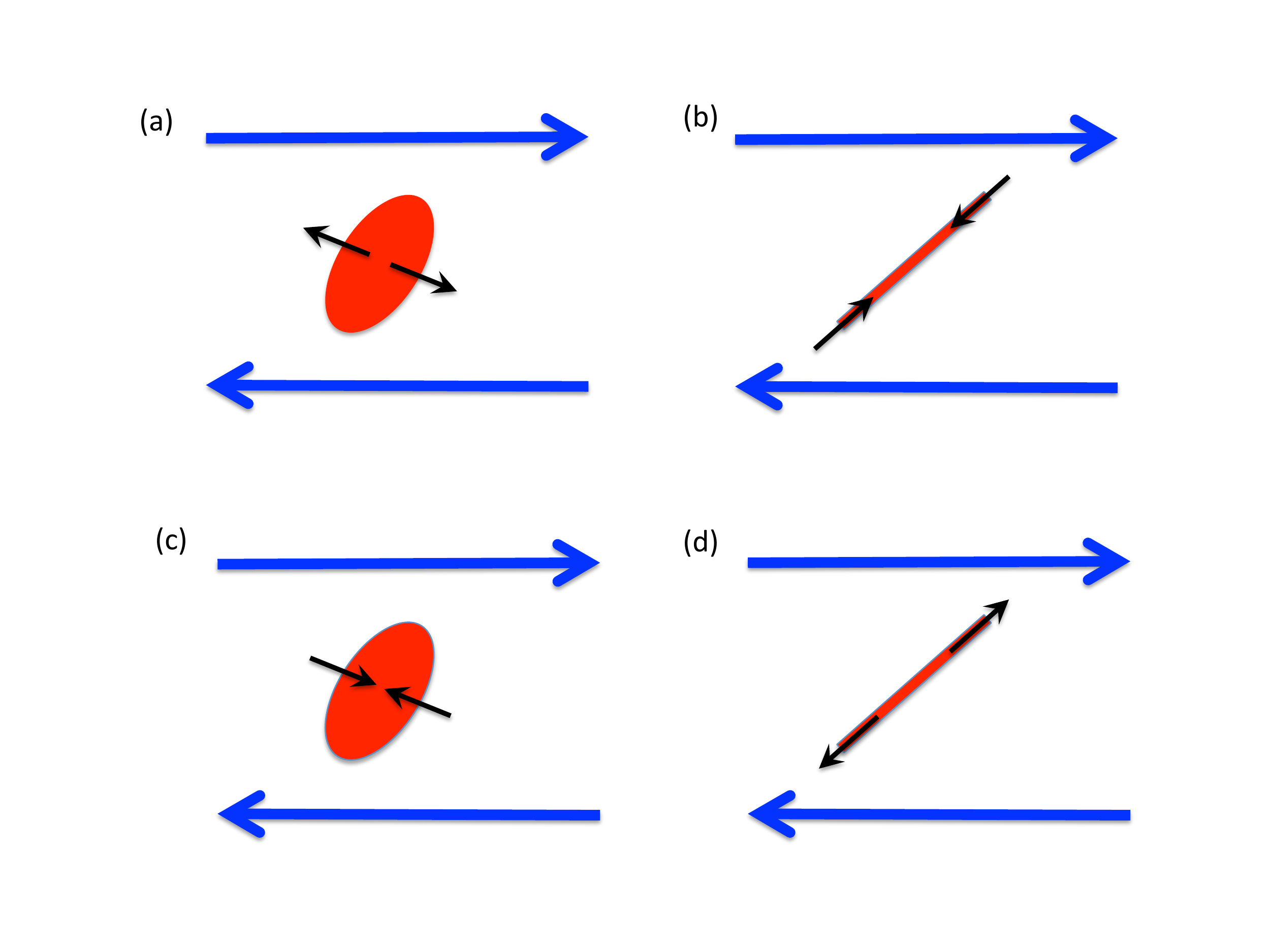}
\caption{\label{rmpfig-rheo1} (color online) Discs (a) and (c), and rods (b) and
(d) with active
force densities attached along their symmetry axes, under shear (horizontal
arrows).
The parameter $\zeta \Delta \mu <0$ in (a) and (b)
and $>0$ in (c) and (d).}
\end{center}
\end{figure}

\subsubsection{Linear rheology of active isotropic matter}
\label{rheoisosubsec}

To appreciate what is unique about active matter rheology,
it is useful to recall the linear rheology of passive nematogens.
The stresses arising from distortions of the orientational order
parameter $\mathbf{Q}$ and concentration $\rho$ are derived from a
free-energy
functional
$F_Q[\mathbf{Q}, \rho]$ (Eq.~\eqref{eq:nemfree}), giving
rise to a passive deviatoric
order-parameter stress~\citep{Forster1974},
\beq
\label{passiveopstress}
\bm\sigma^{op} =
 3 \mathbf{H} - \mathbf{H} \cdot \mathbf{Q} - \mathbf{Q} \cdot \mathbf{H}
\eeq
where $\mathbf{H}\equiv -\delta F_Q / \delta \mathbf{Q} + (1/3) \mathbf{I}\, \Tr
\delta F_Q / \delta \mathbf{Q}$
is the nematic molecular field.
The {\em mean} deviatoric
passive stress (\ref{passiveopstress}) is zero in the isotropic phase
at equilibrium (and in the nematic phase as well).
In addition, fluctuations of the deviatoric stress are small as
one nears the transition to the nematic phase, as can be seen by
making small
perturbations $\delta \mathbf{Q}$ in the isotropic phase, leading to a change in
the
 free energy $F_Q \propto (\alpha_Q/2) \int  \, Tr (\delta \mathbf{Q})^2$,  which
in turn
give rise to
stress fluctuations $\sim \alpha_Q  \, \delta \mathbf{Q}$ with a coefficient
$\alpha_Q$ that
{\it decreases}
on approaching the isotropic-nematic (IN)
transition to the ordered phase.
Thus even though fluctuations
of $\mathbf{Q}$ are large as one approaches the IN transition, their
contribution to rheology is small,
resulting simply in a renormalisation of $\tau$, the order parameter relaxation
time.
As shown below, this pretransitional feature is fundamentally different in
active isotropic systems.
In the following we take a purely coarse-grained approach and follow closely the
work
of \textcite{Hatwalne2004,Liverpool2006,Giomi2010}; for
a more microscopic treatment see \textcite{Haines2009,Saintillan2010a}.

For the active system, the deviatoric reactive stress has to be obtained from
the equations of motion for the order parameter and concentration,
rather than simply from the free energy functional. The
linearized equations for $\mathbf{Q}$ in
the isotropic phase,
 are of the form,
\beq
\label{opeom}
{\partial Q_{\alpha\beta} \over \partial t}
  =    - {1 \over \tau} Q_{\alpha\beta} + D \nabla^2Q_{\alpha\beta}
              + \nu_1 v_{\alpha\beta} + ....
+ f_{\alpha\beta},
\eeq
where
Eq.
\eqref{opeom} can be regarded as the modification of Eq,~\eqref{eq:nem_gen} to
include the effect
of shear flow to linear order. Here
$\tau$ is the orientational relaxation time, which could for
example be the run time in a collection of run-and-tumble bacteria, or the
rotational diffusion time, perhaps modified by collective and/or active
effects in an actomyosin extract, $D$ is a diffusivity ($\sim \ell^2/\tau_0$)
related to the ratio of a Frank constant to a viscosity, $\nu_1$ is a
``reversible'' kinetic coefficient or flow coupling parameter, taken for simplicity to be of the same order in the one entering 
the equation for the polarization ${\bf p}$~\citep{Forster1974}, $f_{\alpha\beta}$ is a
traceless,
symmetric, spatiotemporally white tensor noise representing active fluctuations,
and the dots include the coupling of orientation to flow.  Note that this
form of the linearized equation is valid for the passive system too, with the
time scale $\tau$ given by the order parameter relaxation time proportional to
$1/\alpha_Q$, which gets larger as one approaches the IN transition.

One can now proceed to calculate the linear viscoelastic properties of the
active suspension. In addition to the active deviatoric stress
(\ref{sigproptoq}),
we include the contribution from the
viscous dissipative stress $\sigma^{d}_{\alpha\beta}$ given in Eqs.~\eqref{eq:diss-fluxes}. For simplicity we neglect the tensorial nature of the 
viscosity of fluids with orientational order. 
Using (\ref{sigproptoq}), (\ref{passiveopstress}) and (\ref{opeom}), and
applying them to
spatially uniform ($q=0$) oscillatory shear flow at frequency $\omega$ in the
$xy$ plane one
obtains, to linear order in the fields, 
\bea
\label{stressvsrate}
\sigma_{xy}(\omega) &=& \left[\eta + {(\alpha_Q  -\zeta \Delta \mu ) \nu_1
\over
-i \omega +
\tau^{-1}}\right] v_{xy}\;.
\eea

The rheological response is defined by the complex modulus
$G(\omega) = \frac{\sigma_{xy}(\omega)}{\epsilon_{xy}(\omega)}$, with $\epsilon_{xy}(\omega)=v_{xy}(\omega)/(- i \omega)\, $ the strain.
The corresponding  storage (in-phase) and loss (out-of-phase) moduli $G'(\omega)$
and $G''(\omega)$,
defined by $G(\omega)=G'(\omega)+i G''(\omega)$,
characterize the elastic and viscous response of the system to an oscillatory
shear flow.
These moduli can be read out from  Eq. (\ref{stressvsrate}).

The active isotropic system is rheologically a Maxwell fluid to linear order.
This is best highlighted by the behavior of the apparent shear viscosity $\eta_{app} =
\lim_{\omega\to 0}G(\omega)/(-i \omega)$, which
shows an active excess viscosity  $\eta_{act} \propto -\zeta \Delta
\mu\tau\rho$, corresponding to either an enhancement or reduction, depending on the sign of $\zeta$.
 This active thickening (thinning) can be understood as follows
(Fig. \ref{rmpfig-rheo1}): in an imposed flow, in the {\em absence} of activity,
discs (rods) tend to align their symmetry axis along the compression (extension)
axis of the flow \citep{Forster1974}. When activity is switched on, the flow
induced by the intrinsic force dipoles clearly opposes the imposed flow in Fig.
\ref{rmpfig-rheo1} (a) and (b), and enhances it in (c) and (d). Activity thus
enhances viscosity in Fig. \ref{rmpfig-rheo1} (a) and (b), since
$\zeta \Delta \mu<0$,
and reduces it in (c) and (d) ($\zeta \Delta \mu>0$).
For $\zeta \Delta \mu < 0$ (\ref{stressvsrate}) shows that the viscosity grows
substantially as the system approaches a transition to orientational order as
$\tau$ is increased, and in fact should diverge if $\tau$ could
grow without bound.  Even more strikingly, a system of pushers,
i.e., extensile swimmers, should show a prodigious reduction in viscosity as
$\tau$ grows; indeed, nothing rules out a negative viscosity, that is,
spontaneous flow in an initially quiescent isotropic system. 
Experiments by Sokolov and coworkers~\citep{Sokolov2009}
have indeed shown that the
extensile activity  of  \emph{Bacillus subtilis}, a``pusher'' swimmer, can
substantially lower the viscosity of a suspension. 
By contrast, in a
passive, i.e., thermal equilibrium, system approaching a continuous or weak
first-order transition to a nematic phase, the excess viscosity $\sim \alpha_Q
\tau$ is
roughly constant since $\tau \propto 1/\alpha_Q$, as required by the
constraints of thermal equilibrium.

\begin{figure}
\begin{center}
\includegraphics[width=0.90\columnwidth]{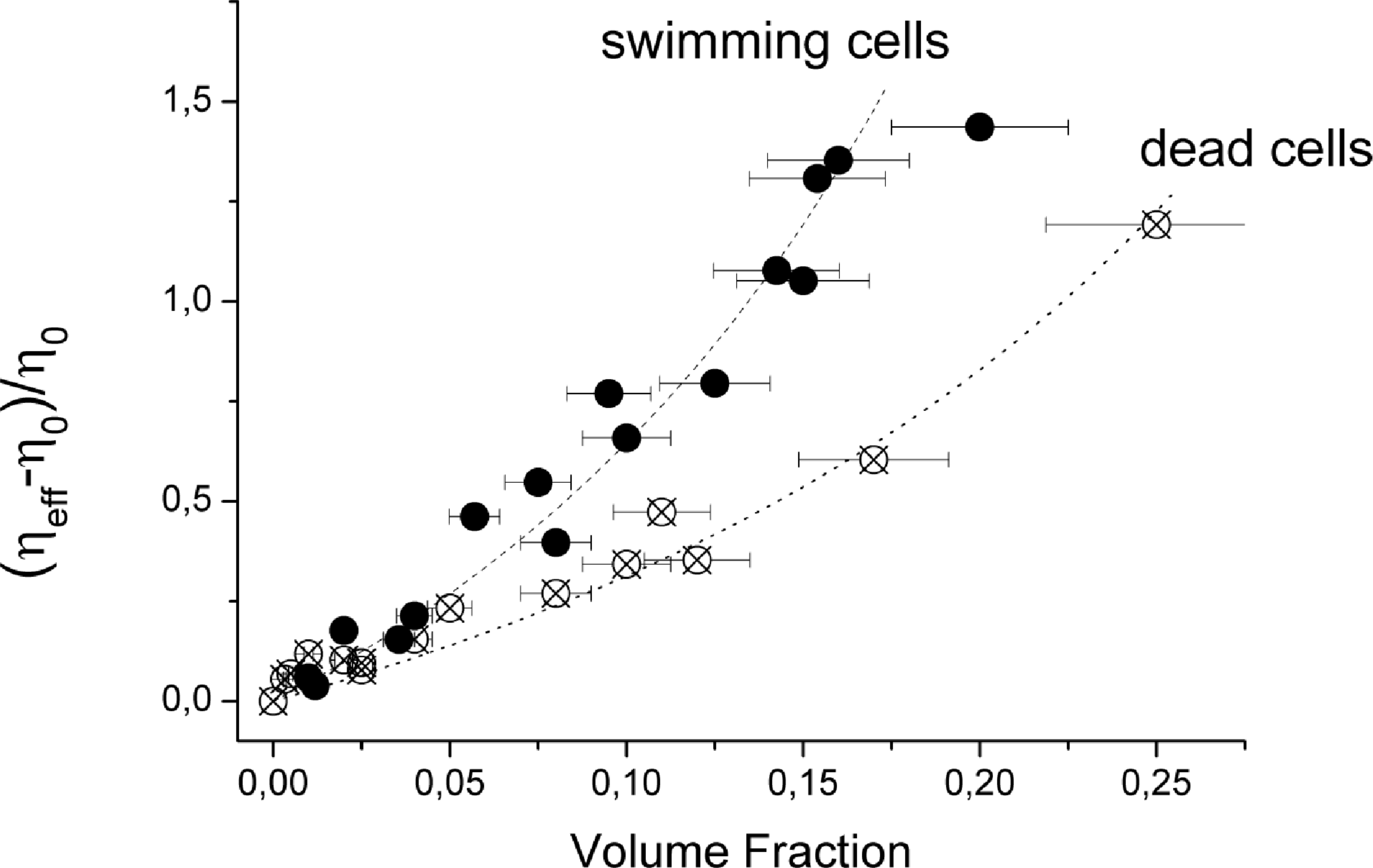}
\label{fig:rafai}
\caption{ The
contractile activity  of ``puller'' swimmers can increase  the viscosity of a
suspension. The data represent the effective viscosity of chlamydomonas
suspensions  relative to the viscosity $\eta_0$ of the culture medium as a
function of the volume fraction of bacteria. Solid symbols represent live cell
data and crossed symbols represent dead cell data. The mechanism producing
flow-orientation, however, remains
unclear.
Adapted with permission from \citep{Rafai2010}.}
\end{center}
\end{figure}

The active excess viscosity obtained within
the linear theory  should be compared to the well-known
result of \textcite{Einstein1906,Einstein1911}  that the fractional excess
viscosity due to the
addition of passive particles to a fluid is proportional to the particle volume
fraction $\phi=\pi\rho\ell^3/6$, to lowest order in $\phi$, with a coefficient $5/2$ for
spheres. For this purpose it is convenient to take
the active coupling
proportional to the concentration $\rho$ and define an active stress per particle $W$ as
\beq
\label{zeta_to_W}
\zeta \Delta \mu = f \ell \rho \equiv W \phi\;,
\eeq
where
$W = f / \ell^2$.
Then
\begin{equation}
\label{eta_act_zero_freq}
\eta_{act} = -W \tau \phi. 
\end{equation}
Equation~\eqref{eta_act_zero_freq} can be viewed as an
additive correction to the $5/2$,
proportional to $W \tau$, which of course can be of either sign. Behavior
consistent with these predictions is seen in recent experiments
measuring the activity-induced thickening in a system of \emph{Chlamydomonas}
algae (pullers,$W < 0$), as shown in
Fig.~\ref{fig:rafai}~\citep{Rafai2010},
and extreme thinning in a
system of \emph{Bacillus subtilis} bacteria (pushers, $W >
0$)~\citep{Sokolov2009}. In the case of chlamydomonas, there is some question
as to the direct applicability of the mechanism discussed above, as recent
experiments have shown that the flow field generated by these microorganisms,
although contractile, is more complex than
dipolar~\citep{Guasto2010,Drescher2010}. 
In general the linear rheology is controlled by the interplay of the nature of the active stresses, determined by $\zeta$, 
and the flow alignment coefficient $\nu_1$.
A remarkable duality has been identified that shows that shows that tensile ($\zeta>0$) rod-shaped flow-aligning particles ($|\nu_1|>1$) are rheologically equivalent (to linear order in the strain rate) to contractile ($\zeta<0$) discotic flow-tumbling particles  ($|\nu_1|>1$) \cite{Giomi2010}.

Equation (\ref{stressvsrate}) also predicts strong viscoelasticity as $\tau$
increases. For {\em passive} systems, $\zeta \Delta \mu = 0$ and $\alpha_Q
\propto
\tau^{-1}$, and so
$G'(\omega \tau \gg 1)$ decreases as  $\nu_1  \eta / \tau$, suggesting as
earlier, that there is little viscoelasticity near an {\em equilibrium}
IN transition. For {\em active}  contractile ($\zeta
\Delta \mu < 0$) systems, by contrast, the active contribution is of $O(1)$ and
independent of $\tau$ even close to the IN transition. Thus, as $\tau$ grows,
\beq
\label{gpact}
G'(\omega \tau \gg 1) \simeq -\zeta \Delta \mu
\eeq
independent of $\tau$! In addition to this enhanced elasticity, the dynamic
range over which elastic behavior is seen increases.  This is quite a dramatic
rheological manifestation of activity, since at {\em equilibrium}, one would
expect such strong viscoelastic behavior from a fluid or suspension near {\em
translational} freezing, as at a glass transition, not near {\em
orientational} ordering. Put more radically, the orientationally ordered phase
of contractile active particles is a peculiar yield-stress material, with
nonzero shear and normal stresses in the limit of zero shear rate
\citep{Hatwalne2004,Liverpool2006,Marenduzzo2007,Giomi2010} as discussed in
section
\ref{nonlinearrheosubsec}.

Another observable manifestation of activity is an enhanced noise temperature
as inferred for example from tracer diffusion measurements in bacterial
suspensions \citep{Wu2000}, and frequency-dependent shear viscosity arising from
fluctuations in stress and concentration \citep{Chen2007,Hatwalne2004}. Activity
is the transduction of chemical energy, say in the form of ATP hydrolysis equal
to about $20 k_BT$ per ATP molecule. This suggests that fluctuations from this
athermal noise would be significant. Following \textcite{Hatwalne2004}, we may
estimate the strength of nonequilibrium stress fluctuations that result from
fluctuations of the force generation of active particles, through the variance
of the active stress at zero wavenumber and frequency. The divergence of the
active stress $\bm\sigma^a = - \zeta \Delta \mu \bsf{Q}$ appears as a forcing
term
in the momentum equation. Assuming the active objects, whether biofilaments or
bacteria, are collectively in a spatially isotropic state, this forcing can be
viewed as a noise on scales large compared to the correlation length and time
$\xi$ and $\tau$ defined at the start of this section. Recall that in
fluctuating hydrodynamics \citep{Landau1998}, for a thermal equilibrium fluid
with shear viscosity $\eta$ and temperature $T$, the variance of the random
stresses at zero frequency is $ \eta T$. For an active fluid, then, the
\textit{apparent} temperature as probed by the motion of a tracer particle can
be estimated by the zero-frequency, zero-wavenumber variance of the active
stress $\eta T_{eff} \sim (\zeta \Delta \mu)^2\int d^3r dt \langle
\bsf{Q}(\mathbf{0},0){\bf :} \bsf{Q}(\mathbf{r},t) \rangle$. The simplest
dimensional
argument would then give $T_{eff} \sim (\zeta \Delta \mu)^2 \xi^3 \tau/\eta$. An
Ornstein-Zernike form $(\ell/r) \exp(-r/\xi)$ for the equal-time correlations of
$\bsf{Q}$, which requires the introduction of a microscopic length which we take
to be the active-particle size $\ell$, gives instead $T_{eff} \sim (\zeta \Delta
\mu)^2 \xi^2 \ell \tau/\eta$. From (\ref{zeta_to_W}), using
rough estimates for the bacterial system of \textcite{Wu2000} $f \sim v/\eta \ell$
with $\eta \simeq 10^{-2}$ Poise, $\ell \sim 1 \, \mu$m, $\phi \sim 0.1$ $\xi
\simeq 20 \, \mu$m corresponding to a ``run'' for a time $\tau = 1$s and speed
$v \simeq 20 \, \mu$m/s yields $T_{eff}$ about $400$ times room temperature,
consistent with the measurements in \citep{Wu2000}. Note that this enhancement
is
{\it independent of the sign of} $\zeta \Delta \mu$, i.e., the diffusivity
increases regardless of whether the viscosity decreases or increases, a
tell-tale sign of the nonequilibrium nature of the system. A further
interesting consequence \citep{Hatwalne2004} of this excess noise is a huge
enhancement of the amplitude of the well-known $t^{-d/2}$ long-time tails in the
autocorrelation of tagged-particle velocities. We know of no experiment that has
probed this last feature.

More extended analyses of fluctuations in active systems, demarcating the 
dynamical regimes lying within and beyond the conventional
fluctuation-dissipation theorem include work by \textcite{Mizuno2007,Kikuchi2009}.

We now turn to nonlinear fluctuation effects \citep{Chen2007,Hatwalne2004}. From
(\ref{sigproptoq}), the fluctuations in the deviatoric stress gets contributions
from fluctuations bilinear in the orientation $\mathbf{Q}$ and the concentration
$\rho$. The stress autocorrelation is therefore a convolution of $\bsf{Q}$ and
$\rho$ correlations. The former is evaluated from (\ref{opeom}), while the
latter can be calculated from the the linearized equations for the
concentration, $\partial_t \delta \rho = - \nabla \cdot {\bf J}$, where the
current $J_{\alpha} = -D \partial_{\alpha} \delta \rho - W^{\prime} \rho_0
\partial_{\beta} Q_{\alpha \beta} +  f^{\rho}_{\alpha}$, where $D$ is the diffusion
constant, $\rho_0$ the mean concentration, $f^{\rho}_{\alpha}$ a random noise, and
$W^{\prime}$ is an activity parameter.  The resulting key findings of
\textcite{Chen2007,Lau2009} are an excess fluctuation $\phi
\omega^{-1/2}$ in the stress \textit{fluctuations}, and no excess
\textit{response}, i.e., viscosity, in the range studied, again a sign of
nonequilibrium behaviour. \textcite{Hatwalne2004} consider stress contributions
nonlinear in $\bsf{Q}$ and show they should lead to excess viscosity as well of
similar form, but the effect is presumably below detectable levels in the
\textcite{Chen2007} experiment.

\subsubsection{Linear rheology of active oriented matter}
\label{rheoorientsubsec}
In extending the study of rheology to active oriented matter, we immediately
encounter a problem. As remarked in Section \ref{spontflowsubsec},
long-range uniaxial orientational order, whether polar or apolar, in active
Stokesian suspensions of polar particles is {\em always} hydrodynamically
unstable to the growth of long wavelength splay or bend fluctuations, depending
on the sign of $\zeta \Delta \mu$. This instability has no threshold in a
spatially unbounded system \citep{Simha2002}, and the growth rate is highest at
wavenumber $q=0$. Frank elasticity with stiffness $K$ stabilizes modes with $q$
greater than
\beq
\label{q_cross}
q_0 \propto \sqrt{|\zeta \Delta \mu| / K}, 
\eeq
so that there is a band of unstable modes from $0$ to $q_0$. Note that $q_0$ is simply an approximate form of
the length scale defined by Eq.~\eqref{L2}. In any case, in the
limit of infinite system size there is no stable reference state, no ideal
active nematic or polar liquid crystal whose rheology one can study as
a geometry-independent material property. We must therefore ask what
suppresses this generic instability \citep{Ramaswamy2007}. A mechanism for suppression of the instability is confinement by
boundary walls. The existence of
 a crossover wavenumber (\ref{q_cross}) implies that the
instability exists only if the sample's narrowest dimension $h > \sqrt{K/|\zeta
\Delta \mu|}$; equivalently, for fixed $h$ the activity must cross a threshold
$\sim K/h^2$. This is the essential content of the treatment of
\textcite{Voituriez2005} discussed in section \ref{spontflowsubsec}.
Confinement along $y$, with the director spontaneously aligned along $x$
and free to turn in an unbounded $xz$ plane, shows \citep{Ramaswamy2007} a
similar threshold but with a $q^2$ dependence of the growth rate at small
in-plane $q$.
In either case, it is clear that confinement can produce a stable active liquid
crystal whose linear rheology one can study. Alternatively, the instability can be suppressed by imposing an external shear
flow or by the presence of
partial translational order, either columnar or lamellar.
Orientational stabilization can also be achieved in  flow aligning systems
 ($|\nu_1| >1$).
by imposing a uniform shear flow with shear rate ${\dot \gamma}$.
This stabilizes those wavevectors whose growth rate is smaller than
the shear rate ${\dot \gamma}$. As ${\dot \gamma}$ is increased more and more
modes are stabilized, until at a threshold ${\dot \gamma}_c$, the oriented phase
is completely stabilized by the shear flow, yielding a stability diagram
controlled by two variables, the flow alignment parameter $\nu_1$
and the ratio
of the shear to active stress \citep{Muhuri2007,Giomi2010}. Translational order, both partial as in
smectic or columnar liquid crystal,  or full as in three-dimensional crystalline systems can also yield
a
stable active system, whose properties have been the subject of recent
studies \citep{Adhyapak2012}.

Stabilizing the orientational phase of active matter now sets the stage for a
study of its unusual rheology \citep{Hatwalne2004,Liverpool2006,Giomi2010}.
We have already seen that on approaching the orientationally ordered state from
the isotropic fluid, a suspension of active {\it contractile} elements ($W<0$)
exhibits solid-like behaviour without translational arrest. In the
orientationally ordered phase, the orientational order parameter $\langle
\mathbf{Q}\rangle \neq 0$, which by (\ref{sigproptoq}) immediately leads to a
nonzero steady-state average of the deviatoric stress, {in the absence of any
external deformation}. This {\it prestress} is a truly nonequilibrium effect,
and has no analogue in a passive equilibrium nematic fluid which has a purely
{\em isotropic} mean stress, i.e., a pressure, despite
its orientational order.
 This prestress implies that in a flow experiment, the
shear stress will not vanish at zero deformation rate.
The same features of being first order in shear rate and having a nonzero value
at zero shear rate are exhibited by the normal stresses
$\sigma_{yy}-\sigma_{xx}$. We emphasize that this strange kind of yield
stress is a manifestation of the rheology of an active oriented fluid without
any form of {\it translational arrest}.

We end this section on the rheological properties of active oriented
matter by drawing the attention of readers to an interesting analogy between
active stress of contractile filaments with macroscopic orientational order and
fragile jammed granular matter, as discussed by \textcite{Ramaswamy2007}.

\subsubsection{Nonlinear rheology of active nematics}
\label{nonlinearrheosubsec}

We now turn to the rheological properties of orientationally ordered active
fluids beyond the linear regime.  As detailed in the previous section, the
generic instability of orientationally ordered active suspensions
\cite{Simha2002} can be suppressed by confinement or by imposing an external
shear \cite{Muhuri2007,Marenduzzo2007}. It is therefore meaningful to explore the dynamics and
rheology of these phases in confined geometries. Extensive numerical studies of
active nematic and polar films \cite{Cates2008,Marenduzzo2007,Fielding2011,Giomi2010} under shear reveal a rich variety of phenomena influenced by boundary conditions and geometry. This complex behavior results again from the interplay between local stresses generated by activity (quantified by the parameter $\zeta$), the flow aligning property of these particles (characterized by the parameter
$\nu_1$) and the typical self-propulsion velocity $v_0$ in polar active fluids. A complete understanding of their response to shear necessitates exploring the space spanned by these parameters under various boundary conditions. 


We present only a summary of the main results obtained in the literature. 
While detailed theoretical and numerical studies of linear active rheology including the effect of
polarity can be found in \cite{Giomi2008,Giomi2010},
the nonlinear rheology has so far been studied mainly for apolar systems.  For contractile active fluids in an orientationally ordered
state, numerical studies of the hydrodynamic equations \cite{Cates2008} when
only one-dimensional variation is allowed show the onset of solid-like
behavior of \cite{Hatwalne2004} and the existence of a yield shear stress as
in \cite{Liverpool2006,Ramaswamy2007}.  Studies of {\em extensile} fluids,
allowing variation in one \cite{Cates2008} and two \cite{Fielding2011} spatial
directions display the onset of spontaneous flowing states, with complex flow
states including bands in $1d$ and rolls and turbulence in $2d$.  Related
experimental findings include large scale chaotic flows in bacterial systems
\cite{Aranson2007,Dombrowski2004}, as remarked earlier. An important general
feature \cite{Cates2008,Fielding2011} is that dimensionality matters: imposed
restrictions to $1d$ spatial variation lead to significantly different rheology
from that seen when $2d$ variation is allowed, for example.  Active stresses
generally appear to stabilize shear bands, but have the opposite effect close
to the isotropic-nematic transition.  In systems with free boundary
conditions, the approach to the isotropic-nematic contractile transition shows
\cite{Cates2008} an active enhancement of viscosity. As claimed in
\cite{Hatwalne2004}, orientational ordering  in active systems indeed resembles
translational arrest in equilibrium systems.   

The relation between the rheology of active fluids
in external shear and the onset of spontaneous flow in absence of shear is discussed in \cite{Giomi2010}.
This work has also analyzed in detail the nonlinear rheology of active fluids, revealing 
strongly non-monotonic stress vs strain-rate curves beyond a
threshold value of activity, as suggested in \textcite{Hatwalne2004}, with  macroscopic  yield-stress behavior, and hysteresis. Finally, at even higher activity, the theoretical stress-strain curve has a discontinuous jump at zero strain rate, corresponding to a finite ``spontaneous stress'' in the absence of applied shear. 

Given the complexity and richness of the nonlinear rheology of active fluids, we refer the reader
to the literature for further details.

\subsection{Applying the hydrodynamic theory to phenomena in living cells}
\label{subsec:cells}
To demonstrate that the hydrodynamic theory of active gels described in the
previous sections  indeed provides generic tools for addressing questions 
relevant to living cells, we briefly summarize  here a few examples of concrete
successes of the theory. A more complete review can be found in
\textcite{Joanny2010,Grill2011}.

The first example is directly relevant to cell motility, in particular to the
migration of fish keratocytes, eukaryotic cells extracted from fish scales that
are among the fastest
moving cells, migrating on a substrate at a steady speed of about $10\mu m/min$.
This cells have a characteristic fan shape, with a large flat region extending
in front of the nuclei, towards the direction of
forward motion, known as the lamellipodium. This region is filled with a
cross-linked actin gel, where myosin motor complexes use the energy from ATP
hydrolysis 
to grab on neighboring actin filaments and exert stress.  
\begin{figure}
\centering
\includegraphics[width=0.95\columnwidth]{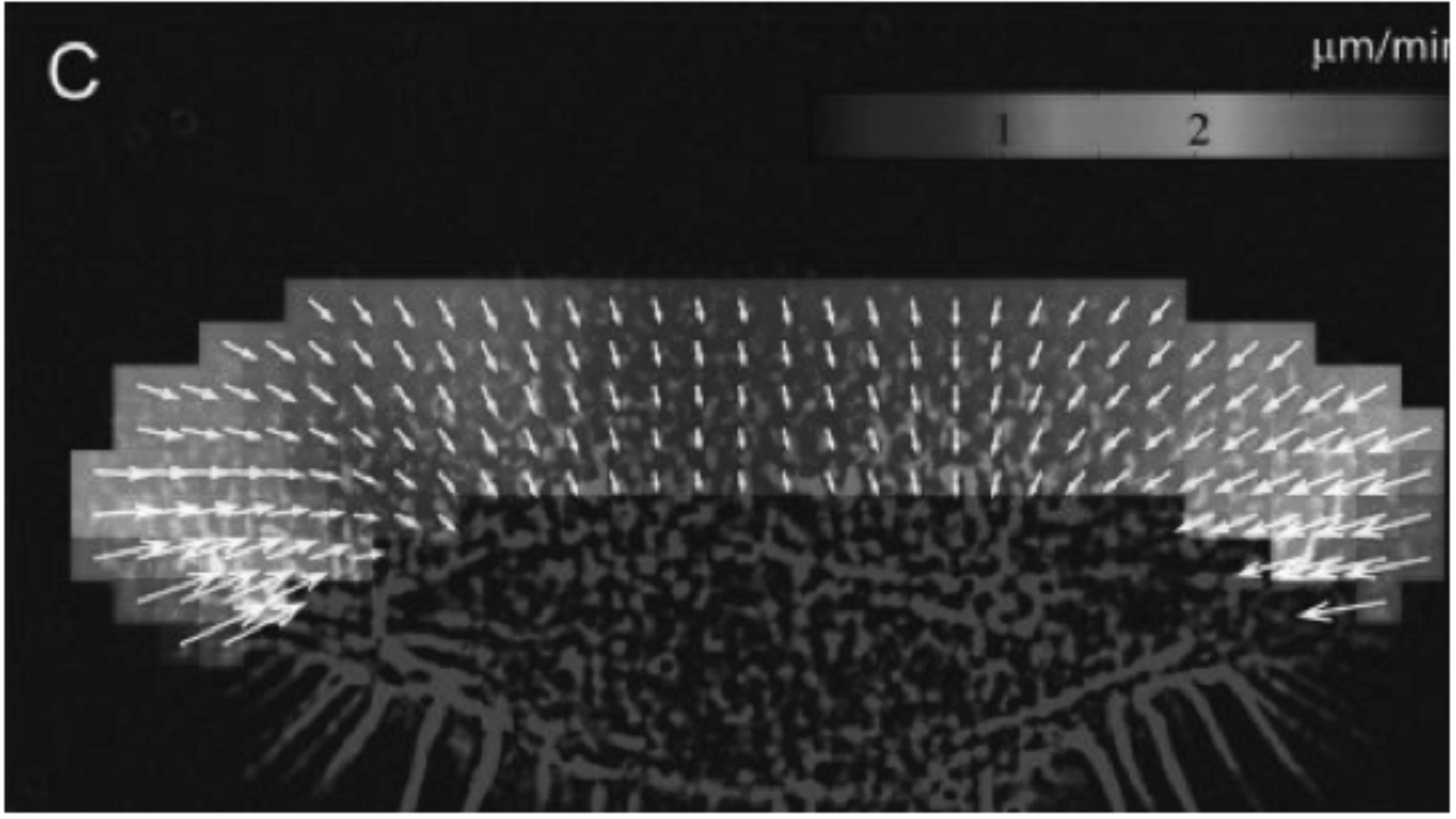}
\caption{Velocity field determined by speckle microscopy in a keratocyte
lamellipodium. Adapted with permission from \textcite{Vallotton2005}.}
\label{fig:keratocyte}
\end{figure}
Actin polymerization takes place at the leading edge of the lamellipodium, while
the filaments disassemble in the rear region, in a process that plays a crucial
role in driving the motility. Using active gel theory to model the
lamellipodium, Kruse and collaborators~\citep{Kruse2006a} were
able to
evaluate quantitatively the profile of the retrograde actin flow that
accompanies the forward motion of the lamellipodium. Such a retrograde flow has
been seen in experiments, as shown in
Fig.~\ref{fig:keratocyte}~\citep{Vallotton2005}.

\begin{figure}
\centering
\includegraphics[width=0.95\columnwidth]{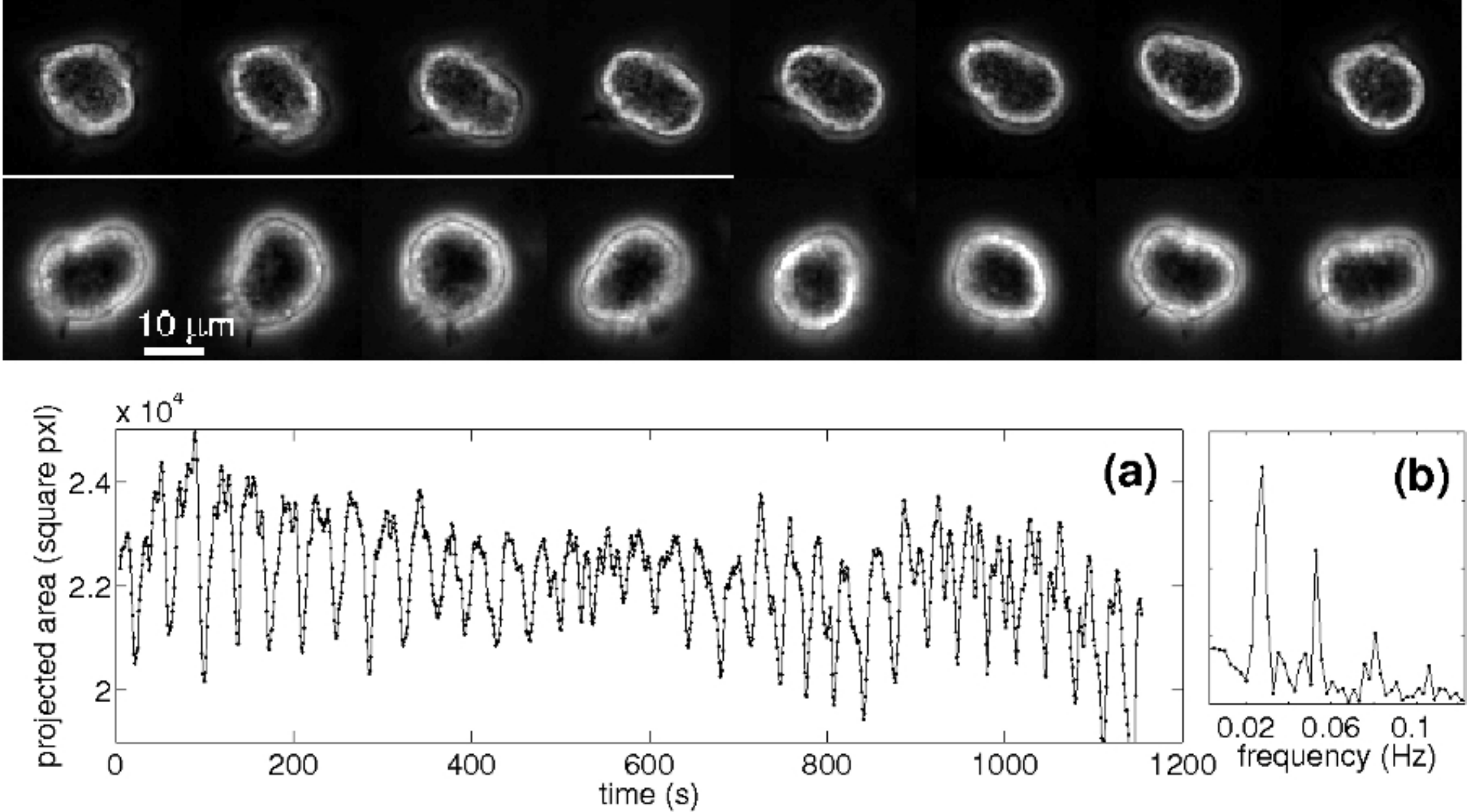}
\caption{Shape oscillations of non-adhering fibroblasts. The second frame shows
the periodic oscillation of the projected area of the cell and the associated
Fourier spectrum. The period of the oscillation of these cells  is very well
defined and of the order of $30s$. The oscillation period is found
to decrease
when myosin activity increases. The latter can in turn be modulated by the
addition of various drugs. Adapted with permission from \textcite{Salbreux2007}.}
\label{fig:cell-oscillation}
\end{figure}

A second example is that of  shape oscillations, observed ubiquitously in many
cells. An example is shown in Fig.~\ref{fig:cell-oscillation}.
This phenomenon has been described successfully by the active gel theory,
augmented by assuming a coupling to calcium channels in the cell membrane that
in turn are gated by the deformations in the actin cortical layer. Calcium
concentration couples to local myosin activity that in turn controls the
stretching and compression of the actin layer, in a feedback loop that results
in sustained oscillations~\citep{Salbreux2007}.

\begin{figure}
\centering
\includegraphics[width=0.25\columnwidth]{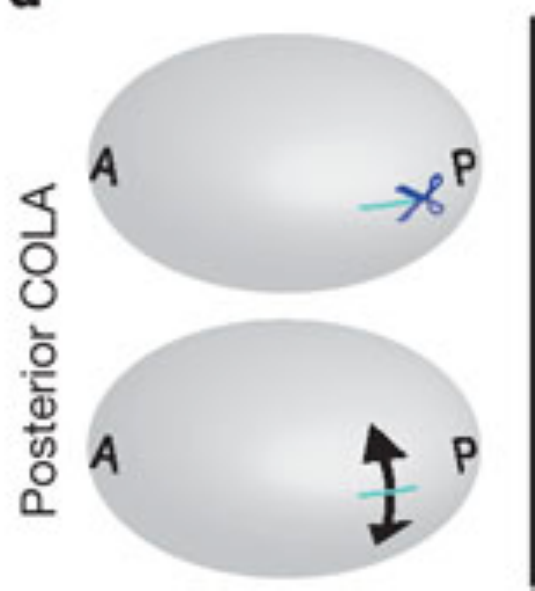}
\includegraphics[width=0.7\columnwidth]{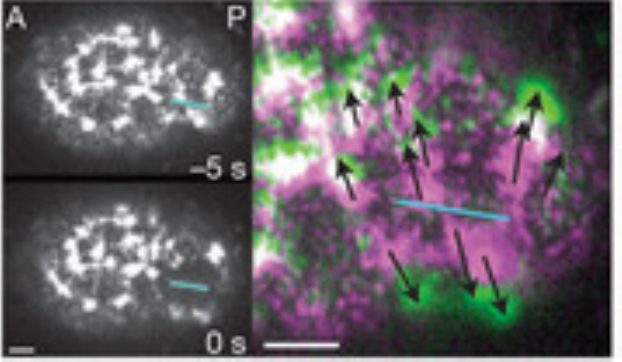}
\caption{(color online) This figure displays the imaging of tension in the
cortex of a \emph{C. elegans }zygote obtained before and after  cortical laser
ablation (COLA)  performed on the actomyosin meshwork. The left image shows a
schematic of COLA performed  with a pulsed ultraviolet laser along a $6\mu m$
line (light blue in both images) along the anterior-posterior (AP) axis,  in the
posterior of the cell. The center frames show the pre- (top) and post-cut image
(bottom) of posterior COLA.  The image to the right shows the tension
measurement in the zygote's actomyosin cortex upon laser ablation along the blue
line. The black arrows are the displacements between pre-cut (purple) and
post-cut images (green). The average speed of this initial recoil measured in a
direction orthogonal to the cut line is proportional to the normal component of
the 2D tension tensor. The white bar is $4\mu m$. Adapted  with permission from
\textcite{Mayer2010}.}
\label{fig:Grill}
\end{figure}

The active gel theory has also been used recently to model cortical flow in the
\emph{ C. elegans} zygote~\citep{Mayer2010}, as shown in Fig.~\ref{fig:Grill}.
The theory supported experiments in successfully identifying two prerequisites
for large-scale cortical flow necessary to initiate the anteroposterior cell
polarization which directs the asymmetry of the first mitotic division: a
gradient in actomyosin contractility to drive flow, and a sufficiently large
viscosity of the cortex to allow flow to be long-ranged.

Another recent success of active hydrodynamics has come from its application
to the organization and dynamics of molecules on the \textit{surface} of living
metazoan cells \citep{Gowrishankar2012a}. The lateral compositional
heterogeneity of the plasma membrane at 
sub-micron scales, termed ``lipid rafts'',  has been the subject of intense
research \citep{Lingwood2010}. Most attempts to understand this heterogeneity
are based on equilibrium thermodynamics. Using a variety of fluorescence
microscopy techniques, and confirmed by other methods, it has been shown that
the dynamics, spatial distribution and statistics of clustering of a key `raft'
component, viz., GPI-anchored proteins, is 
regulated by the active dynamics of cortical actin filaments
\citep{Goswami2008}. Following this, Gowrishankar and collaborators
\citep{Gowrishankar2012a} have developed a
model for an active composite 
cell membrane based on active hydrodynamics,
which shows how the dynamics of transient aster-like regions formed by active
polar filaments (actin) can drive passive advective scalars (molecules such as
GPI-anchored proteins) to form dynamic clusters. In addition to successfully
explaining the many striking features of the dynamics and statistics of
clustering of GPI-anchored proteins on the cell surface,  this model makes
several
predictions, including the existence of giant number fluctuations which have
been verified using fluorescence microscopy \citep{Gowrishankar2012a}.
These studies suggest that rafts are actively constructed and that the active
mechanics of the cortical cytoskeleton regulates local composition at the cell
surface via active currents
and stresses.

\begin{figure}
\centering
\includegraphics[width=0.95\columnwidth]{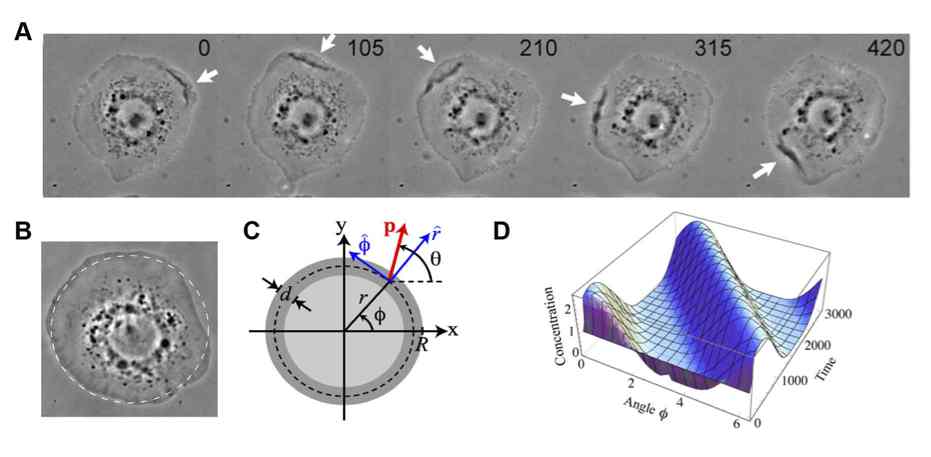}
\caption{(color online) The top images (A) show a  a circular actin wave in the
lamellipodium in a  Drosophila cell fixed to a substrate and treated to enhance
polarization of actin filaments. 
The wave travels along the cell's  periphery, with the arrow marking the region
of maximum actin density at various times. (B) Phase contrast microscopy picture
of a control
Drosophila cell fixed to a substrate with a typical circular shape. (C) In the
theoretical model, the
cell perimeter is represented by a circle of radius R. (D) The calculated
concentration of active filaments as a
function of the polar angle $\phi$ and time. Adapted with permission from
\textcite{Asano2009}.}
\label{fig:Asano}
\end{figure}

Finally, Fig.~\ref{fig:Asano} shows polarized actin waves propagating around the
periphery of a Drosophila cell that has been fixed to a substrate, preventing it
from moving, under condition of enhanced actin polymerization~\citep{Asano2009}.
In this case the hydrodynamic theory of a polar active gel successfully
reproduces the existence of a critical value of polarity  above which the
polarization waves will occur, as observed in experiments.

\section{Microscopic descriptions of active matter}
\label{microscopic}

\subsection{Review of microscopic models}

While  the macroscopic equations of motion of active matter can be obtained
from 
general considerations of symmetry, determination of the magnitude and often
even  the 
sign of the coefficients requires additional physical assumptions. Unlike for
systems near 
equilibrium, the assumptions are not easily identified.  Furthermore the
relaxation of the 
constraints required by equilibrium allows the possibility of a much larger
number of  terms. 
Physical insight  can be provided by using the tools and methods of
nonequilibrium
statistical mechanics~\citep{Zwanzig2001} to derive the continuum equations via
systematic coarse-graining of
simplified microscopic models of the active processes driving the system 
\citep{Kruse2000,Kruse2001,Liverpool2001,Liverpool2003,Aranson2005,
Kraikivski2006,Baskaran2008a,Bertin2006,Bertin2009}. One approach is to start
from stochastic equations for
the microscopic dynamics  and then systematically  project the microscopic
degrees of 
freedom  on to macroscopic variables, such as the density $\rho(\bfr,t)$ and
polarization 
$\bfp (\bfr, t)$, defined in  Eq.( \ref{rho},\ref{p}). This procedure yields
hydrodynamic 
equations  for the macroscopic variables  on length scales long compared to the
size $\ell$ of 
the individual active elements. For simplicity, we restrict ourselves here to
overdamped 
systems, where the medium through which the active particles may move (fluid or
substrate) 
is inert and only provides friction. In this case the momentum of the active
particles is not 
conserved and the microscopic degrees of freedom are  the positions $\bfr_n(t)$
and 
orientations $\bnu_n(t)$ of the active elements, plus (possibly) other
microscopic degrees of 
freedom describing internal active processes. Of course medium-mediated
hydrodynamic 
interactions can play a crucial role in controlling the dynamics of collections
of swimmers. 
To incorporate such effects one needs to consider a two-component system and
explicitly 
describe the exchange of momentum between particles and solvent. Both the 
translational and rotational velocities of each particle must be incorporated in
the 
microscopic model, as well as the solvent degrees of freedom. This more general
case is 
outlined briefly in section \ref{Hydro-int}.
In general  approximations  are required both to identify a tractable
microscopic 
model and to carry out  the coarse-graining procedure to derive equations for
the macroscopic 
variables.  The form of the resulting equations is general, as it is dictated by
symmetry, 
and the microscopic description allows a calculation of the phenomenological
parameters and
transport coefficients involved in the hydrodynamic equations. This is important
for example for biological systems because it allows to predict the variation
of the transport 
coefficients with the biological parameters that can be changed in the
experiments. 

 The purpose of this 
section is to review a number of such derivations and  summarize their
similarities and 
differences.  In equilibrium statistical mechanics there is a long tradition of
simplified  
microscopic models which have eventually led to a number of paradigmatic minimal
models 
(Ising, XY, Heisenberg) that capture the essentials of the behavior of variety
of equilibrium 
systems. Similarly here we describe some 
minimal microscopic descriptions of active matter that have provided insight
into the 
complex physics of these systems.  An interesting aspect of these microscopic
realizations is 
that they allow one to identify 
the common behavior of different experimental systems. Two examples which we
will focus 
on are (1) mixtures of cross-linking motor constructs and protein filaments 
\citep{Nedelec1997,Nedelec1998,Surrey2001}  and (2) collections of
self-propelled particles.
Experimental realizations of the latter may be actin filaments in 
 motility assays ~\citep{Kron1986,Schaller2010,Butt2010} or  suspensions of
swimming micro-organisms \citep{Dombrowski2004,Zhang2010}.

\subsubsection{Self-propelled particles}

A microscopic realization of active systems is provided by  interacting
self-propelled particles (SPP). 
While these can be thought of as simplified agent based models of flocks of
birds and shoals 
of fish~\citep{Vicsek1995}, they can also act as realistic models for less
complex  systems,
such as polar protein filaments (e.g. F-actin) in gliding motility assays on
surfaces decorated 
with molecular motors (e.g. myosin)  \citep{Kron1986,Schaller2010}, suspensions
of
swimming micro-organisms  \citep{Dombrowski2004,Zhang2010}, or even layers of
vibrated granular rods~\citep{Narayan2007}.
 It should be stressed, 
however, that there have been suggestions that hydrodynamic interactions may be
important 
in motility assays at high filament density~\citep{Schaller2011}

Each self propelled particle has a position and an orientation.  
It also has an individual 
self -propulsion velocity of magnitude $v_0$ and direction specified by the
particle's  orientation.
SPP interact with 
neighboring particles  either via a local rule~\citep{Vicsek1995} or via
physical steric or
other interactions~\citep{Baskaran2008a,Baskaran2008,Baskaran2009,Leoni2010}.
Mean field models for
the stochastic dynamics of the particles can be expressed in terms of the one
particle 
density, $c(\bfr,\bnu,t)$. This measures the probability of finding a
self-propelled particle  with  position $\bfr$ and 
orientation $\bnu$ at time $t$.   The equation of motion for $c(\bfr,\bnu,t)$
takes into
account the interplay of fluctuations (diffusion), interactions and
self-propulsion. These mean-field models are 
valid in the regime of 
weak interactions, low density and weak density correlations. The kinetic
equation for the one-particle density can then be solved directly either
analytically or numerically for specific geometries and initial
condition~\citep{Saintillan2012}
Alternatively, to describe  macroscopic behavior, the kinetic theory can also
be 
further coarse-grained by projecting on to macroscopic fields such as
$\rho(\bfr,t)$ and $\bfp(\bfr,t)$ to 
obtain the equivalent of Eq. \eqref{P-eq-2}, where the parameters
$\alpha,\beta,K,v_1,
\lambda_{1,2,3}$ are expressed in terms of the microscopic parameters of the
system~\citep{Baskaran2008a,Bertin2009,Baskaran2010,Ihle2011}.
A concrete example of this procedure is given in section \ref{micro-example}
below.

\subsubsection{Motors and filaments}

\begin{figure}
\centering
\includegraphics[width=0.65\columnwidth]{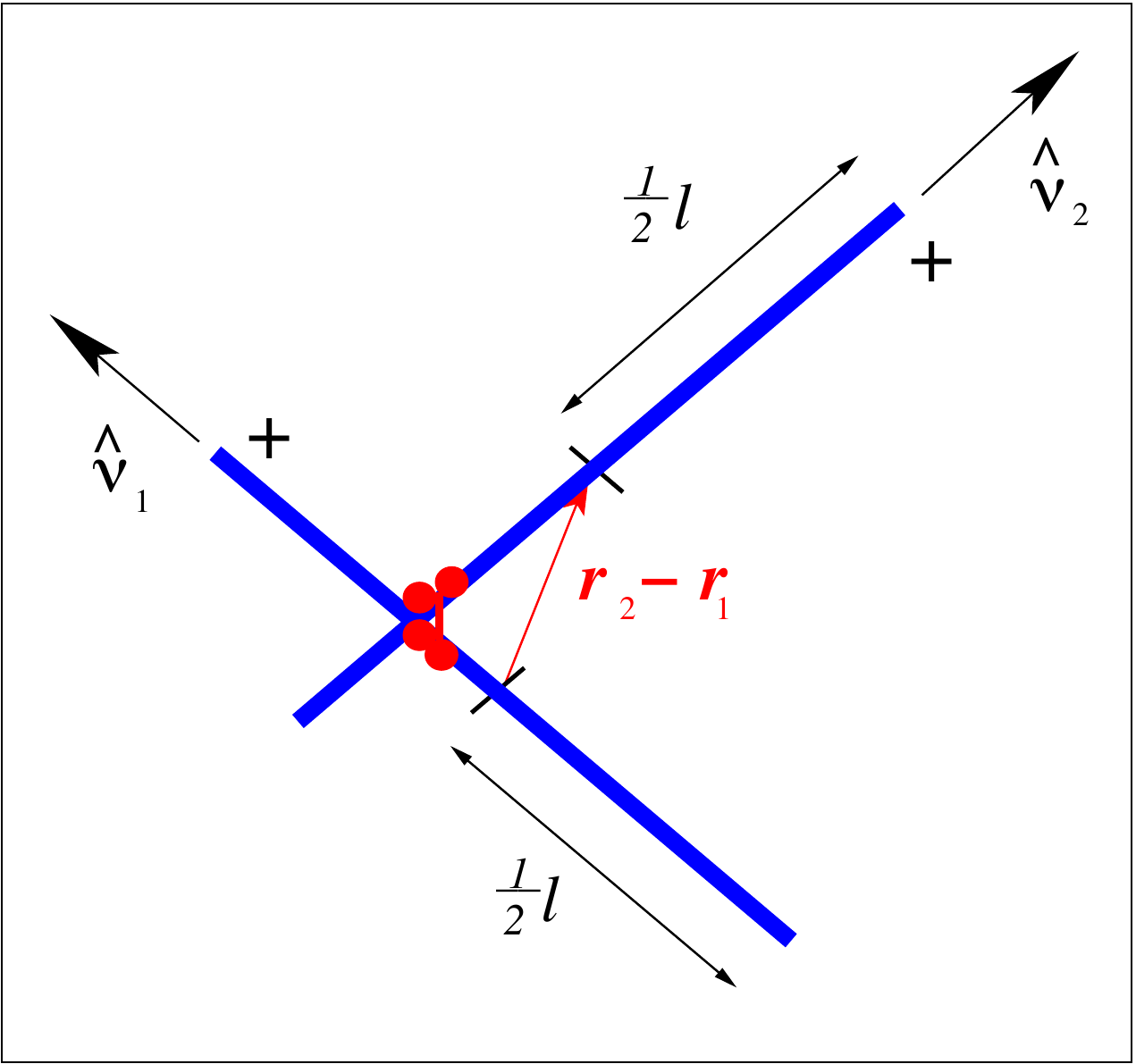}
\caption{(color online) The figure shows a schematic of a pair of filaments of
length $\ell$ with mid-point positions and orientations 
$\bfr_1,\bnu_1$ and $\bfr_2,\bnu_2$ driven by an active cross-link.}
\label{fig:motor-2filam}
\end{figure}

Another microscopic realization of active matter is provided by suspensions of
polar protein 
filaments cross-linked by active cross-links consisting of clusters of molecular
motors 
\citep{Nedelec1997,Nedelec1998,Surrey2001,Backouche2006,Mizuno2007}. The
motivation
for this \emph{in vitro} work is to provide an understanding of the mechanics of
the eukaryotic cell cytoskeleton  from the 
bottom up. Molecular motors are proteins that are able to  convert stored
chemical energy  
into mechanical work by hydrolyzing ATP molecules. The 
mechanical work is done by the motors moving in a uni-directional manner  along
the polar 
filaments. Particular motor proteins are associated with specific polar
filaments, e.g. kinesins  
`walk' on microtubules while myosins `walk' on filamentous actin.  Since the
filaments are 
`polar, they can be characterized by a position and an orientation. Some
authors~
\citep{Kruse2000,Kruse2003,Kruse2006b,Liverpool2003,Liverpool2005,Ahmadi2006}
have modeled the motor
clusters as active cross linkers (see schematic in Fig.~\ref{fig:motor-2filam})
capable of walking along the filaments and exchange forces and torques among
filament pairs, hence yielding additional (active) contributions to the
translational and 
rotational velocities of filaments. Another model has been proposed by Aranson
and coworkers~\citep{Aranson2005,Aranson2006} who have described the filament
dynamics via a stochastic master equations for polar rigid rods interacting via
instantaneous inelastic ``collisions".  Both models  express the stochastic
dynamics of the active system in terms of the one particle density of
filaments: 
$c(\bfr,\bnu,t)$. The kinetic equation  for $c(\bfr,\bnu,t)$  takes account of
the 
interplay of diffusion and  active contributions induced by the cross linking
motor clusters or the inelastic collisions. The kinetic equations can again be
further coarse-grained by projecting on the continuum fields,  such as 
$\rho(\bfr,t)$ and $\bfp(\bfr,t)$ to obtain the equivalent of Eq. \ref{P-eq-2}
where the parameters 
$\alpha,\beta,K,v_1,\lambda_{1,2,3}$ can again be related to microscopic
parameters
characterizing the filaments and motor clusters. Although the parameter values
depend on the microscopic model, the continuum equations obtained by these two
approaches have the same structure. The work by ~\citep{Aranson2005,Aranson2006}
incorporates, however, terms of higher order in the gradients of the continuum
fields neglected in \citep{Kruse2000,Kruse2003,Liverpool2003,Ahmadi2006}. These
models give rise to
contracted states and propagating density/polarization waves in
both one
\citep{Kruse2000,Kruse2001} and  higher
dimensions~\citep{Liverpool2003,Liverpool2005}, as well as aster and spiral
patterns~\citep{Aranson2005,Aranson2006}.
Finally, a number of other mean-field implementations of the dynamics of motor
filament suspensions, some including explicitly motor dynamics or additional
passive cross linkers, have been used by other authors to understand pattern
formation in these
systems~\citep{Sankararaman2004,Lee2001,Ziebert2004,Ziebert2007,Ziebert2005}.

In the next sections we briefly outline  one  procedure for going from the
fluctuating 
microscopic dynamics to the macroscopic equations of motion. 
We restrict our discussion to dry systems. For wet systems we refer the reader
to published work  \citep{Baskaran2009,Leoni2010,Liverpool2006,Marchetti2007}.

\subsection{From stochastic dynamics to macroscopic equations}

We consider a collection of $N$ identical  (a simplification)  anisotropic
particles, each 
described by a position $\bfr_n(t)$ and orientation $\bnu_n(t)$, for particle
$n$.
 The microscopic stochastic  equations of motion for the positions and
orientations are given by
\begin{subequations}
\label{micro-dyn}
\begin{gather}
\partial_t \bfr_n ={ \bf v}_{n}(\bfr_n,\bnu_n)  + {\bm \xi}_n (t)\;, \\
\partial_t \bnu_n = \bomega_i(\bfr_n,\bnu_n) \times \bnu_n   + {\bTheta}_n
(t)\;.
\end{gather}
\end{subequations}
The first terms on the rhs  are the deterministic contributions to the motion
arising from {\em 
both} passive interactions (e.g., steric repulsion, attractive interactions,
etc.) and active 
velocities. Their specific form  depends on the particular model system
considered.
The second terms on the rhs of Eqs.~\eqref{micro-dyn} are stochastic forces
arising from a 
variety of noise sources, including but not limited to thermal noise. They are
assumed to be  
Gaussian and white, with zero mean and correlations 
\begin{subequations}
\label{micro-noise}
\begin{gather}
 \aver{ \xi_{n\alpha}(t) \xi_{m\beta} (t')} = 2\Delta_{\alpha\beta}(\bnu_n) \,
\delta_{nm}  \, \delta (t-t') \;, \\
 \aver{ \Theta_{n\alpha}(t) \Theta_{m\beta} (t')} = 2\Delta_R
 \,  \delta_{nm} \,  \delta_{\alpha\beta} \, \delta (t-t') \; .
\end{gather}
\end{subequations}
The translational noise correlation tensor is of the form $\bm\Delta(\bnu)=
\Delta_\| \bnu \bnu + 
\Delta_\perp \left( \bdelta -\bnu\bnu\right)$. In a thermal system
$\Delta_\|=D_\|$ and $\Delta_\perp=D_\perp$,
with $D_\|$ and $D_\perp$ describing Brownian diffusion along the long direction
of the particle and 
normal to it, respectively. Also in this case $\Delta_R=D_R$, with $D_R$ is the
rotational diffusion rate.
In an active system the noise strengths are  in general an independent
quantities.

For particles with fixed self-propulsion speed $v_0$ along their long axis the
deterministic 
part of the velocities in Eqs.~\eqref{micro-dyn} has the form
\begin{subequations}
\label{drift}
\begin{gather}
{ \bf
v}_{n}(\bfr_n,\bnu_n)=v_0\bnu_n+[\bm\zeta(\bnu_n)]^{-1}\cdot\sum_m\mathbf{f}
(\bfr_n,\bfr_m;\bnu_n, \bnu_m)\;,\\
\bomega_n(\bfr_n,\bnu_n)=\zeta_R^{-1}\sum_m\bm\tau(\bfr_n,\bfr_m;\bnu_n,
\bnu_m)\;,
\end{gather}
\end{subequations}
with $\bm\zeta(\bnu)=\zeta_\|\bnu\bnu+\zeta_\perp(\bm\delta-\bnu\bnu)$  a
friction tensor and $\zeta^R$ a rotational friction coefficient.
Again, in a Brownian system in thermal equilibrium at temperature $T$ friction
and diffusion (which in this case is also the strength of the noise) are related
by the Stokes-Einstein relation,
$D_{\alpha\beta}=k_BT\left[\bm\zeta^{-1}\right]_{\alpha\beta}$ and
$D_R=k_BT/\zeta_R$. In active systems these relations do not in general hold and
noise and friction should be teated as independent.

The pairwise forces and torques in Eqs.~\eqref{drift} can be expressed in term
of the total passive and 
active interactions in the system.
The case of particles propelled by internal torques has also been considered in
the literature, 
but will not be discussed here~\citep{Fily2012}.

\subsubsection{Smoluchowski dynamics}
\label{Smoluchowski}

For simplicity we again restrict ourselves to the case of  particles with
overdamped 
dynamics.  Starting from the Langevin equations \eqref{micro-dyn} for the
individual 
filaments, standard techniques can be used to derive an equation for the
one-particle probability distribution function~\citep{Zwanzig2001}.
First one obtains an infinite hierarchy of equations for the $M$ ($\le 
N)$
particle distribution functions, $c_M(\bfr_1, \bnu_1, \bfr_2,\bnu_2, \ldots,
\bfr_M, 
\bnu_M,t)$, the probability of finding $M$ particles  with positions and
orientations $\{ 
\bfr_1, \bnu_1, \bfr_2,\bnu_2, \ldots, \bfr_M, \bnu_M\}$ at time $t$, regardless
of the 
position and orientation of the other $N-M$ particles~\citep{Baskaran2010}.  A
closure approximation is then
required 
to truncate the hierarchy. This is implemented  by expressing a higher order
distribution 
function in terms of the lower ones. The traditional method is to  write the
2-particle 
distribution function as a product of two 1-particle 
densities~\citep{Kruse2000,Liverpool2003,Aranson2005,Bertin2009}, as in the
familiar molecular chaos approximation used to obtain the Boltzmann equation.

This procedure gives a nonlinear  equation for the 1-particle density, 
$ c(\bfr,\bnu,t) =   \sum_{n} \aver{ \delta (\bfr-\bfr_n)
\delta(\bnu-\bnu_n)}$, where 
the bracket denotes  a trace over all other degrees of freedom and an average
over the noise, 
in the form of a conservation law, given by
 \begin{equation}
\partial_t c
+ \bnabla \cdot {\bf
{J}}_c + {\bf {\cal {R}}} \cdot {\cal {J}}_c =0\;, \label{smoluchowski}
\end{equation}
where ${\bf{\cal {R}}}=\bnu\times\partial_{\bnu}$ is the rotation
operator. The {\em translational} probability current, ${\bf
{J}}_c (\bfr, \bnu, t)$, and the {\em rotational} probability current,
$\bm{\mathcal {{J}}}_c (\bfr, \bnu, t)$, are given by \citep{Doi1986}
\beq
 {\bf J}_c =  \bfv \, c- \bm\Delta \cdot  \nabla c , \left. \right. \left.
\right.  
 \bm{\mathcal{J}}_c  = \bomega c- \Delta_R \bm{\mathcal{R} } c ,  \nonumber
\eeq
where $\bfv$ and $\bomega$ are given in Eqs.~\eqref{drift} and $\bm\Delta$ and
$\Delta_R$  are respectively the translational  and rotational noise strengths
introduced in Eqs.~\eqref{micro-noise}.

It should be stressed that this closure scheme may yield different physical
approximations for 
different classes of microscopic dynamics~\citep{Bialke2012,Fily2012a}.
An illustrative example can be found in ~\textcite{Baskaran2010}   which
systematically
derives a kinetic equation for self-propelled hard rods starting with a
microscopic Langevin 
dynamics  that includes inertia by first obtaining a Fokker-Planck equation for
the joint 
probability distribution of both position/orientation and velocities, and
finally derive the 
Smoluchowski limit where friction is large relative to inertia at the level of
the kinetic 
equation by making a non-thermal assumption on the local distribution of
velocities. This 
procedure yields a different Smoluchowski equation  from that  obtained by
simply taking 
the overdamped limit at the level of the Langevin equation.  
While the structure of the hydrodynamic equations obtained by coarse-graining
the one-
particle kinetic equation is the same in both cases (as it is dictated by
symmetry 
considerations), the values of the parameters in the continuum equations and
particularly 
their dependence on $v_0$ and noise strength depend on whether the overdamped
limit is 
taken right at the outset, i.e., at the level of the microscopic dynamics,  or
at the level of the 
kinetic equation. Much work remains to be done to understand this subtle point
and 
therefore we will not discuss it further in this review. 
A further interesting open question is how to systematically obtain a stochastic
equation for 
the 1-particle density which  properly includes the noisy dynamics of the
density fluctuations 
for active systems~\citep{Dean1996}.

We leave some of these very interesting open questions for the reader and
discuss  below 
how to use  the Smoluchowski dynamics as described by Eqs.~\eqref{smoluchowski}
in the 
derivation of hydrodynamic equations.

\begin{figure}
\centering
\includegraphics[width=0.80\columnwidth]{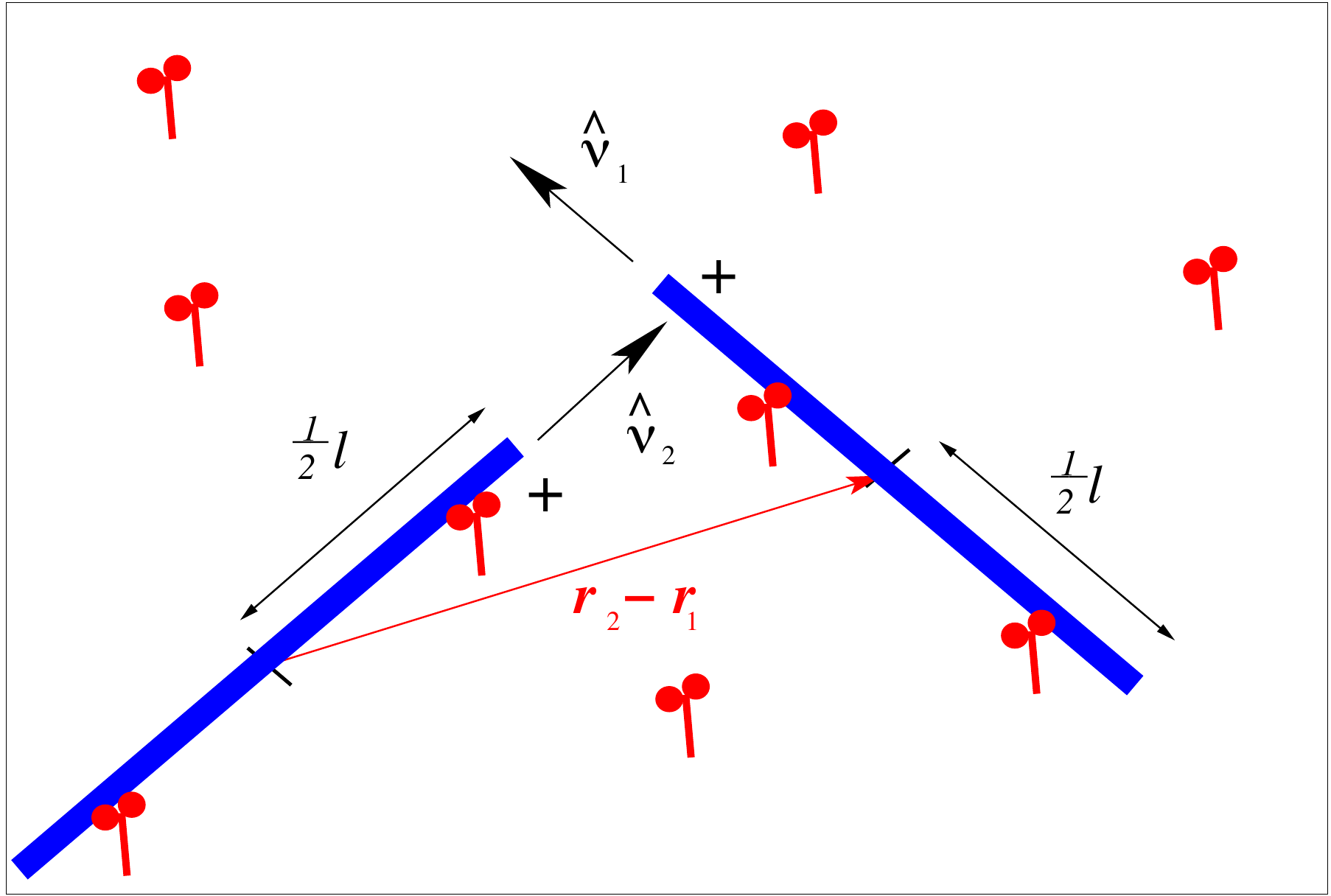}
\caption{(color online) An example of self-propelled particles:  cartoon of a
motility assay with  
filaments of length $\ell$ with positions and orientations $\bfr_1,\bnu_1$ and
$\bfr_2,\bnu_2$ 
driven by motors  tethered to a plane.}
\label{fig:filament-motor-spp}
\end{figure}

\subsubsection{From Smoluchowski to hydrodynamics}
\label{Smol-hydro}

One approach, explored extensively by Saintillan, Shelley and
collaborators~\citep{Saintillan2007,Saintillan2008,Saintillan2008a}, is to solve
directly the Smoluchowski equations, either analytically or numerically. This
work has been used to investigate the stability of both aligned and isotropic
suspensions of active particles with hydrodynamic interactions (see also below)
and has revealed a rich dynamics with strong density fluctuations. This approach
will not be discussed further here. A recent review can be found
in~\textcite{Saintillan2012}.

Here we focus instead  on obtaining the description of the dynamics of the
system in terms of a few macroscopic fields introduced phenomenologically in the
first part of this review. A crucial assumption in deriving this continuum or
hydrodynamic theory is the choice of the continuum fields as those whose
fluctuations are long lived on large length scales. They include fields
associated with conserved quantities and possible broken symmetries of the
system. In a collection of active particles with 
no
momentum conservation, the only conserved quantity is the number
of particles and hence the density, defined in
Eq.~\eqref{rho} is a slow variable. In addition, to allow for the possibility
of broken
orientational order, with either polar or nematic symmetry, we consider the
dynamics of a polarization field (Eq.~\eqref{p}) and an alignment tensor
(Eq.~\eqref{Q}). The fields are defined as moments of the one particle
distribution function as.
\begin{subequations}
 \label{orient_moments}
\begin{gather}
\rho(\bfr,t)=\int d\bnu ~c(\bfr,\bnu,t)\;,\\
\rho (\bfr,t){\bf p}(\bfr,t)=\int d\bnu ~\bnu~c(\bfr,\bnu,t)\;,\\
\rho  (\bfr,t)\bsf{Q}(\bfr,t)=\int d\bnu
~\mathbf{\hat{Q}}(\bnu)~c(\bfr,\bnu,t)\;.
\end{gather}
\end{subequations}
with $\mathbf{\hat{Q}}(\bnu)=\bnu\bnu-\frac{1}{d}\mathbf{1}$ and $d$ the
system's dimensionality. In general one could write an exact expansion of
$c(\bfr,\bnu,t)$ in terms of all its moments and transform the Smoluchowski
equation into an infinite hierarchy of equations for the moments. The
hydrodynamic description is obtained by assuming that higher order moments relax
quickly on the time scales of interest and by truncating this expansion to
include only conserved quantities and order parameter fields.  Equivalently,
considering for simplicity the case of $d=2$, one can simply write
\begin{equation}
c(\bfr,\bnu,t)=\frac{ \rho (\bfr,t)}{2 \pi}\Big\{1+2 {\bf
p}(\bfr,t)\cdot\bnu+4\bsf{Q}(\bfr,t):\mathbf{\hat{Q}}(\bnu) \Big\}\;,
\label{tens_expand}\end{equation}
insert this ansatz into the Smoluchowski  equation, Eq.~(\ref{smoluchowski}),
and obtain
the hydrodynamic equations for filament concentration,
polarization and alignment tensor. 
For the details of the calculation, which also involves  using a small
gradient expansion for the filament probability distribution and
evaluating angular averages, we refer the reader to the literature
\citep{Liverpool2003,Bertin2009,Ihle2011,Ahmadi2006,Baskaran2008a}.
The result is a set of coupled equations for the hydrodynamic variables, $\rho
(\bfr,t), \bfp (\bfr,t), {\bf Q}(\bfr,t)$ that has the form given in section
\ref{dry}, with parameter values that are calculated explicitly in terms of
microscopic properties of the system.

In particular, one finds that the continuum equations governing the dynamics of
motor-filament systems and collections of self-propelled particles have the same
structure, but important qualitative differences arise in the hydrodynamic
parameters. In  models  of SPP the activating internal processes drive each
individual unit (as for instance in bacterial suspensions or actin filaments in
motility assays, where the filaments are driven by the action of a carpet of
myosins tethered to a plane) and  active contributions to the dynamics arise
even at the single-particle level.  In cell extracts of cytoskeletal filaments
and associated motor proteins  activity arises from interactions among the
filaments mediated by motor clusters that act as active crosslinkers, exchanging
forces among the filaments.  As a result, the various active parameters in the
hydrodynamic equations are proportional to the square of the density of
filaments, resulting in different behavior at large scales. 
For this reason motor-filament suspensions have also been referred to as systems
of ``mutually propelled particles" (MPP) \citep{Giomi2012a}, to
distinguish them from SPPs.

\subsubsection{An example: derivation of continuum equations for aligning
Vicsek-type particles}
\label{micro-example}

To illustrate the method, we show in this section the derivation of the
continuum equations for a simple model of self-propelled point particles  on a
substrate in two dimensions with a polar angular interaction that tends to align
particles as in the Vicsek model. We consider $N$ particles with polarity
defined by an axis $\hat{\bm\nu}_n=(\cos\theta_n,\sin\theta_n)$. The
dynamics is governed by Eqs.~\eqref{micro-dyn} that now take the explicit form
\begin{subequations}
\begin{gather}
\label{micro-Vicsek}
\partial_t{\bf r}_n=v_0\hat{\bm\nu}_n-\zeta^{-1}\sum_{m}\frac{\partial
V}{\partial{\bf r}_n}+\bm\xi_n(t)\;,\\
\partial_t\theta_n=\zeta^{-1}_R\sum_{m}\frac{\partial
V}{\partial\theta_n}+\Theta_n(t)\;,
\end{gather}
\end{subequations}
where $v_0$ is the fixed self-propulsion speed and  we have taken
$\zeta_{\alpha\beta}=\zeta\delta_{\alpha\beta}$. The noise  is Gaussian and
white, with zero mean, as given in Eqs.~\eqref{micro-noise} and
$\Delta_{\alpha\beta}=\Delta\delta_{\alpha\beta}$. Forces and torques in
Eqs.~\eqref{micro-Vicsek} are expressed here as derivatives of a pair potential,
given by $\zeta_R^{-1}V(x_n,x_m)=-\frac{\gamma}{\pi R^2}\Theta(R-|{\bf r}_n-{\bf
r}_m|)\cos(\theta_m-\theta_n)$,   that tends to align particles of the same
polarity, with $\Theta(x)$ the Heavside step function and $R$ the range of the
interaction. Here we use the shorthand  $x_n=({\bf r}_n,\theta_n)$ and $\gamma$
is the  interaction strength with dimensions of inverse time.   As we are
interested in a hydrodynamic model that describes spatial variations on length
scales large compared to the range $R$ of the interaction, we assume the
interaction to be local. This polar aligning interaction was used recently in
\textcite{Farrell2012} in a  related model that highlights the role of caging
effects. As outlined in section \ref{Smoluchowski}, one can then use standard
methods~\citep{Zwanzig2001} to transform the coupled Langevin equations into a
Smoluchowski equation for the one-particle distribution $c({\bf r},\theta,t)$,
given by
\begin{widetext}
\begin{equation}
\partial_tc+v_0{\bm\nu}_1\cdot\bm\nabla_1 c=\Delta\nabla_1^2
c+\Delta_R\partial_{\theta_1}^2 c
+\frac{1}{\zeta}\bm\nabla_1\cdot c(x_1,t)\int_{x_2}\bm\nabla_1
V(x_1,x_2)c(x_2,t)
+\Delta_R\partial_{\theta_1}^2c
+\frac{1}{\zeta_R}\partial_{\theta_1}c(x_1,t)\int_{x_2}\partial_{\theta_1}
V(x_1,x_2)c(x_2,t)
\end{equation}
\end{widetext}
To obtain hydrodynamic equations, we now proceed as outlined in section
\ref{Smol-hydro} and transform the kinetic equation into a hierarchy of
equations for the angular moments of the one-particle probability density, $c$.
In two dimensions the moments are most simply written by introducing an angular
Fourier transform~\citep{Bertin2009}   in terms of the Fourier components
$f_k({\bf r},t)=\int_0^{2\pi}
c({\bf r},\theta,t) e^{i k \theta} d\theta$. By comparing with
Eqs.~\eqref{orient_moments} it is easy to see that
$f_0=\rho$, $f_1=w_x+iw_y$, with ${\bf w}=\rho{\bf p}$ the polarization density,
and the real and imaginary parts of $f_2$ are proportional to the two
independent components of the alignment tensor, ${\bf Q}$. Using $2\pi c({\bf
r},\theta,t)=\sum_k f_k e^{-ik\theta}$ and retaining for simplicity only terms
up to linear order in the gradients,
 we obtain a
hierarchy of equations
\begin{equation}
  \label{eqn:FT}
  \begin{aligned}
    &\partial_t f_k +  \frac{v_0}{2}\partial_x( f_{k+1} +  f_{k-1})  +
\frac{v_0}{2i}\partial_y( f_{k+1} -  f_{k-1})  \\  
& = -k^2\Delta_R f_k +  \frac{ik\gamma}{2\pi} \sum_q f_q V_{-q} f_{k-q}+{\cal
O}(\nabla^2)
  \end{aligned}
\end{equation}
where  all discrete sums run from $-\infty$ to $+\infty$ and 
$V_q=\int d\theta
e^{iq\theta}\sin\theta=i\pi\left(\delta_{q,1}-\delta_{q,-1}\right)$.
We consider equations for $f_0$ and $f_1$, assume that $f_2$ is a fast variable,
so the $\partial_t f_2\simeq 0$, and neglect all higher order Fourier
components, i. e., assume $f_k=0$ for $k\ge 3$. Eliminating $f_2$ in favor of
$f_0$ and $f_1$, it is then straightforward to obtain~\citep{Farrell2012}
\begin{widetext}
\begin{subequations}
\begin{gather}
\label{micro-hyd}
\partial_t\rho+v_0\bm\nabla\cdot{\bf w}=0\;,\\
\partial_t{\bf w}+\frac{3 v_0\gamma}{16\Delta_R}({\bf w}\cdot\bm\nabla){\bf
w}=\left(\frac12\gamma\rho-\Delta_R\right){\bf w}
-\frac{\gamma^2}{8\Delta_R}w^2{\bf
w}-\frac{v_0}{2}\bm\nabla\rho+\frac{5v_0\gamma}{32\Delta_R}\bm\nabla w^2
-\frac{5v_0\gamma}{16\Delta_R}{\bf w}(\bm\nabla\cdot{\bf w})+{\cal O}(\nabla^2)
\end{gather}
\end{subequations}
\end{widetext}
Microscopic derivations of continuum equations of the type presented here
naturally yield an equation for the \emph{polarization density} ${\bf
w}=\rho{\bf p}$ rather than an equation for the \emph{order parameter field}
${\bf p}$ as given in Eq.~\eqref{P-eq-2}. Of course it is straightforward to use
the density equation to transform the equation for ${\bf w}$ into an equation
for ${\bf p}$, with the result,
\begin{eqnarray}
\label{P-eq-micro}
\partial_t{\bf p}+&&\lambda_1({\bf p}\cdot\bm\nabla){\bf
p}=-\left[a(\rho)+\beta|{\bf p}|^2\right]{\bf p}-{\bf
v}_1\cdot\bm\nabla\rho\notag\\
&&+\frac{\lambda_3}{2}\bm\nabla |{\bf p}|^2
+\lambda_2{\bf p}(\bm\nabla\cdot{\bf p})+{\cal O}(\nabla^2)\;,
\end{eqnarray}
with
\begin{subequations}
\begin{gather}
a(\rho)=\Delta_R-\frac12\gamma\rho\;,\\
\beta=\frac{\gamma^2\rho^2}{8\Delta_R}\;,\\
\lambda_1=\frac{3v_0\gamma\rho}{16\Delta_R}\;,\\
\lambda_2=-\lambda_3=-\frac{5v_0\gamma\rho}{16\Delta_R}\;,\\
{\bf
v}_1=\frac{v_0}{2\rho}\left(1-\frac{5\gamma\rho}{8\Delta_R}p^2\right)\bm\delta
-\frac{v_0}{\rho}\left(1-\frac{\gamma\rho}{2\Delta_R}\right){\bf p}{\bf p}\;.
\end{gather}
\end{subequations}
Eq.~\eqref{P-eq-micro} has the same structure as Eq.~\eqref{P-eq-2}, although
noise has been neglected here. The various coefficients are expressed in terms
of microscopic parameters and are in general found to depend on density and
order parameter. In particular, the coefficient ${\bf v}_1$ of the
$\bm\nabla\rho$ term is now a tensor, describing anisotropic pressure gradients
that can play a role in the ordered state. These effects were neglected for
simplicity in the phenomenological model.
We also stress that there is an important difference between the parameter
values obtained in the present model and those obtained by \textcite{Bertin2009,Baskaran2008a} as in both the latter models the effective interaction
strength (here $\gamma$) depends linearly on $v_0$.  As a result, both
~\textcite{Bertin2009,Baskaran2008a} find that $\lambda_i\sim v_0^2$,
while here $\lambda_i\sim v_0$. This difference arises because in
\textcite{Bertin2009,Baskaran2008a} the authors systematically describe binary
collisions, yielding an instantaneous change of direction with probability one
in a small time interval, while the model of \textcite{Farrell2012} presented here
(and chosen for illustrative purpose because of its simplicity) considers a
continuous evolution of the position due to the application of a force in the
Langevin equation. These differences may of course be important
for large values
of $v_0$.

This example shows how the derivation of hydrodynamics from a microscopic model
yields explicit values for the various parameters in the continuum equations in
terms of microscopic parameters. Of course, as stressed above, these values are
model dependent. In addition, there is a price to be paid in that in the
derivation one has made two important assumptions. The first is the assumption
of low density, that allows us to replace the two-particle probability
distribution by the product of two one-particle distribution functions in the
kinetic equation.  The second is effectively an assumption of weak interaction
that enters in the moment closure approximation. The two assumptions are not
independent and essentially amount to the so-called assumption of ``molecular
chaos" used for instance in the derivation of the familiar Boltzmann equation.

\subsubsection{Hydrodynamic interactions}
\label{Hydro-int}

The examples presented above ignore the momentum exchange between the active
particles and the solvent that one would have for instance in 
suspensions of  swimming organisms. 
This physical effect is important for understanding the properties of a wide
variety of experiments and should be included for a complete 
microscopic description of active matter.

The coupling to the solvent has been  treated in the literature in at least two
ways.

One approach is to include an additional equation for the solvent velocity field
$\bfv(\bfr,t)$
which is coupled to the microscopic equations of motion
\citep{Liverpool2006,Marchetti2007} or to the corresponding Smoluchowski
equation
~\citep{Saintillan2008,Pahlavan2011}.   The active particles are convected and
rotated by the local  fluid flow while also imparting  additional forces
(stresses) onto  the fluid.   This gives rise to a coupled equation of motion
for the concentration of filaments $c(\bfr,\bnu,t)$ and the velocity field
$\bfv(\bfr,t)$ which can be solved directly~\citep{Saintillan2008,Pahlavan2011}
or coarse-grained as described above to give coupled equations for
$\rho(\bfr,t),\bfp(\bfr,t),\bsf{Q}(\bfr,t),\bfv(\bfr,t)$. For the case where the
suspension is incompressible ($\rho={\rm constant}$), the momentum ($\bfv$)
equation has the form
$\rho( \partial_t +\bfv\cdot\bm\nabla)\bfv=\bm\nabla\cdot\bm\sigma$.  This
equation may also be often treated in the Stokes approximation,
$\bm\nabla\cdot\bm\sigma=0$, appropriate for systems at low Reynolds number. The
active forces due to the swimmers are incorporated as an additional (active)
contribution to the stress tensor, so that
$\sigma_{\alpha\beta}=\sigma_{\alpha\beta}^{p}+\sigma_{\alpha\beta}^a$, where
the passive contribution $\sigma_{\alpha\beta}^p$ has the form
obtained in
equilibrium liquid crystals~\citep{Gennes1993}.  The active contribution
$\sigma_{\alpha\beta}^a$ is evaluated by noting that in the absence of external
body  forces, the force distribution exerted by the active particles on the
fluid can be written as  a multipole expansion with lowest order
non vanishing
term being a dipole.  The active stress $\sigma^a_{\alpha\beta}$   can then be
identified with the active reactive fluxes proportional to $\zeta \Delta \mu$
(see Eq. \ref{eq:reac-stress}). To leading order in a gradient expansion it
contains two additive contributions. The first, $\sigma^{a1}_{\alpha\beta}
\propto \zeta\Delta\mu q_{\alpha\beta}$, has nematic symmetry  and is present
in both polar and nematic systems, where it is written as
$\sigma^{a1}_{\alpha\beta} \propto \zeta\Delta\mu Q_{\alpha\beta}$.  In
addition, for polar systems only, the active stress also contains terms
$\sigma^{a2}_{\alpha\beta} \propto  \zeta'\Delta\mu\partial_\alpha p_\beta
+ \zeta''\Delta\mu \partial_\beta p_\alpha$~\citep{Liverpool2006,Marchetti2007}. When the
hydrodynamic equations are written on the basis of the entropy production
formulation discussed in section \ref{nematicgels}, this term is discarded as
one of higher order in the driving forces. It is also of higher order in the
gradients as compared to $\sigma_{\alpha\beta}^{a1}$, although only of first
order, rather than quadratic, in the polarization field.
Finally, it is the leading non-vanishing contribution to the active stress that
has polar (as opposed to nematic) symmetry and it does play a role in
controlling the onset of oscillatory states in these
systems~\citep{Giomi2008,Giomi2012}.  This shows that a microscopic derivation
is needed to go beyond linear hydrodynamics.

Alternatively, one may integrate out the solvent velocity field to generate
effective hydrodynamic interactions at the two-body level which give rise to
additional long-range pair-wise contribution to the deterministic velocities and
angular velocities 
$\bfv_n,\bomega_n$ 
in Eqs.~ \eqref{micro-dyn} \citep{Baskaran2009,Leoni2010}.  These interactions
arise from the forces generated on the fluid by the active particles. The local
force distribution due to the active elements can be expanded in a multipole
expansion. The lowest non vanishing term in this expansion is a dipole, which
also gives  the longest range interactions.   The simplest static models of
swimming organisms then are force dipoles which can be characterized as
contractile or extensile depending on
the direction of the forces making up the dipole. Coarse-graining the equations
leads  to a set of coupled integral equations for
$\rho(\bfr,t),\bfp(\bfr,t),\bsf{Q}(\bfr,t)$ which in  extended (large) domains 
can be used to obtain generalized linear hydrodynamic modes whose behavior is
identical to that obtained when keeping an explicit velocity field. 
  Hydrodynamic equations for collections of static force dipoles have been 
obtained \citep{Baskaran2009}  which show all the phenomenology described
before.
Swimming objects however are dynamic, undergoing an internal cyclical motion
that leads to self-propulsion.
The collective behavior of 
simplified dynamic models of swimmers with such internal cycles has also been
studied and 
used to generate effective hydrodynamic equations~\citep{Leoni2010}.
Averaged over an internal cycle such dynamic models have an average force
distribution that also has a multipolar expansion whose lowest term is
generically a force dipole.
The hydrodynamic equations obtained are then similar to those obtained for
static dipoles.
However by tuning internal parameters one can also study self-propelled swimmers
whose averaged force distribution starts at quadrupolar order. 
Here the symmetry broken between contractile and extensile
objects is restored and
only nematic phases are possible~\citep{Leoni2010,Leoni2012}.

\subsection{Current Status of Microscopic Theories of Active Matter}
Perhaps the most important open question in deriving active theories from
microscopic or mesoscopic models concerns understanding the nature of the
noise. 
Thermal noise is often negligible in active systems at low frequency, but noise is nonetheless
ubiquitous and plays a crucial role in controlling the large scale behavior, as
it can both act to destroy large scale coherence or drive synchronization,
depending on the specific situation.  So far practically all work on active
systems has either neglected noise or modeled it as a Gaussian, white random
force, akin to thermal noise, but of unknown strength. In general one expects
the noise to depend on activity.  Multiplicative or non-Markovian random forces
could also be at play in active systems. For instance,
the multiplicative character of the noise introduces \citep{Mishra2012} singular
features into the fluctuation-driven phase-separation \citep{Mishra2006} that
characterizes a stable active nematic. Microscopic models are needed that
attempt to take into
account the stochastic nature of active forces and to derive a coarse-grained
model
where the effective noise amplitude is expressed in terms of local
nonequilibrium processes that may lead to temporal or spatial correlations at
large
scales. An example is the  work of ~\textcite{Lacoste2005} that couples the shot
noise associated with the on/off switching of energy-dissipating  pumps in
active membranes to the membrane curvature and  demonstrates that pump
stochasticity plays a crucial role in controlling super diffusive behavior in
the membrane.

Of great interest is also the detailed understanding of hydrodynamic
interactions in bacterial suspensions and their role in
controlling the
large-scale behavior. Recent experiments have probed for the first time the
flow fields induced by swimming unicellular organisms. Measurements on
Chlamydomonas have  revealed qualitative differences as compared to the 
puller stresslet configuration that had been used in the
literature~\citep{Drescher2010} and have further shown complex time-dependent oscillatory patterns.
In contrast it was found that the flow field of E.coli is well described by a
pusher stresslet, but its strength is very small and it is washed out by
rotational diffusion of the swimming direction~\citep{Drescher2011}. These
results have important implications for the behavior of microorganisms near
surfaces and open the way to new quantitative investigations of the role of
hydrodynamic interactions.

Finally, an area that is receiving increasing attention is the study of active
matter at high density,  where active glassy or solid states may emerge. As
mentioned in section~\ref{subsec:current-dry}
the collective dynamics of  confluent layers of epithelial cells has been
likened to that of glassy and supercooled systems. Mesoscopic models of
interacting cell layers and tissues are beginning to emerge~\citep{Henkes2011},
although much more work remains to be done to understand the complex interplay
of contractile stresses and substrate adhesion in controlling the build up of
cellular stresses in collective cell migration.

\section{Conclusions, Outlook and Future Directions}
\label{conclusions}

In this review we have discussed two dimensional and three dimensional active
systems,
that are maintained out of equilibrium by a permanent energy consumption that
takes place 
locally in each active unit. 
In these systems local polarity, when present, yields  spontaneous motion of the
polar  entities
with respect to the center of mass motion.
Typically, models of active matter describe the collective behavior of 
 systems like fish
shoals, birds flock or animals herds as well as vibrated granular matter,
bacterial colonies or the cellular cytoskeleton.
From a spatial symmetry point of view these systems are not different from
ferroelectric and nematic liquid crystals.
The fundamental difference stems from the fact that active systems consume and
dissipate energy at all times.
This feature has a number of profound consequences which we have addressed in
this review:
\begin{itemize}
\item Polar flocks on a substrate, e.g., animal herds, keratocytes on a
glass slide, display long-range order in two dimensions. This behavior is fundamentally different form that of equilibrium systems
that are bound to obey the Mermin-Wagner theorem and can at best exhibit
quasi-long range order;
\item nematic as well as polar ordered phases systems on a substrate
exhibit giant density fluctuations, breaking the familiar $\sqrt{N}$ equilibrium
scaling of number fluctuations in subregions with $N$ particles on average; 
\item the uniform ordered state of bulk momentum-conserving systems of polar and
nematic symmetry is generically unstable in the Stokesian regime; 
\item the end result of this instability for the typical active
suspension appears to be turbulence at low Reynolds number
\citep{Dombrowski2004,Wolgemuth2008} driven by the competition between forcing
by
active stress and relaxation by orientational diffusion.  
\item all these systems can support a new type of 
sound-like propagating waves
that can have different propagation laws in opposite directions due to
polarity\end{itemize}

We have described how
macroscopic hydrodynamic equations 
for active systems can be obtained either from symmetry arguments,
from generalized thermodynamics close to equilibrium, or from microscopic models.
In each 
case we have presented 
the simplest description that highlights 
the generic features uniquely associated with 
activity.
Our analysis
is restricted to two and three dimensions and we have deliberately left aside 
one-dimensional and quasi 
one-dimensional systems, which have already been investigated
extensively~\citep{Helbing2001a,Reichenbach2006}.
We note, however, that the 
mechanisms leading to the onset of traffic jams 
are probably related to the giant density
fluctuations predicted  in active systems in two dimensions [see also 
\citep{Yang2010,Fily2012}].

The rewarding feature is that all these unusual features have either been
observed 
experimentally
or confirmed in careful 
numerical simulations. It is very difficult to establish the existence of long
range order
experimentally, but simulations with a large number of active particles show
unambiguously
its existence. This is an important result that the statistical physics
community was not 
inclined to believe. 
Giant density fluctuations have now been observed in
simulations~\citep{Chate2008},
experimentally in vibrated granular media~\citep{Narayan2007} and in bacterial
suspensions~\citep{Zhang2010}.
These observations have confirmed the theoretically predicted scaling.  We note,
however, that this scaling has recently been questioned by
~\textcite{Toner2012}.
These large fluctuations  are also present in the cellular cytoskeleton 
and their role is starting to be recognized ~\citep{Gowrishankar2012a}
and their potential physiological relevance is beginning to be
explored~\citep{Goswami2008,Gowrishankar2012,Chaudhuri2011}.

The long
wavelength instability of momentum conserving systems is confirmed by numerical 
simulation and in fact can
give rise to "low Reynolds number turbulence"  in 
some instances~\citep{Dombrowski2004}.  It is a good
paradigm for explaining bacterial swirls but also the rotating microtubule
spindles observed 
both
\emph{in vivo} and \emph{in vitro}~\citep{Nedelec1997,Nedelec1998}.
Furthermore, the existence of
"low Reynolds number" waves have also been predicted in crystals moving without
inertia through a dissipative medium, such as sedimenting colloids~
\citep{Lahiri1997} or drifting
Abrikosov lattices~\citep{Balents1998,Ling1998,Simha1999},  and also
observed and understood in microfluidic drifting drop
arrays~\citep{Beatus2006,Beatus2007}.
Such waves propagating in only one direction have been observed in 
cell lamellipodia, as expected from active gel theory~\citep{Giannone2004}.

Eventually,
one may be able to discuss with order of magnitude accuracy phenomena such as
cell wound healing~\citep{Petitjean2010},
cell division ~\citep{Salbreux2009}, and cell oscillations ~\citep{Salbreux2007}.
One can
thus say that we currently have a good qualitative understanding of active
systems. What do 
we need
to get to the next stage, that is to get to real quantitative understanding? One
ingredient 
which we have 
omitted on purpose up to now is signaling: biochemical signaling for cell
biology or bacterial 
colonies external
factors like food, smell or sun in flocks, shoals and herds. Such factors have
been discussed 
in bacterial colonies
and shown to give rise to beautifully ordered structures \citep{Cates2010}. 
A direction that looks particularly promising, is 
coupling of active gel physics
and chemical reactions in the context of cells and tissues, resulting for instance, in increasing considerably the range of existence of Turing
structures \citep{Bois2011,Giomi2010,Giomi2012a}, and in the spatiotemporal control and enhancement of chemical reaction
rates in signaling and sorting \citep{Chaudhuri2011}.
A complete understanding of cell biology will obviously require including
detailed biochemical 
signaling networks. However this cannot 
be achieved without an extensive experimental input. Symmetry considerations
and 
conservation laws are no longer sufficient and this task 
requires the cooperation of several disciplines. It is worth the effort though,
since it  could be extended to tissue 
dynamics and developmental
biology. 

Another natural extension in the case of herds, flocks and shoal is the 
investigation of a putative leader
role~\citep{Couzin2011,Leonard2012}.
Up to now, completely ``democratic" rules have been assumed. Within this
approach, 
a leader
could appear
like a delta function, and its role involve the response function of the
collection of 
individuals. 
In another promising direction, \textcite{Guttal2010} treat the parameters in a Vicsek-style model of flocking and migration as heritable, selectable attributes, and study their dynamics on evolutionary timescales, finding a remarkable range of evolutionarily stable strategies including coexistence of distinct behaviors in a herd. 

Another frontier concerns
active quantum systems. A beautiful classical mimic has been introduced
recently~\citep{Couder2006} and some
Al-Ga-Se heterostructures give rise to zero resistivity
behavior described by
equations 
very similar to those of Toner-Tu 
\citep{Alicea2005}. A general investigation of the type we present here would be
very useful. 
Eventually working with real systems is necessary and important, but it is often
difficult. 
Artificial systems such as provided by layers of vibrated granular
matter~\citep{Narayan2007,Kudrolli2008,Deseigne2010,Kumar2011},
noisy walkers \citep{Kumar2008},
artificial swimmers~\citep{Dreyfus2005,Bartolo2010} and colloids propelled by
catalytic
reactions~\citep{Paxton2004,Golestanian2005,Golestanian2009,
Howse2007,Palacci2010,Gibbs2011} are very useful but are often
restricted to two dimensions (granular layers) or have been studied only at
relatively low density (artificial self-propelled particles). It would be 
nice to have three dimensional
active systems driven by light or chemical reactions~\citep{Buttinoni2011}.
Attempts have been made~\citep{Prost2010} to raise interest among chemists, but
so far progress towards realizing artificial active gels has been limited.
Such systems  would allow
one to vary parameters
over a wide range and in a controlled manner and would therefore
provide extremely valuable artificial  realizations of active matter.

\begin{acknowledgements}
We would like to thank many colleagues for invaluable discussions and 
suggestions, and collaboration on some of the work summarized
here, including: Aparna Baskaran, Shiladitya
Banerjee, Andrew Callan-Jones, Yaouen Fily, K. Gowrishankar, Yashodhan
Hatwalne, Silke Henkes, Karsten Kruse, Frank J\"ulicher, Norio
Kikuchi, Gautam Menon, Satyajit Major, Narayanan Menon, Shradha Mishra, Vijay Narayan,
Guillaume Salbreux, Sumithra Sankararaman, Ken Sekimoto,
John Toner and Rafael Voituriez.
MCM was support by the National Science Foundation on grants DMR-0806511 and
DMR-1004789.
JFJ, JP, SR and MR acknowledge support from grant 3504-2 of CEFIPRA, the
Indo-French Centre for the Promotion of Advanced Research. 
TBL acknowledges the support of the EPSRC under grant EP/G026440/1.
\end{acknowledgements}

\bibliography{RMP-References.bib}

\end{document}